\documentclass{aa}  

\usepackage{graphicx}
\usepackage{txfonts}
\usepackage{lipsum}
\usepackage{subcaption}         % necessary for continued figures, example in section 3
\usepackage{lscape}             % to rotate a single page table, example in appendix.
\usepackage{placeins}           % useful with \FloatBarrier, to keep 
\usepackage{xcolor}

\usepackage{hyperref}
\hypersetup{
    colorlinks=true,
    linkcolor=blue,
    citecolor=blue,
    urlcolor=blue
}
\begin{document}

   \title{Inferring stellar metallicity and elemental abundances from kinematic and spectroscopic data using machine learning}

   \subtitle{Implications for exoplanet host stars}

\titlerunning{ML abundances}
%%%%%%%%%%%%%%%%%%%%%%%%%%%%%%%%%%%%%%%%
% Please do not include ORCIDs next to author names.
% Only ORCIDs authenticated by individual authors in EDP Sciences editorial system will be taken into account.
% ORCIDs included here will be removed.
%%%%%%%%%%%%%%%%%%%%%%%%%%%%%%%%%%%%%%%%

   \author{V.~Adibekyan\inst{1,2} \and
    B.~M.~T.~B.~Soares\inst{1,2} \and
    S.~G.~Sousa\inst{1,2} \and
    N.~C.~Santos \inst{1,2} \and
    E.~Delgado-Mena \inst{3} \and
    I.~Minchev \inst{4} \and
   R.~Chertovskih \inst{5} \and
   Zh.~Martirosyan\inst{6} \and
   G.~Israelian\inst{7,8} \and
   A.~A.~Hakobyan \inst{9}
        }

  \institute{
  	  Instituto de Astrof\'isica e Ci\^encias do Espa\c{c}o, Universidade do Porto, CAUP, Rua das Estrelas, 4150-762 Porto, Portugal \\
  	  \email{vadibekyan@astro.up.pt} \and
  	  Departamento de F\'{\i}sica e Astronomia, Faculdade de Ci\^encias, Universidade do Porto, Rua do Campo  Alegre, 4169-007 Porto, Portugal \and
      Centro de Astrobiolog\'{\i}a (CAB), CSIC--INTA, Camino Bajo del Castillo s/n, 28692 Villanueva de la Ca\~nada (Madrid), Spain \and
	  Leibniz-Instit\"ut f\"ur Astrophysik Potsdam (AIP), An der Sternwarte 16, D-14482, Potsdam \and
	  Research Center for Systems and Technologies (SYSTEC/ARISE), Faculdade de Engenharia, Universidade do Porto, Rua Dr. Roberto Frias, 4200-465 Porto, Portugal \and
	  Institute of Physics, Yerevan State University, A. Manukyan str. 1, 0025 Yerevan, Armenia \and
	  Instituto de Astrof\'{i}sica de Canarias, E-38205 La Laguna, Tenerife, Spain \and
  	  Departamento de Astrof\`{i}sica, Universidad de La Laguna, E-38206 La Laguna, Tenerife, Spain \and
	  Center for Cosmology and Astrophysics, Alikhanian National Science Laboratory, 2 Alikhanian Brothers Str., 0036 Yerevan, Armenia
  	  }

   \date{Date}

% \abstract{}{}{}{}{}
% 5 {} token are mandatory
 
  \abstract
   {Elemental abundances of FGK stars can be derived routinely from high-resolution optical spectra, but this remains considerably more difficult for cooler stars. Yet, even estimates of the chemical composition of planet-host stars is central to studies of planet formation and interior composition. Machine-learning methods offer a practical route to infer otherwise inaccessible abundances from more widely available stellar data.}
   {We investigate how much information on stellar metallicity and selected elemental abundances is encoded in kinematic and orbital properties, and whether these empirical relations can be used to estimate abundances and abundance ratios relevant for exoplanet-host characterisation.}
   {We use a large APOGEE DR17 sample of red giant stars as the main training set and an independent HARPS sample of nearby FGK dwarfs for external validation. From astrometry and radial velocities we derive Galactic velocities, orbital parameters, and stellar population-membership probabilities. We benchmark several machine-learning regressors, optimise the strongest models, and analyse feature importance using gain-based metrics, permutation importance, single-feature models, and SHAP values. We also explored the prediction of C and O from Mg, Si, and [Fe/H], and derived simple empirical relations between selected abundance ratios (Fe/Si, Mg/Si, C/O, and Fe/O) and metallicity.}
   {Kinematic information alone recovers only a limited fraction of the variance in stellar metallicity, with a clear performance ceiling at RMSE $\sim$0.20 dex. The most informative predictor is the maximum vertical orbital excursion, $Z_{\max}$, followed by radial orbital parameters. When [Fe/H] is combined with kinematic information, the abundances of C, O, Mg, and Si are predicted significantly more accurately than with the baseline approximation $\mathrm{[X/H]}=\mathrm{[Fe/H]}$. In contrast, when predicting C and O from Mg, Si, and [Fe/H], most of the predictive power is already contained in the elemental abundances themselves, with Mg being the dominant contributor, and the addition of kinematic information provides little improvement. The trained models reproduce the main abundance trends associated with Galactic chemical evolution. We find that the slopes of the relations between Fe/Si, Mg/Si, C/O, and Fe/O and metallicity differ slightly between the HARPS and APOGEE samples, with fractional differences generally below 17\%.}
   {Stellar kinematics and orbital parameters contain only limited information on stellar metallicity, with predictive performance primarily constrained by the intrinsic information content of the input features rather than by the choice of model architecture. Nevertheless, when combined with metallicity, these parameters enable accurate predictions of elemental abundances, including those that are difficult to determine directly. This provides a practical framework for extending chemical characterisation to stars for which detailed abundance determinations are challenging.}

 \keywords{stars: abundances -- stars: kinematics and dynamics -- methods: data analysis -- methods: statistical -- planetary systems}

\maketitle
\nolinenumbers

%%%%%%%%%%%%%%%%%%%%%%%%%%%%%%%%%%%%%%%%%%%%%%%%%%%%%%%%%%%%%%
\section{Introduction}

Determining stellar chemical abundances is fundamental for a wide range of astrophysical studies, from Galactic chemical evolution to the formation and characterisation of planetary systems.  In the context of exoplanets, the chemical composition of the host star is of particular interest because it reflects, to first order, the composition of the protoplanetary disk. Stellar metallicity is known to correlate with the occurrence of giant planets \citep{Santos-04, FischerValenti-05, Adibekyan-19}, while abundance ratios such as Mg/Si, Fe/Si, and C/O are relevant for the mineralogy, internal structure, and bulk composition of planets \citep{Bond-10, Dorn-15, Wang-19, Adibekyan-21, DelgadoMena-21}.

Large surveys such as the Apache Point Observatory Galactic Evolution Experiment \citep[APOGEE;][]{Abdurrouf-22} and the GALactic Archaeology with HERMES survey \citep[GALAH;][]{Buder-25} have enabled precise determinations of chemical abundances for hundreds of thousands of stars. However, accurate abundance determinations remain challenging for several stellar regimes and chemical species. In cool stars, particularly at optical wavelengths, spectra are strongly affected by molecular absorption and severe line blending, which complicate both the determination of stellar parameters and detailed abundance analyses \citep{Woolf-05, Passegger-18, Maldonado-20, Marfil-21}. This problem is especially acute for M dwarfs, for which even metallicity determinations remain difficult and are often based on indirect calibrations or limited spectral diagnostics \citep{Neves-12, Rojas-Ayala-12, Mann-13, Antoniadis-Karnavas-24, Duque-Arribas-25}. Even when high-resolution near-infrared spectra are available, multi-element abundance determinations for M dwarfs still require careful modelling and remain limited to relatively small samples \citep{Souto-22, Tabernero-24, Vilar-25, Olander-25}. The determination of C and O abundances from optical spectra is also challenging for stars cooler than 5200\,K \citep{BertrandeLis-15, Suarez-Andres-17, DelgadoMena-21, Biazzo-22}, because the lines become weak at these temperatures and blends become increasingly important.

The growing volume of stellar datasets motivate the use of data-driven methods. Machine-learning (ML) techniques have been widely applied to derive stellar parameters and abundances from spectra and photometric data, capturing complex and non-linear relations that are difficult to model explicitly \citep{Ball-10,  Ness-15, Fabbro-18, Leung-19}. Recent studies have shown that ML models can recover metallicity for cool stars from photometric data in M dwarfs \citep{Duque-Arribas-25}, further illustrating the potential of these methods in regimes where classical analyses face limitations.

The connection between stellar chemistry and dynamics is an important ingredient of Galactic archaeology, since stars with different birth radii, ages, and evolutionary histories populate different regions of chemo-dynamical space \citep{Adibekyan-11, Boeche-13, Minchev-14, Ratcliffe-23, Hackshaw-24}. In particular, metallicity and abundance ratios have been shown to correlate with orbital properties reflecting the coupled chemical and dynamical evolution of the Galactic disc \citep{Boeche-13}. Recent studies have also highlighted the strong connection between stellar abundances, age, and Galactic dynamics. Using APOGEE red clump stars in the low $\alpha$ disk, \citet{Ness-19} showed that many elemental abundance ratios can be predicted with high precision from stellar age and metallicity alone, implying that much of the chemical information is encoded in these two quantities. Their results also emphasize the strong coupling between the chemical and dynamical evolution of the Galactic disk and provide additional motivation for exploring data driven relations between abundances and stellar dynamical properties.

In this work, we investigate how well stellar metallicity and individual elemental abundances can be inferred from kinematic and orbital information, and whether these relations can be exploited to estimate abundances that are particularly relevant for exoplanet studies. We first examine the predictive power of stellar dynamics for metallicity (relevant for M dwarfs), and then explore the extent to which metallicity and kinematics can recover the abundances of C, O, Mg, and Si. Finally, we test whether C and O can be predicted from Mg, Si, and [Fe/H], with optional inclusion of kinematic information. This approach is particularly relevant for extending chemical characterisation to stars for which direct abundance determinations, especially of C and O, are difficult.

Finally, we apply our models to stars hosting planets and investigate the resulting abundance distributions. This allows us to extend chemical studies of planet-host stars to samples where direct abundance determinations are unavailable, providing new insights into the chemical conditions associated with planet formation.

This article is organized as follows. In Sect.~\ref{sample} we describe the samples and the derivation of the kinematic and orbital parameters used throughout this work. In Sect.~\ref{feh_from_kinematics} we investigate how well stellar metallicity can be inferred from kinematic and orbital information alone. In Sect.~\ref{abundances_from_feh} we explore the prediction of individual elemental abundances from [Fe/H] combined with kinematic information. In Sect.~\ref{main_CO_from_MgSiFe} we focus on the prediction of C and O abundances from Mg, Si, and [Fe/H], and assess the extent to which additional kinematic information improves those predictions. In Sect.~\ref{sec:ratio_metallicity_trends} we examine simple empirical relations between abundance ratios and metallicity that are relevant for exoplanet studies. Finally, in Sect.~\ref{conclusion} we summarize our main results and conclusions.

%%%%%%%%%%%%%%%%%%%%%%%%%%%%%%%%%%%%%%%%%%%%%%%%%%%%%%%%%%%%%%
\section{The samples} \label{sample}

In this study, we use stellar data from the APOGEE, complemented by a smaller sample of nearby FGK-type dwarf stars from the High Accuracy Radial velocity Planet Searcher (HARPS) exoplanet survey \citep{Mayor-03, Adibekyan-12}. APOGEE~DR17 provides astrometric information from \textit{Gaia}~EDR3 \citep{Gaia-21}, while for the HARPS sample the astrometry was taken from \textit{Gaia}~DR3 \citep{Gaia-23}.

From APOGEE we selected red giant stars with high-quality spectra and reliable abundance determinations, and we restricted the metallicity range to $-1.0 < [\mathrm{Fe/H}] < 0.6$ dex. The HARPS sample spans a similar metallicity regime and provides precise spectroscopic parameters and abundances for well-characterised stars in the solar neighbourhood, many of which are relevant for exoplanet studies.

Using astrometry, positions, and radial velocities, we computed Galactic space velocitiy components and population-membership probabilities with \texttt{GalVel\_Pop}\footnote{The code is available at \url{https://github.com/vadibekyan/GalVel_Pop}} and derived orbital parameters with \texttt{galpy}\footnote{The code is available at \url{http://github.com/jobovy/galpy}}.

Together, these two datasets allow us to examine the links between metallicity, dynamics, and detailed abundance patterns, while also testing how well models trained on APOGEE transfer to an independent, chemically precise HARPS sample. A full description of the target selection and parameter derivation is given in Sect.~\ref{target_selection}.

\begin{figure}[ht]
    \centering
    \includegraphics[width=1\linewidth]{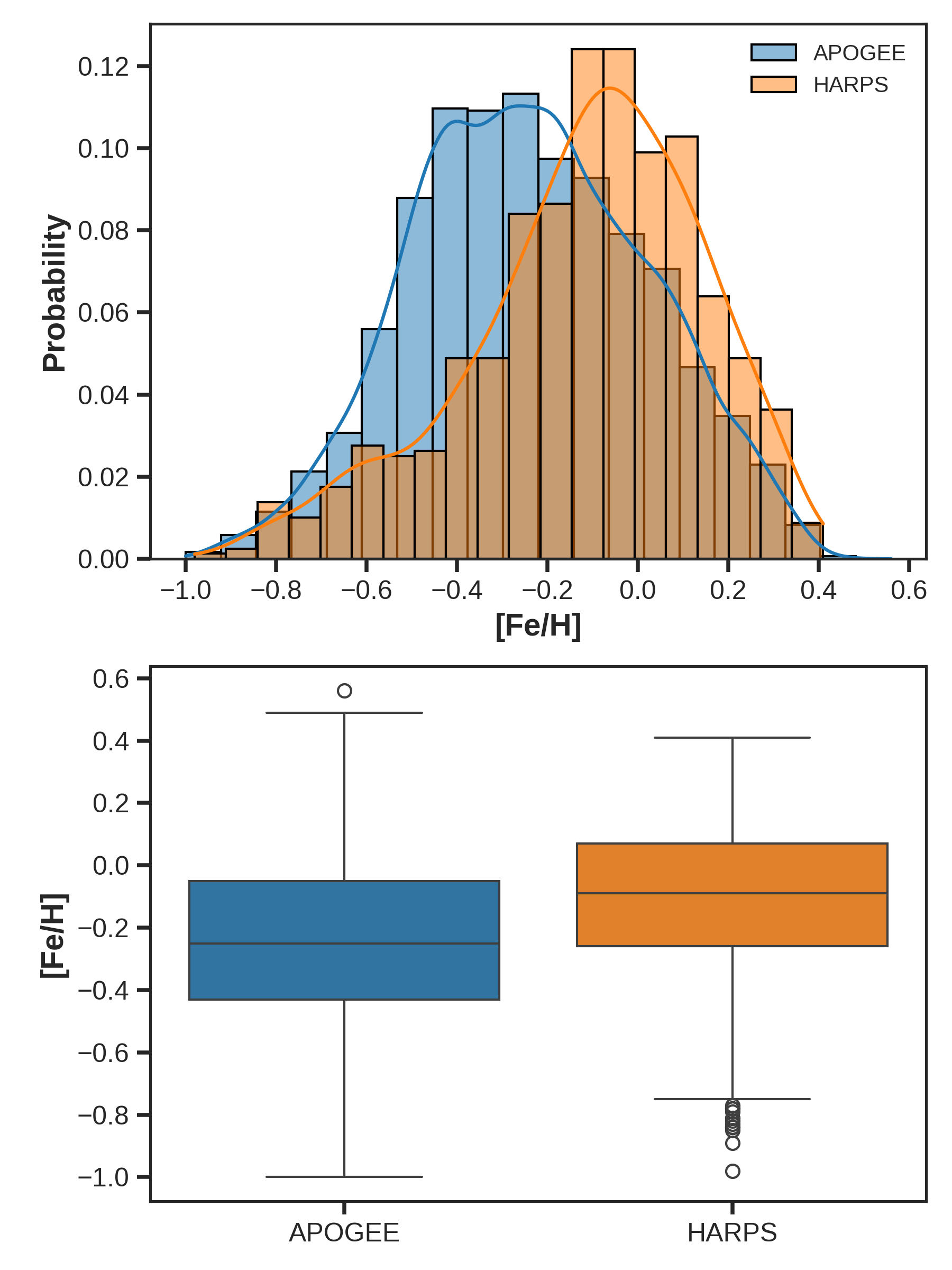}
    \caption{
    Distributions of [Fe/H] for the two samples used in this work: APOGEE (blue)  and HARPS (green). The top panel shows the normalised probability density functions, while the bottom panel presents boxplots summarizing the median and spread of the [Fe/H] distributions.
    }
    \label{fig:feh_distributions}
\end{figure}

The distributions of stellar metallicities ([Fe/H]) for the two samples are displayed in Fig.~\ref{fig:feh_distributions}. The samples exhibit distinct metallicity distributions, reflecting the different stellar types and selection criteria applied. The APOGEE red giants show, on average, lower metallicities than the HARPS stars, which are primarily nearby solar-type dwarfs targeted for exoplanet searches. This difference is consistent with the underlying population biases and survey selection functions.

\section{Predicting metallicity from kinematics} \label{feh_from_kinematics}

To predict stellar metallicity, [Fe/H], we initially considered nine kinematic and orbital parameters: the Galactic velocity components ($\mathrm{U_{LSR}}$, $\mathrm{V_{LSR}}$, and $\mathrm{W_{LSR}}$), the total space velocity $\mathrm{UVW_{LSR}} = \sqrt{U_{\rm LSR}^2 + V_{\rm LSR}^2 + W_{\rm LSR}^2}$, orbital parameters derived from orbit integration (the maximum vertical excursion $Z_{\max}$, the pericentric and apocentric distances $R_{\rm peri}$ and $R_{\rm apo}$, and the orbital eccentricity $e$), and the probability of belonging to the Galactic thin-disk population, $P_{\rm gal}$.

To quantify the statistical relationships between the kinematic properties and metallicity, we evaluated monotonic correlations using Spearman’s rank coefficient \citep{spearman1904} and nonlinear dependencies using mutual information (MI). MI quantifies the general statistical dependence between two variables, including nonlinear relationships, without assuming monotonicity or linearity \citep{shannon1948}. It therefore complements Spearman’s rank coefficient and allows us to identify features that may contain predictive information beyond simple monotonic trends.

Among all parameters, $Z_{\max}$ and $P_{\rm gal}$ show the strongest correlations with metallicity, but with opposite signs. The maximum vertical excursion $Z_{\max}$ exhibits a clear negative correlation, indicating that stars reaching larger distances from the Galactic plane tend to be more metal-poor. In contrast, the thin-disk membership probability $P_{\rm gal}$ correlates positively with [Fe/H], consistent with the higher metallicities characteristic of thin-disk stars.

While some quantitative differences are observed between the APOGEE and HARPS samples—reflecting their distinct stellar populations and selection functions—the overall trends remain consistent, with $Z_{\max}$ and $P_{\rm gal}$ emerging as the dominant kinematic predictors of metallicity.

To assess multicollinearity among the input features, we computed the variance inflation factor (VIF), a standard diagnostic for linear dependence among predictors in regression analysis \citep{farrar1967}. Large VIF values indicate strong linear dependence among variables.

As expected, several parameters exhibit large VIF values. Interestingly, the two strongest predictors identified in the correlation analysis, $Z_{\max}$ and $P_{\rm gal}$, also display comparatively low VIF values, indicating that their predictive power is not merely a consequence of linear redundancy with other orbital parameters.

Although substantial multicollinearity is present (VIF $>$ 10 is typically considered as substantial), particularly among the derived velocity quantities, we retained all features in the subsequent analysis, as tree-based ensemble models are largely insensitive to correlated predictors. 

Supplementary analyses related to the prediction of stellar metallicity from kinematic and orbital parameters are provided in Sect.~\ref{appendix:metallicity}.

\subsection{Regression models}

We benchmarked a diverse set of regression algorithms to predict stellar metallicity from these parameters. The tested models include linear regression (LR), random forest (RF; \citealt{breiman2001}), extra trees (ET; \citealt{geurts2006}), gradient boosting methods including XGBoost \citep{chen2016}, LightGBM \citep{ke2017}, and CatBoost \citep{prokhorenkova2018}. All models were evaluated using identical 5-fold cross-validation splits.

To provide a consistent baseline comparison, we adopted standard implementations of each algorithm with default hyperparameters and did not perform hyperparameter tuning. The goal of this experiment was to assess the intrinsic predictive power of the kinematic features
rather than to optimise individual models.

Linear regression yields RMSE values of approximately 0.21 dex. Given that the standard deviation of the metallicity distribution is $\sigma_{\rm [Fe/H]} \approx 0.26$ dex (corresponding to the RMSE of a trivial mean predictor), these models explain roughly 30\% of the total variance in [Fe/H]. This indicates that the metallicity–kinematics relation is intrinsically nonlinear. Tree-based ensemble models significantly outperform these approaches, with LightGBM achieving the lowest RMSE of $0.199 \pm 0.001$ dex ($R^2 \approx 0.41$). CatBoost and XGBoost produce nearly identical performance, while random forest and extra trees are slightly less accurate.

We further explored ensemble strategies. A weighted voting regressor combining the strongest tree-based models did not yield a measurable improvement over the best individual learner. A stacked ensemble based on out-of-fold predictions and a ridge meta-learner provided a marginal gain, reducing the RMSE only at the fourth decimal level.

This limited improvement reflects the high correlation between the predictions of the base models and indicates that performance is primarily constrained by the information content of the kinematic features rather than by algorithmic choice.

We additionally explored a quantile-based transformation of the target variable, mapping [Fe/H] to a Gaussian distribution prior to training. While such transformations can stabilize variance and improve regression performance in skewed distributions, they did not yield any measurable improvement in our case. This result reflects both the approximately symmetric metallicity distribution
and the robustness of tree-based ensemble methods to monotonic target transformations.

Overall, the convergence of multiple algorithms, the absence of improvement from target transformation, and the limited gain from ensembling indicate that predictive performance is primarily limited by the information content of the kinematic features rather than by
model architecture.

\subsection{Exploring model architectures and feature engineering}

We performed a series of additional experiments exploring alternative modelling strategies, automated ML frameworks, and extensive feature transformations with the aim of improving the predictive performance of our models.

First, we employed two automated machine-learning frameworks, AutoGluon \citep{Erickson2020} and H2O AutoML \citep{LeDell2020}. AutoML systems automatically explore large spaces of ML pipelines by training and evaluating multiple algorithms, tuning hyperparameters, and constructing model ensembles. Such approaches are designed to approximate the best achievable performance on tabular datasets by systematically searching the model space and leveraging ensemble learning. We ran both AutoML frameworks for extended periods (of order $\sim$20 hours in total), allowing them to evaluate hundreds of candidate models and ensemble configurations.

Second, we investigated whether additional nonlinear interactions between the input variables could improve the predictive power. Starting from the original set of parameters, we randomly generated 300 sets of interaction features using simple algebraic transformations (multiplication, subtraction, addition, and squared terms). These engineered features were appended to the base feature set and evaluated using the tuned LightGBM model.

Third, to assess the potential impact of feature selection, we constructed 2000 random feature subsets containing between 3 and 20 parameters drawn from the expanded pool of original and engineered features. Each subset was used to train and evaluate the model in order to probe whether particular combinations of features could capture additional predictive information.

Finally, we tested whether performance could be improved by modifying the sampling of the target variable. Since extreme metallicity values are relatively underrepresented, we attempted to balance the metallicity distribution using two approaches: (i) binning the metallicity distribution and sampling more uniformly across bins, and (ii) applying sample weights to emphasize underrepresented metallicity ranges during training.

Despite these extensive experiments, all models and configurations yielded predictive performances comparable to our baseline models, with RMSE values consistently remaining above $\sim$0.198 dex. None of the explored architectures, feature transformations, feature subsets, or sampling strategies produced a measurable improvement beyond this level. This convergence strongly suggests that the achievable predictive accuracy is primarily constrained by the intrinsic information contained in the kinematic features themselves rather than by limitations of the machine-learning methodology.

\subsection{Hyperparameter optimisation and final models}

Given the strong performance of gradient boosting methods in the baseline benchmark, we performed hyperparameter optimisation for LightGBM and XGBoost. LightGBM was selected as the reference model based on its slightly superior performance in the default configuration, while XGBoost was included as a widely used and well-established boosting framework for tabular data applications.

The optimisation was carried out using the Optuna framework \citep{Akiba-19}, which implements efficient Bayesian hyperparameter search. The tuning focused on the main parameters controlling model complexity and regularization. The objective was to minimize the cross-validated root mean squared error (RMSE) using the same 5-fold splits adopted in the baseline analysis.

After optimisation, both LightGBM and XGBoost achieved nearly identical predictive performance, with RMSE $\approx 0.199$ dex. The improvement relative to the default LightGBM configuration was marginal and occurred only at the third decimal level. This convergence of independently optimised models further supports the conclusion that predictive accuracy is primarily constrained by the intrinsic information content of the kinematic features rather than by the specific boosting architecture.

To assess the robustness and transferability of the trained models, we evaluated the optimised LightGBM and XGBoost models on the independent HARPS sample. The HARPS dataset has a metallicity dispersion of $\sigma_{\mathrm{[Fe/H]}} \approx 0.26$ dex, comparable to that of the APOGEE sample. Both models achieve an RMSE of approximately 0.224 dex on HARPS, corresponding to $R^2 \approx 0.26$, indicating that roughly one quarter of the metallicity variance in the HARPS sample is recovered from kinematic information alone.

It is important to note again that, although the APOGEE and HARPS samples exhibit similar metallicity dispersions, they differ in several key aspects. The mean metallicity of HARPS stars is $\langle \mathrm{[Fe/H]} \rangle = -0.115$ dex, while the APOGEE sample has a lower mean value of $-0.239$ dex. Furthermore, the two samples have distinct distributions of orbital and kinematic parameters, reflecting their different stellar populations and selection functions.

\begin{figure}[ht]
\centering
\includegraphics[width=\linewidth]{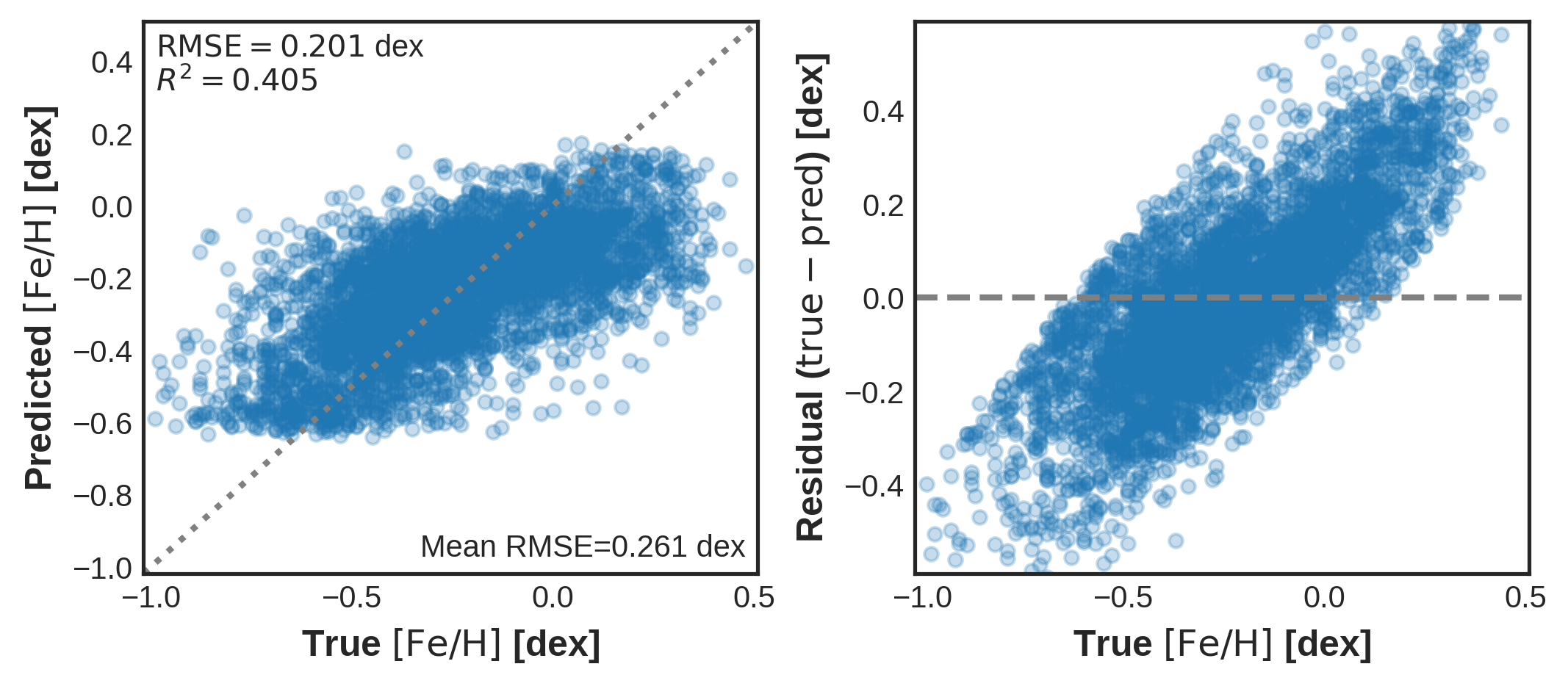}
\vfill
\includegraphics[width=\linewidth]{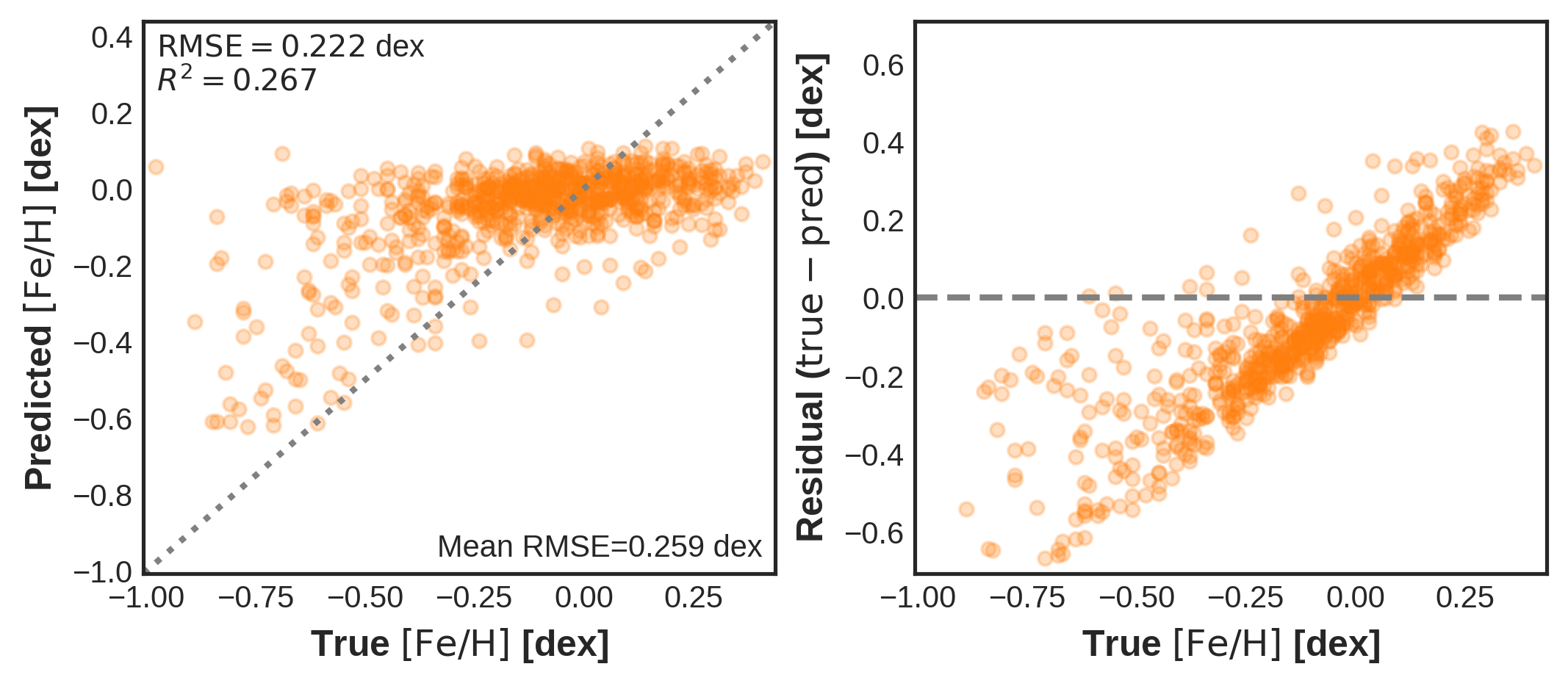}
\caption{Performance of the tuned LightGBM model for metallicity prediction. Top panels: APOGEE test set results. Bottom panels: HARPS results. Left panels show predicted versus true metallicity with the one-to-one relation indicated by the dotted line. Right panels show residuals as a function of true metallicity.}
\label{fig:lgbm_apogee_harps}
\end{figure}

To further illustrate the performance of the LightGBM model, Fig.~\ref{fig:lgbm_apogee_harps} shows the predicted versus true metallicities and the corresponding residuals. For the APOGEE sample, the model was trained on a training subset (80\%) and evaluated on an independent test set. The upper panels of Fig.~\ref{fig:lgbm_apogee_harps} display the results for the test set. The performance of the model, trained on the full APOGEE dataset, when applied to the independent HARPS sample is shown in the lower panels of the figure.

In both datasets, the predictions exhibit a significant regression toward the mean metallicity of the training data, with metal-poor stars predicted slightly too metal-rich and metal-rich stars slightly underestimated. This behaviour reflects the well-known phenomenon of regression toward the mean \citep{Galton-1886} and arises naturally when the predictors do not fully constrain the target variable. Under a squared-error loss function, the optimal prediction in the presence of incomplete information is biased toward the conditional mean of the target \citep{Hastie2009}. In this case, the limited information content of the kinematic features results in a reduced dynamic range of the predictions relative to the true metallicity distribution.

\subsection{Feature importance analysis} \label{main_feat_imp_feh}

To quantify the relative contribution of each kinematic parameter to the metallicity prediction, we performed a comprehensive feature importance analysis using four complementary approaches: built-in gain-based importance, permutation importance \citep{breiman2001}, single-feature predictive performance, and SHAP values \citep{Lundberg-17, Lundberg2020}. Figures are shown in Sect.~\ref{appendix:feature_importance_feh}.

For LightGBM and XGBoost we first considered the built-in feature importance based on the total gain. In tree-based ensembles, the gain measures the total reduction in the loss function (in our case, squared error) produced by splits involving a given feature across all trees in the ensemble. Larger gain values therefore indicate that a feature contributes more strongly to improving model accuracy. Although gain importance is model-dependent and not directly comparable across algorithms, it provides an initial ranking of influential parameters.

In both models, $Z_{\max}$ clearly dominates the ranking, followed by $R_{\mathrm{apo}}$, $R_{\mathrm{peri}}$, and the orbital eccentricity $e$. Galactic space velocity components ($U_{\mathrm{LSR}}$, $V_{\mathrm{LSR}}$, $W_{\mathrm{LSR}}$, and $UVW_{\mathrm{LSR}}$) and $P_{\mathrm{gal}}$ contribute substantially less to the predictive power.

To obtain a model-agnostic estimate of feature relevance, we computed permutation importance. This method evaluates the increase in prediction error when the values of a single feature are randomly shuffled, thereby breaking its statistical relationship with the target variable. The importance is quantified as the change in cross-validated RMSE ($\Delta \mathrm{RMSE}$).

Permutation importance confirms the dominance of $Z_{\max}$, which produces by far the largest degradation in performance when permuted. The next most important parameters are $R_{\mathrm{apo}}$ and $R_{\mathrm{peri}}$, followed by eccentricity.

As an additional diagnostic, we trained each model using only one feature at a time and computed the cross-validated RMSE. This approach quantifies the intrinsic predictive power of each parameter in isolation. Using $Z_{\max}$ alone yields an RMSE of $\sim 0.22$ dex, significantly better than any other single parameter. All other features individually result in RMSE values above $0.24$ dex.

To further investigate both global importance and the direction of feature effects, we computed SHAP (SHapley Additive exPlanations) values for the tuned models. SHAP values decompose each individual prediction into additive contributions from each feature, grounded in cooperative game theory. The global importance of a feature is quantified as the mean absolute SHAP value, $\langle |\mathrm{SHAP}| \rangle$.

Across the two models, the SHAP rankings are remarkably consistent. $Z_{\max}$ exhibits the largest mean absolute contribution, followed by $R_{\mathrm{apo}}$ and $R_{\mathrm{peri}}$. Eccentricity contributes at an intermediate level, while the velocity components have substantially smaller SHAP amplitudes.

All four importance diagnostics consistently identify the same ranking of parameters across LightGBM and XGBoost. The convergence of independent models and independent importance measures indicates that the result is robust and not an artifact of a specific boosting architecture.

In summary, the metallicity–kinematics relation is primarily driven by vertical orbital structure ($Z_{\max}$), with additional contributions from radial orbital extent ($R_{\mathrm{apo}}$ and $R_{\mathrm{peri}}$) and eccentricity. Instantaneous velocity components and the thin-disk membership probability $P_{\rm gal}$ add comparatively little independent information once orbital parameters are included.

\subsection{\texorpdfstring{The $Z_{\max}$--metallicity relation}{The Zmax-metallicity relation}}

\begin{figure}[ht]
\centering
\includegraphics[width=\linewidth]{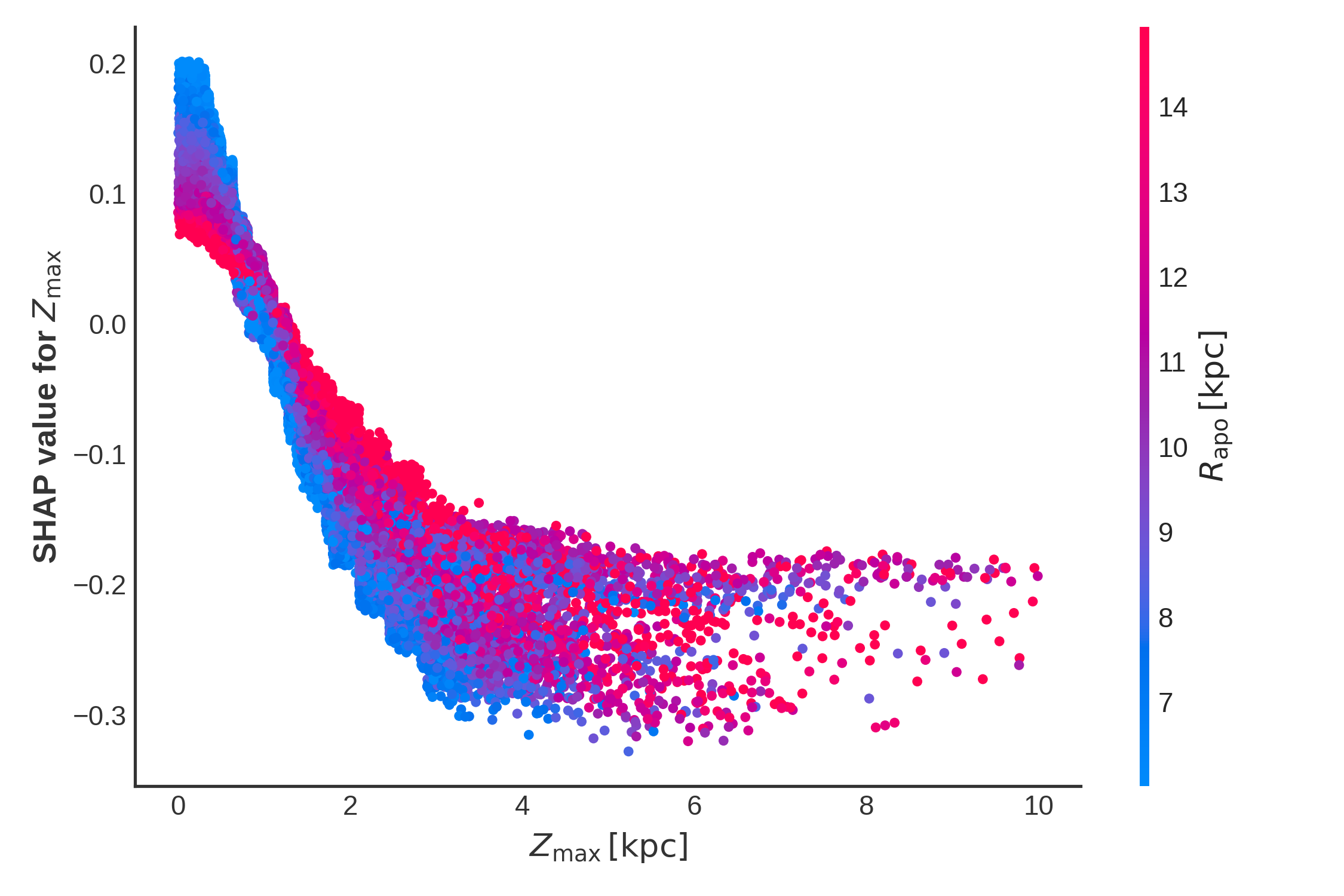}
\vfill
\includegraphics[width=\linewidth]{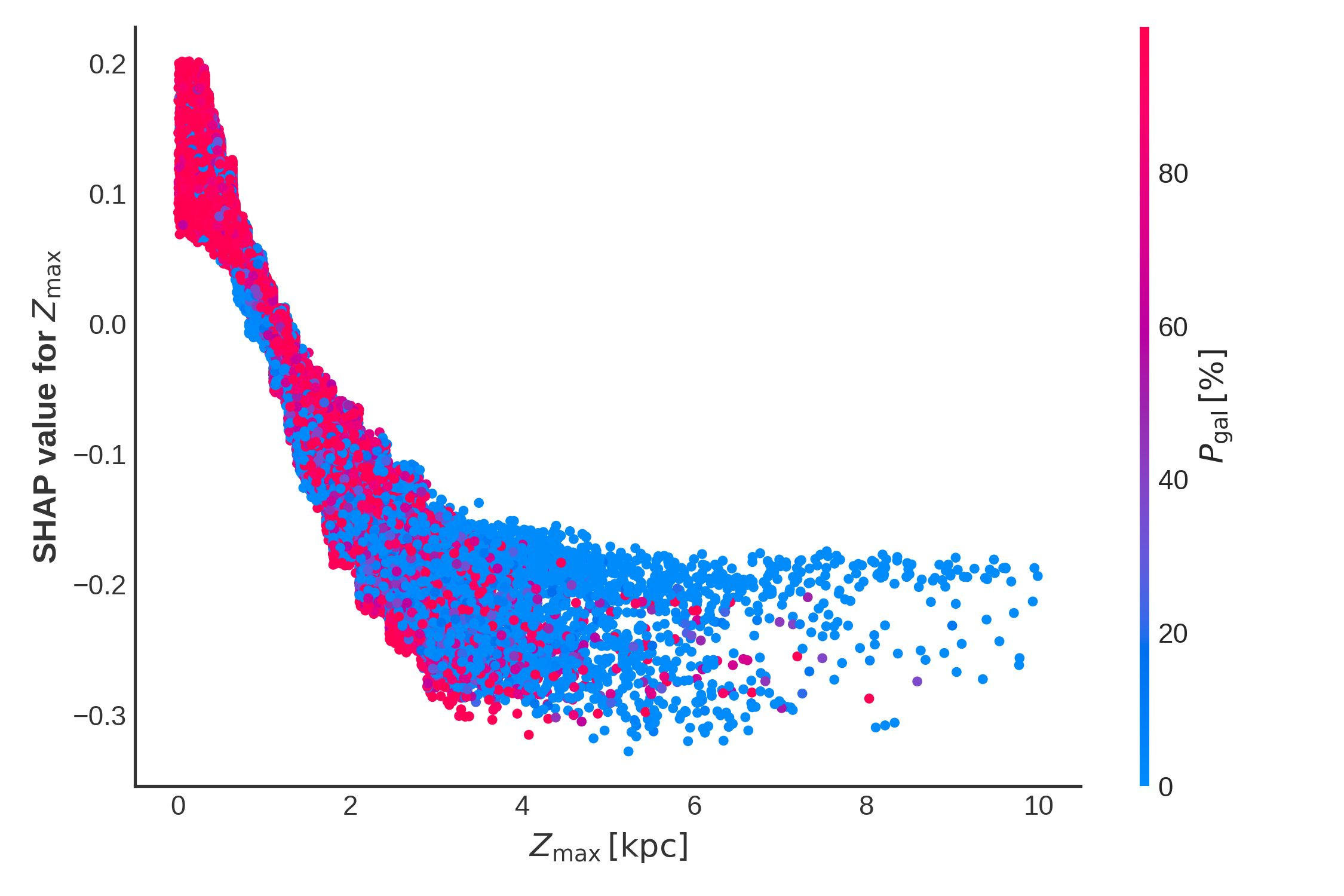}
\caption{SHAP dependence of $Z_{\max}$ for the tuned LightGBM model. The vertical axis shows the SHAP contribution of $Z_{\max}$ to the predicted metallicity, while the horizontal axis shows its physical value in kpc. Points are colour-coded by apocentric radius $R_{\mathrm{apo}}$ (top panel) and by their thin-disk membership probability $P_{\rm gal}$ (bottom panel).}
\label{fig:shap_zmax_apogee_harps}
\end{figure}

Given the dominant role of $Z_{\max}$ in all feature-importance diagnostics, we further examine its detailed relation with metallicity using SHAP dependence plots (Fig.~\ref{fig:shap_zmax_apogee_harps}). These plots show the SHAP contribution of $Z_{\max}$ to the predicted metallicity as a function of its physical value, with points colour-coded by apocentric distance $R_{\mathrm{apo}}$ and thin-disk membership probability $P_{\rm gal}$.

For the APOGEE sample, the SHAP dependence plot reveals a strong and monotonic decrease of the SHAP contribution of $Z_{\max}$ with increasing $Z_{\max}$. Since SHAP values quantify the shift of the prediction relative to the global mean metallicity of the training data, positive values indicate that $Z_{\max}$ increases the predicted metallicity, while negative values indicate that it decreases it. Stars confined close to the Galactic plane ($Z_{\max} \lesssim 1$ kpc) therefore tend to have their predicted metallicities shifted upward relative to the mean, whereas stars reaching large vertical distances shift the prediction downward.

The relation steepens at small $Z_{\max}$ and gradually flattens beyond $\sim$3–4 kpc. Interestingly, the dispersion of SHAP values becomes minimal and approaches zero around $Z_{\max} \sim 1$–1.5 kpc.  In this region, the predictive contribution of $Z_{\mathrm{max}}$ alone becomes less discriminative, and other orbital parameters contribute more strongly to the model predictions. At larger $Z_{\max}$, the increasing dispersion reflects a broader range of metallicities among dynamically hotter populations, consistent with a transition from thin-disk dominated stars to thick-disk and kinematically heated components.

The colour coding by $R_{\mathrm{apo}}$ suggests a secondary modulation related to the radial extent of the orbits. At small $Z_{\max}$, stars with larger apocentric distances tend to show slightly more negative SHAP contributions, indicating lower predicted metallicities. Above $Z_{\max} \sim 1,\mathrm{kpc}$, this trend becomes weaker and partly reverses, suggesting that the relation between vertical and radial orbital properties changes across different disk populations.

This behaviour is broadly consistent with a changing mixture of thin and thick disk stars along the $Z_{\max}$ sequence, as well as with the observed radial and vertical metallicity gradients in the Galaxy \citep{Boeche-13}. Simulations predict a steep radial metallicity gradient close to the Galactic plane that becomes flatter at larger heights \citep{Miranda-16}. Observations also suggest a negative radial metallicity gradient for the thin disk \citep{Katz-21, Imig-23} and a flat or even positive gradient for the thick disk \citep{Recio-Blanco-14, Miranda-16}.

\section{Predicting individual abundances from [Fe/H] and kinematics} \label{abundances_from_feh}

We next investigate whether the abundances of individual elements can be predicted from stellar metallicity combined with kinematic information. We focus on C, O, Mg, and Si, which are of particular interest for exoplanet studies and especially relevant for M-dwarf applications, where reliable determinations of individual elemental abundances are far more difficult than determinations of the overall metallicity. We note that, in this analysis, we use the spectroscopically derived metallicity ([Fe/H]) rather than the metallicity predicted from kinematics.

For each element we first explored which parameters are most strongly associated with its abundance by computing both Spearman's rank correlation coefficient and the MI between the abundance and the input features. For the APOGEE sample, the strongest correlation is found with [Fe/H], as expected from the Galactic chemical evolution, with typical Spearman coefficients of $\rho \sim 0.9$ for all considered elements. The next most informative parameters are related to the stellar orbital properties, in particular $Z_{\max}$, $R_{\mathrm{apo}}$, and $R_{\mathrm{peri}}$, with correlation coefficients typically in the range $0.2 \lesssim \rho \lesssim 0.4$. The MI analysis reveals a very similar picture, indicating that the dominant information about the elemental abundances is carried by [Fe/H], while the orbital parameters provide secondary but non-negligible contributions.

For the HARPS sample, stellar metallicity remains the dominant parameter, with both $\rho$ and MI values significantly larger than those of the remaining features. Among the other parameters, $Z_{\max}$ and $P_{\mathrm{gal}}$ show the strongest associations with the elemental abundances, with typical Spearman coefficients of $\rho \approx -0.2$ and $\rho \approx 0.2$, respectively.
The full set of Spearman correlation and mutual information diagrams for all considered elements is presented in Appendix~\ref{appendix:correlations_mi_abund}. Supplementary analyses related to the prediction of individual elemental abundances using stellar metallicity together with kinematic and orbital parameters are provided in Sect.~\ref{appendix:abundances}.

\subsection{Regression models}

Using the same framework as for metallicity prediction, we benchmarked a broad set of ML regressors to estimate the abundances of C, O, Mg, and Si from stellar metallicity and kinematics. The models were evaluated with 5-fold cross-validation on the APOGEE training sample using the same metrics as before (RMSE, MAE, and $R^2$). The complete set of results is presented in Table~\ref{tab:abundance_cv_scores}.

Overall, the behaviour of the different algorithms is very similar to that found for the metallicity prediction task. Among the considered models, gradient boosting methods consistently provide the best performance without requiring hyperparameter tuning. In particular, the LightGBM and CatBoost regressors achieve the lowest prediction errors for all elements. Given its consistently high predictive performance across all considered elements and its computational efficiency, we adopt the LightGBM regressor as the primary model for the subsequent analysis.

We optimised hyperparameters of LightGBM using Optuna. For each elemental abundance ([C/H], [O/H], [Mg/H], and [Si/H]) the hyperparameters of the LightGBM model were optimised using the APOGEE training sample. The tuned models show only modest improvements relative to the models trained with default parameters. After determining the optimal hyperparameters, the final models were trained using the full APOGEE dataset and subsequently applied to the independent HARPS sample.

Table~\ref{tab:abundance_predictions} summarises the prediction accuracy for both the APOGEE holdout sample and the HARPS dataset. For comparison, we also report the RMSE obtained with a simple baseline predictor assuming that the abundance of a given element is equal to the stellar metallicity (i.e. $\mathrm{[X/H]} = \mathrm{[Fe/H]}$). This baseline model represents the scenario in which no additional information beyond metallicity is used. The results demonstrate that incorporating kinematic information significantly improves the prediction accuracy for all elements relative to this simple approximation.

\begin{table}
\caption{Prediction accuracy of the tuned LightGBM models for elemental abundances.
The APOGEE results correspond to the holdout validation sample, while the HARPS
results represent predictions for the independent test dataset. For comparison,
the RMSE values obtained from a baseline predictor assuming
$\mathrm{[X/H]}=\mathrm{[Fe/H]}$ are also shown.}
\centering
\begin{tabular}{lcccc}
\hline
 & [O/H] & [Mg/H] & [Si/H] & [C/H] \\
\hline
RMSE$_{\mathrm{APOGEE}}$ & 0.054 & 0.049 & 0.037 & 0.080 \\
RMSE$_{\mathrm{HARPS}}$  & 0.098 & 0.070 & 0.043 & 0.097 \\
\hline
RMSE$_{\mathrm{[Fe/H]}}$ (APOGEE) & 0.192 & 0.175 & 0.119 & 0.116 \\
RMSE$_{\mathrm{[Fe/H]}}$ (HARPS)  & 0.170 & 0.107 & 0.078 & 0.106 \\
\hline
\end{tabular}
\label{tab:abundance_predictions}
\end{table}

\begin{figure*}
\centering
\includegraphics[width=0.48\textwidth]{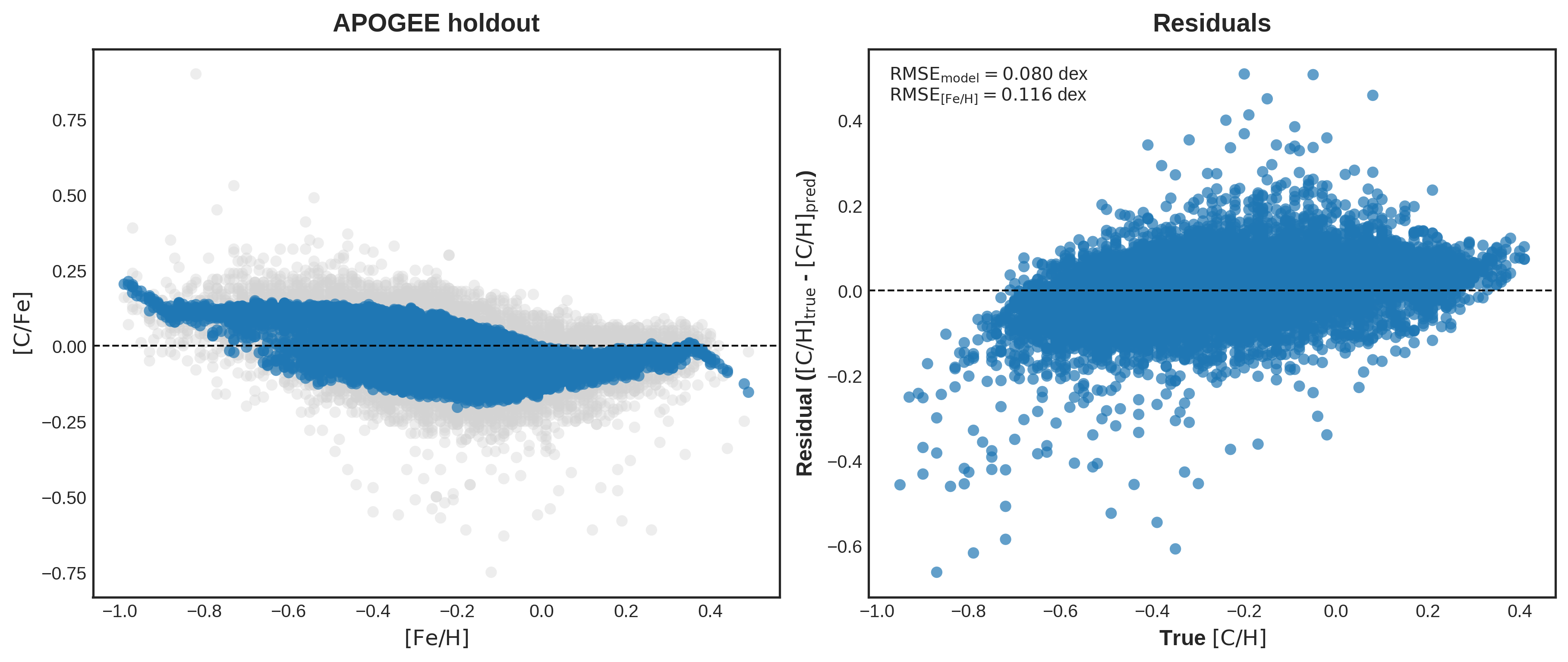}
\includegraphics[width=0.48\textwidth]{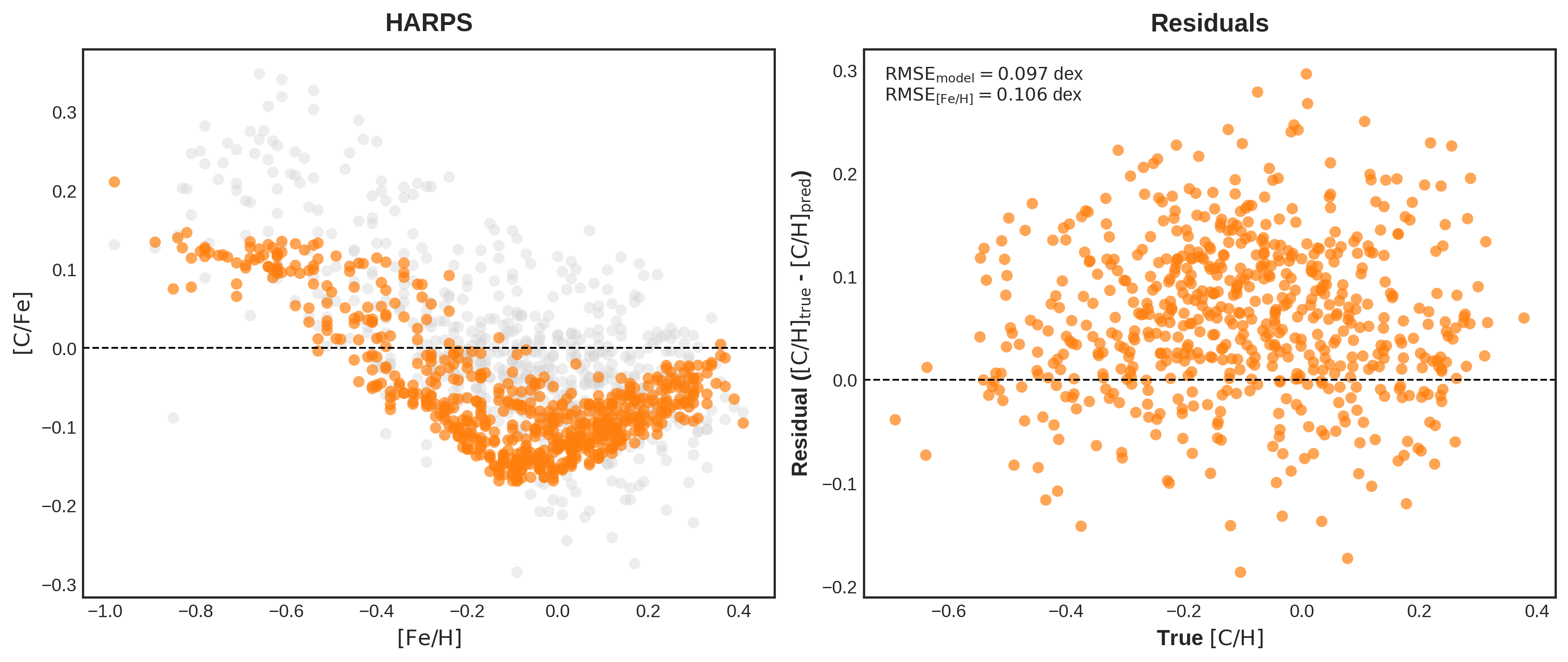}
\hfill
\includegraphics[width=0.48\textwidth]{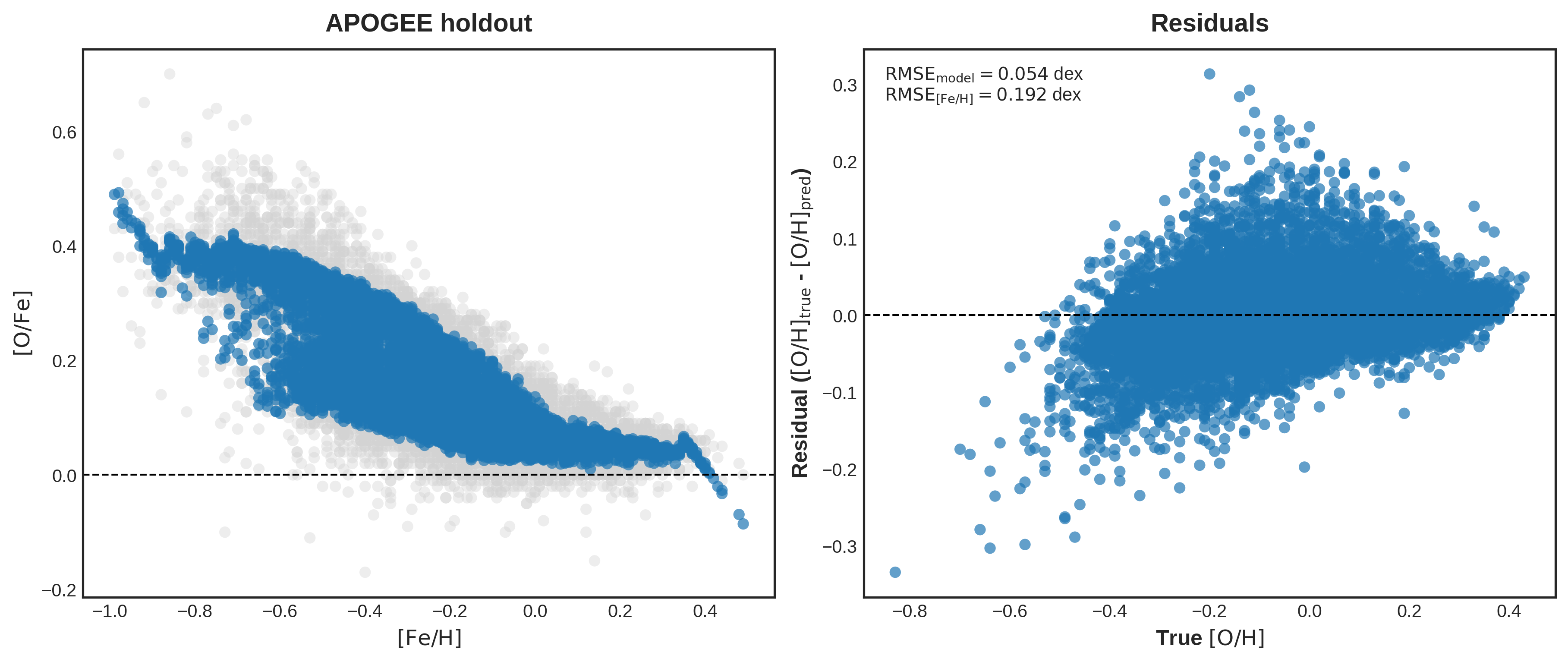}
\includegraphics[width=0.48\textwidth]{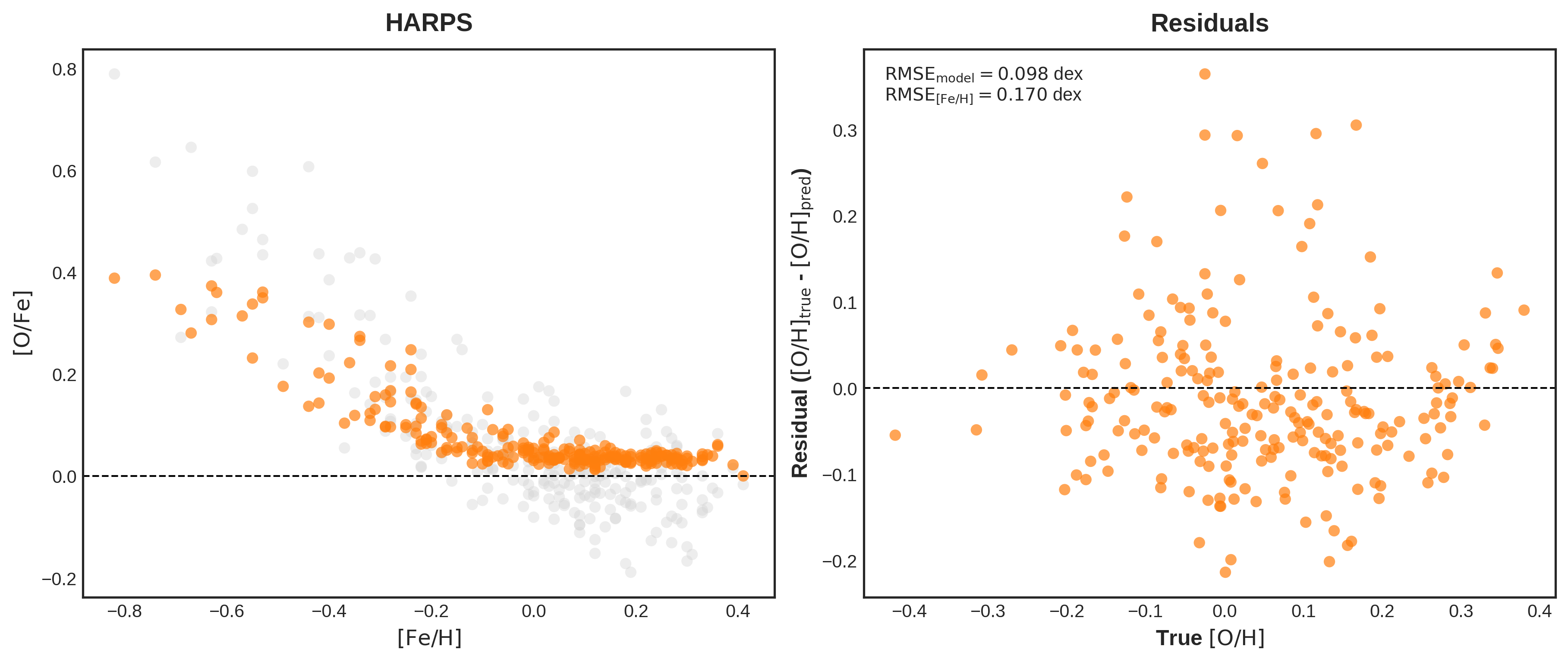}
\hfill
\includegraphics[width=0.48\textwidth]{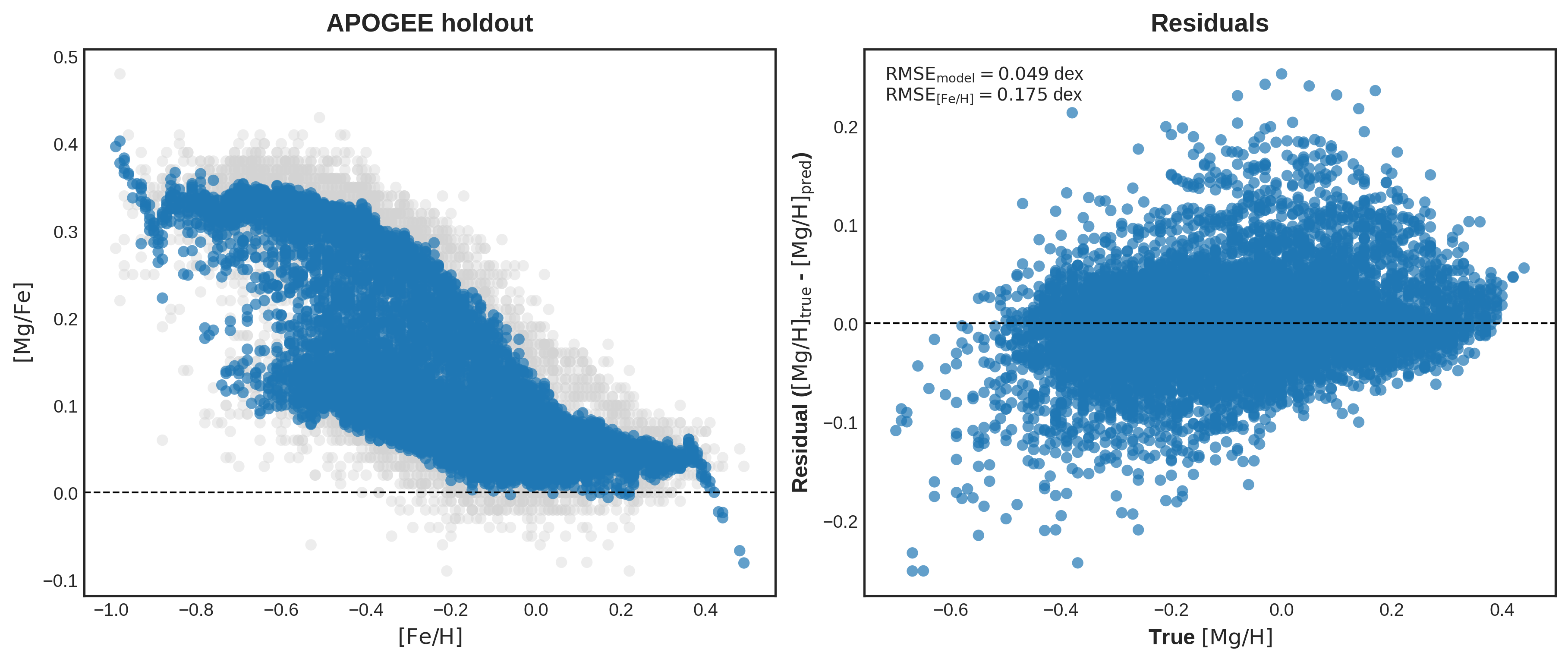}
\includegraphics[width=0.48\textwidth]{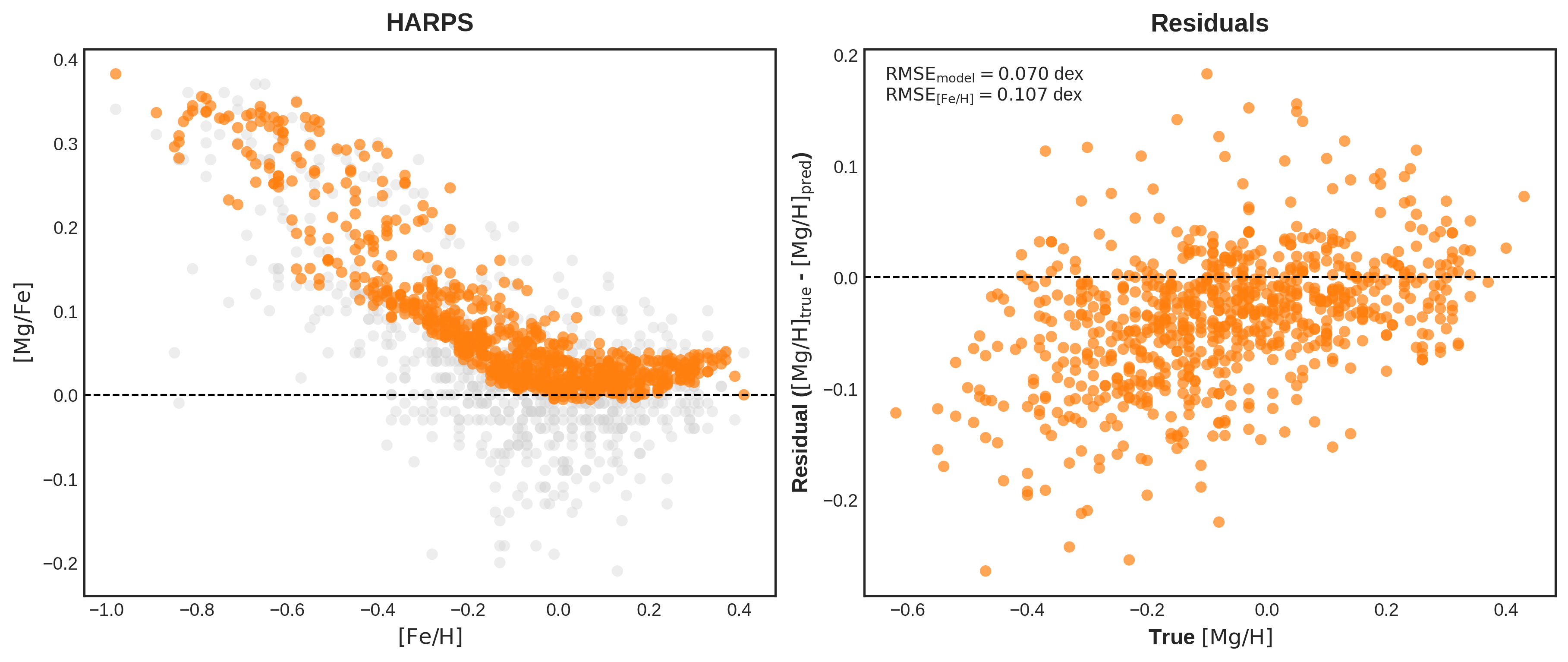}
\hfill
\includegraphics[width=0.48\textwidth]{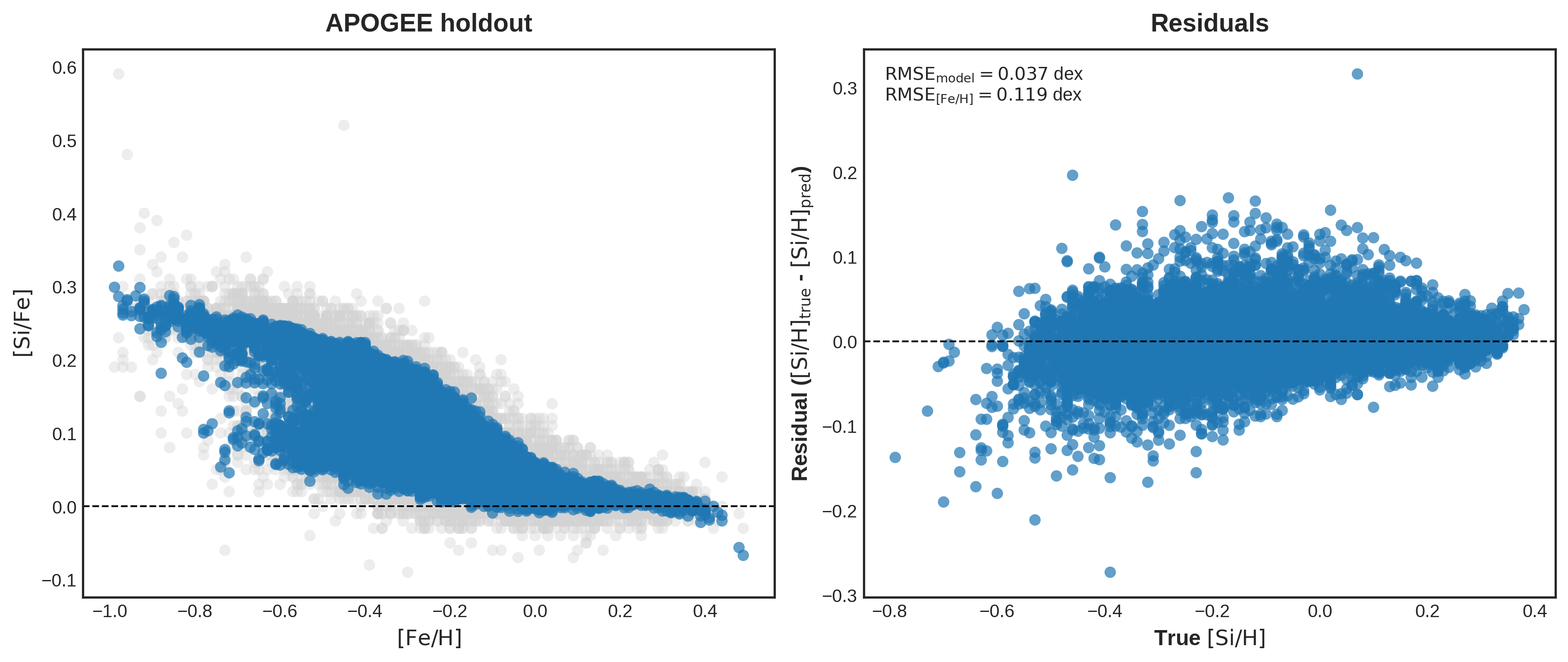}
\includegraphics[width=0.48\textwidth]{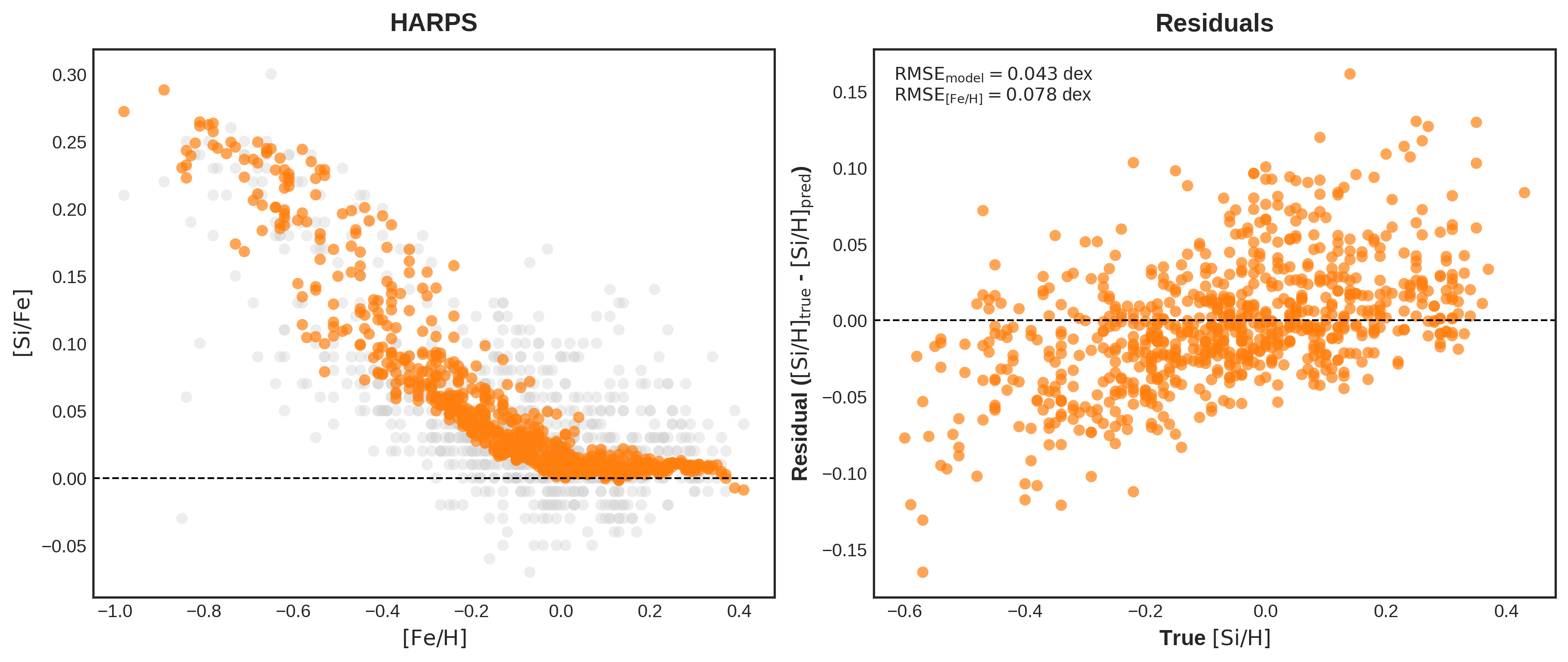}
\caption{
Predictions of elemental abundances using the tuned LightGBM models.
For each element the left panels show the predicted abundance ratios as a function of metallicity for the APOGEE holdout sample, while the right panels correspond to predictions for the independent HARPS dataset. Grey points indicate the true stellar abundances and coloured points show the model predictions. The accompanying panels display the residuals between true and predicted abundances as a function of the true abundance.}
\label{fig:abundance_predictions}
\end{figure*}

Figure~\ref{fig:abundance_predictions} illustrates the performance of the tuned models for both the APOGEE holdout sample and the HARPS dataset. The left panels show the predicted abundance ratios as a function of metallicity, compared with the true stellar abundances. The predicted values closely follow the overall trends expected from Galactic chemical evolution.

However, it is also evident that the predictions become less reliable at the extremes of the metallicity distribution. In particular, stars with $\mathrm{[Fe/H]} < -0.8$ dex and $\mathrm{[Fe/H]} > 0.35$ dex show larger deviations between predicted and true abundances. This behaviour is primarily driven by the scarcity of training data in these metallicity regimes. To focus on the regime where the models are most reliable, the residual distributions shown in the right panels are restricted to stars within the metallicity interval $-0.8 < \mathrm{[Fe/H]} < 0.35$ dex.

Overall, the models reproduce the main Galactic chemical evolution trends remarkably well for all elements considered. Nevertheless, the residual panels reveal small systematic trends, indicating that the models do not fully capture the detailed abundance behaviour across the entire metallicity range. These residual structures likely reflect both the intrinsic complexity of Galactic chemical evolution and the limited information contained in the kinematic features used as predictors.

When applying the trained models to the HARPS dataset, we observe a systematic offset between the predicted and true [C/H] abundances. This discrepancy likely reflects intrinsic differences between the APOGEE red giant and HARPS dwarf samples. Extra mixing processes are expected to modify the original carbon abundance during the red giant branch phase, with the effect depending on metallicity \citep[e.g.][]{Lagarde-19}. Our APOGEE stars mainly correspond to evolved upper RGB stars, for which carbon depletion has been extensively discussed by \citet{Masseron-15}. On average, APOGEE stars exhibit [C/Fe] abundances that are lower by approximately 0.1 dex compared to those in the HARPS sample, with the difference becoming slightly larger at lower metallicities. After accounting for this offset by applying a constant 0.1 dex correction to the predicted values, the RMSE of the [C/H] predictions decreases significantly, reaching $\sim$0.08 dex. This result indicates that the dominant source of the discrepancy is not a limitation of the model itself, but rather a difference in the abundance scales between the two datasets.

\subsection{Feature importance analysis}
\label{main_feat_imp_abund_feh}

As in the metallicity analysis (Sect.~\ref{feh_from_kinematics}), we examined the relative importance of the input parameters when predicting individual abundances from stellar metallicity and kinematic properties. We used the same four complementary diagnostics: built-in gain importance from the LightGBM model, permutation importance, predictive performance of single-feature models, and global SHAP importance. The full set of figures is presented in Appendix~\ref{appendix:feature_importance_abundances_feh}.

As expected, stellar metallicity [Fe/H] emerges as the dominant predictor for all considered elements. This reflects the strong chemical coupling between the abundances of C, O, Mg, and Si and the overall metallicity driven by Galactic chemical evolution. Across all importance diagnostics, [Fe/H] consistently exhibits the largest contribution to the model predictions.

Among the remaining parameters, the orbital pericentric distance $R_{\mathrm{peri}}$ typically provides the next most significant contribution, while  $Z_{\max}$ has a smaller but still measurable impact. The remaining kinematic parameters contribute only marginally once metallicity and the main orbital parameters are included.

\begin{figure*}[ht]
\centering
\includegraphics[width=0.4\textwidth]{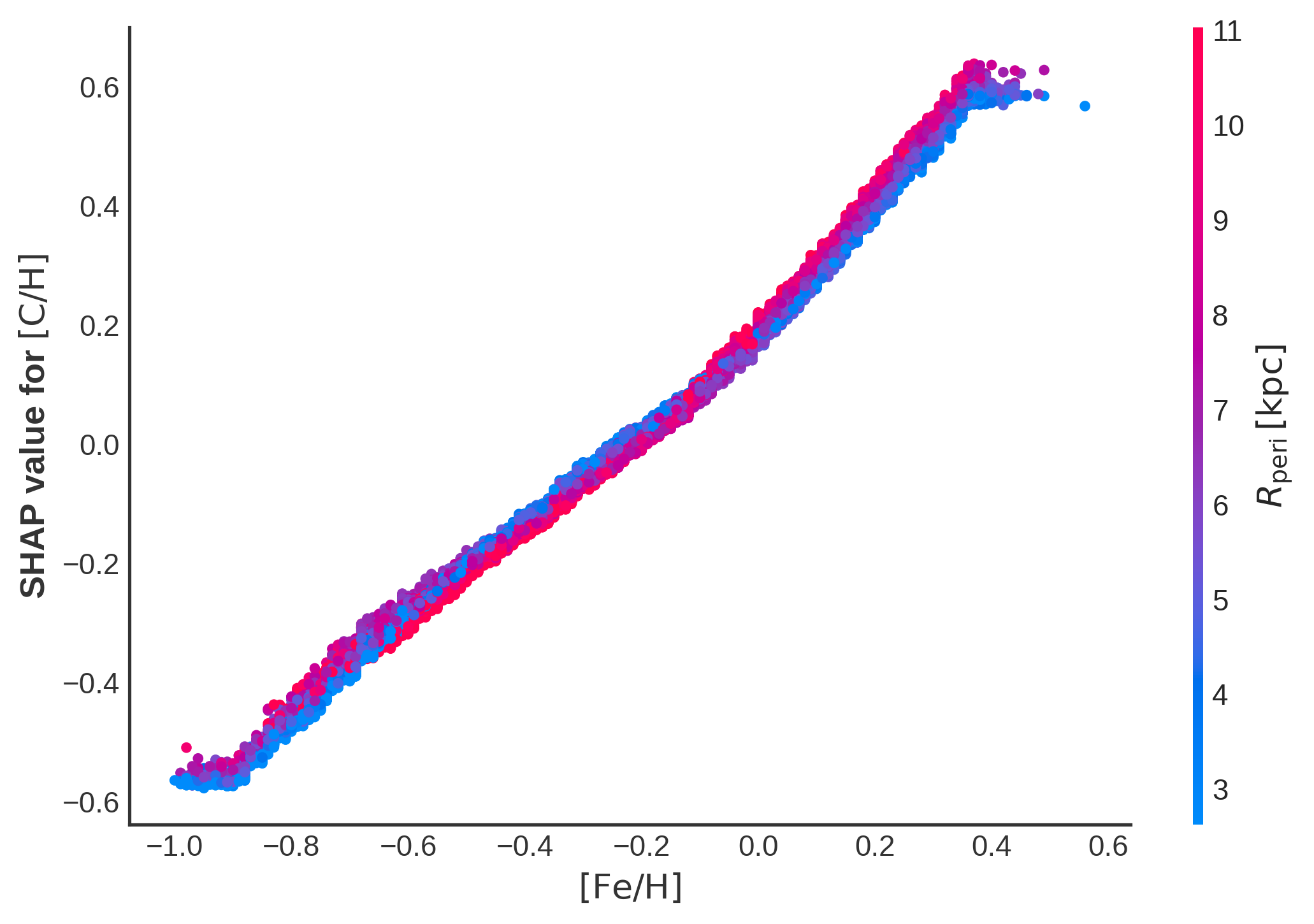}
\includegraphics[width=0.4\textwidth]{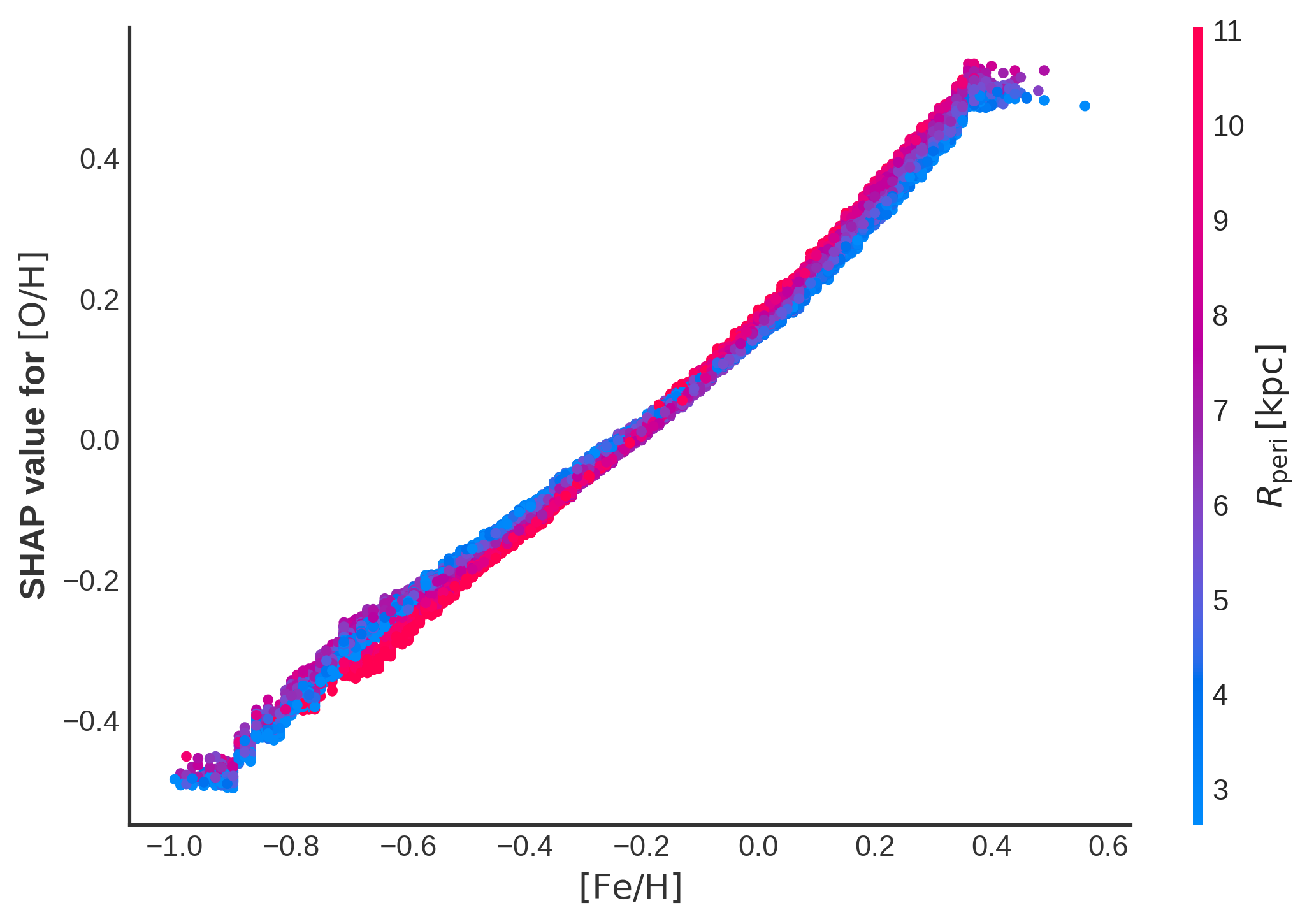}
\vspace{0.4cm}
\includegraphics[width=0.4\textwidth]{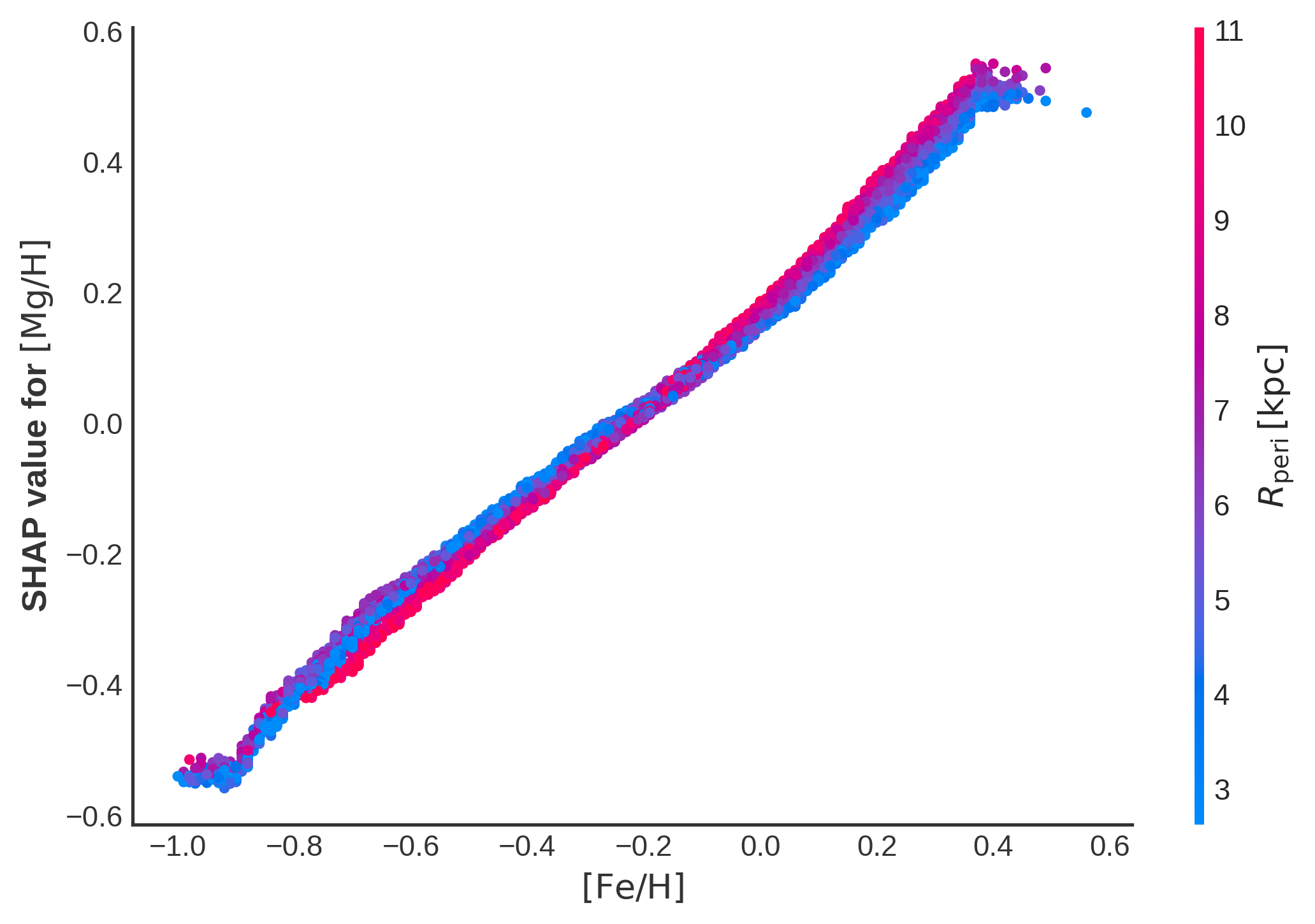}
\includegraphics[width=0.4\textwidth]{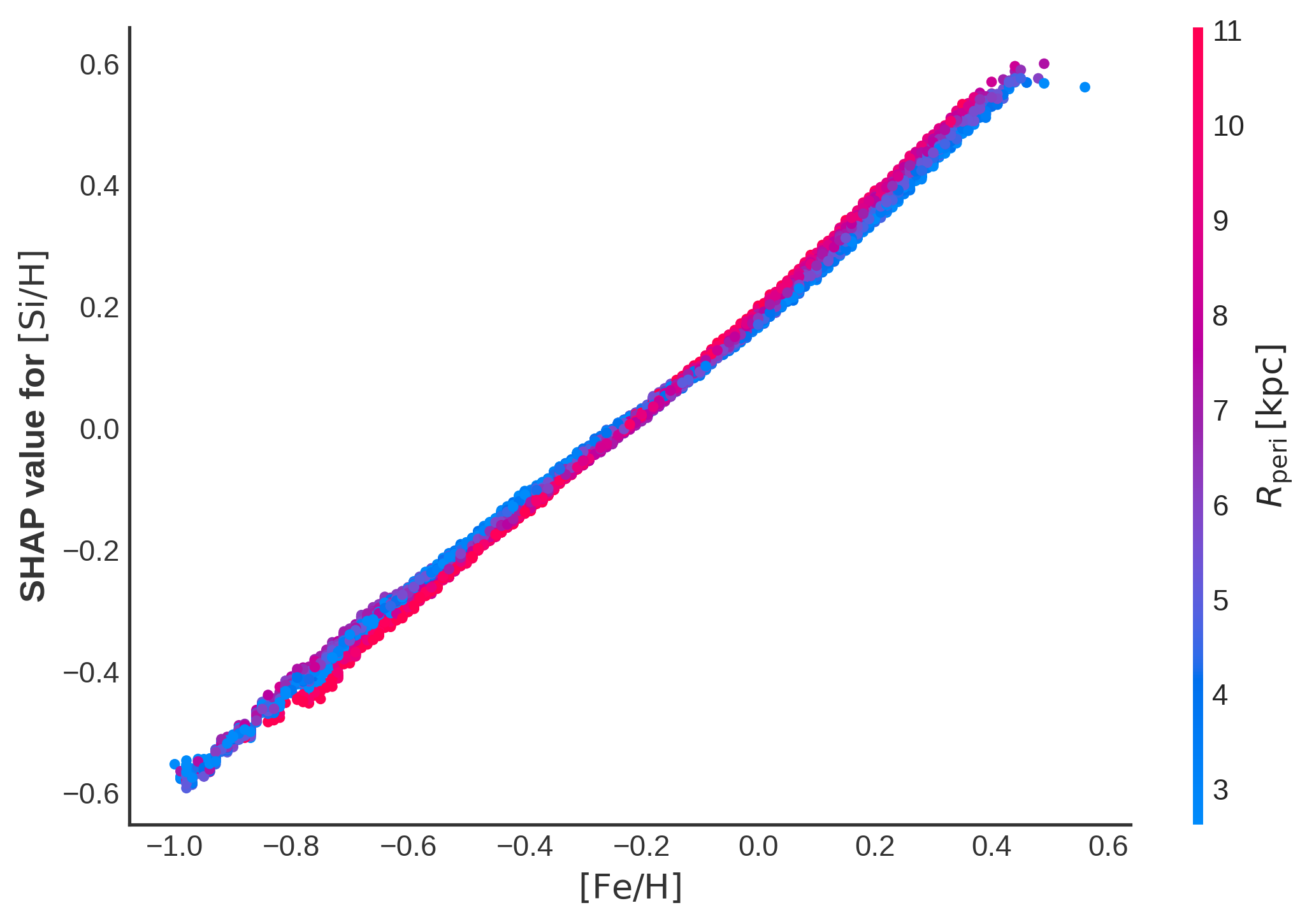}
\caption{SHAP dependence plots showing the contribution of stellar metallicity [Fe/H] to the predicted abundances of individual elements using the tuned LightGBM models. From top left to bottom right the panels correspond to [C/H], [O/H], [Mg/H], and [Si/H]. The vertical axis shows the SHAP value associated with [Fe/H], while the horizontal axis shows the stellar metallicity itself. Points are colour-coded by  $R_{\mathrm{peri}}$.}
\label{fig:shap_dependence_abundances_feh}
\end{figure*}

To further explore the role of stellar metallicity in the abundance predictions, we examined SHAP dependence plots for [Fe/H] in the tuned LightGBM models (Fig.~\ref{fig:shap_dependence_abundances_feh}). The SHAP values show an almost perfectly monotonic relation with [Fe/H] for all considered elements (C, O, Mg, and Si).

The colour-coding by the pericentric orbital distance $R_{\mathrm{peri}}$ reveals a secondary modulation of the predictions. At metallicities above approximately $[\mathrm{Fe/H}] \gtrsim -0.1$ dex, stars with larger $R_{\mathrm{peri}}$ tend to exhibit slightly higher SHAP contributions, indicating somewhat enhanced predicted abundances relative to stars with smaller pericentric distances. At lower metallicities the trend appears to reverse. This modest effect, supported by observations \citep{Katz-21},  suggests that orbital properties encode additional information related to Galactic chemical history that is not fully captured by [Fe/H] alone.

We performed the same analysis using $Z_{\max}$ instead of $R_{\mathrm{peri}}$, but no clear dependence of the SHAP values on $Z_{\max}$ was visually apparent in the corresponding dependence plots.

\section{Predicting C and O abundances from Mg, Si, and Fe}
\label{main_CO_from_MgSiFe}

In this section, we predict the abundances of carbon and oxygen from the abundances of Mg, Si, and [Fe/H], optionally complemented by kinematic and orbital parameters. This is particularly relevant for relatively cool stars, for which the determination of C and O abundances is challenging.

Following the same methodology adopted in the previous sections, we benchmarked a range of regression algorithms. Consistent with the previous results, LightGBM yields the lowest RMSE and is therefore adopted as the reference model.

Baseline models trained using only Mg, Si, and [Fe/H] already achieve strong predictive performance. The inclusion of additional kinematic and orbital parameters does not lead to a measurable improvement in RMSE, indicating that most of the predictive information is already encoded in the elemental abundances themselves.

We also explored a range of feature-engineering strategies aimed at capturing nonlinear interactions between the input abundances. These included combinations such as the mean abundance of Mg and Si, as well as pairwise differences (e.g., Mg$-$Si, Mg$-$Fe, Si$-$Fe) and simple algebraic transformations. However, none of these engineered features produced a measurable improvement in predictive accuracy.

For both C and O abundance predictions, we find that Mg abundance carries the dominant predictive power. In fact, models trained using only Mg abundance achieve RMSE values comparable to those obtained when including Si and [Fe/H]. This result indicates a strong coupling between Mg and the C and O abundances, reflecting their common nucleosynthetic origin in massive stars \citep[e.g.][]{Kobayashi-20},  which induces strong correlations between their abundances. The feature-importance analysis is presented in Appendix~\ref{appendix:CO_from_MgSiFe}, where we show the results of gain-based importance, permutation importance, SHAP values, and single-feature predictive performance.

\begin{figure*}[ht]
\centering
\includegraphics[width=0.48\textwidth]{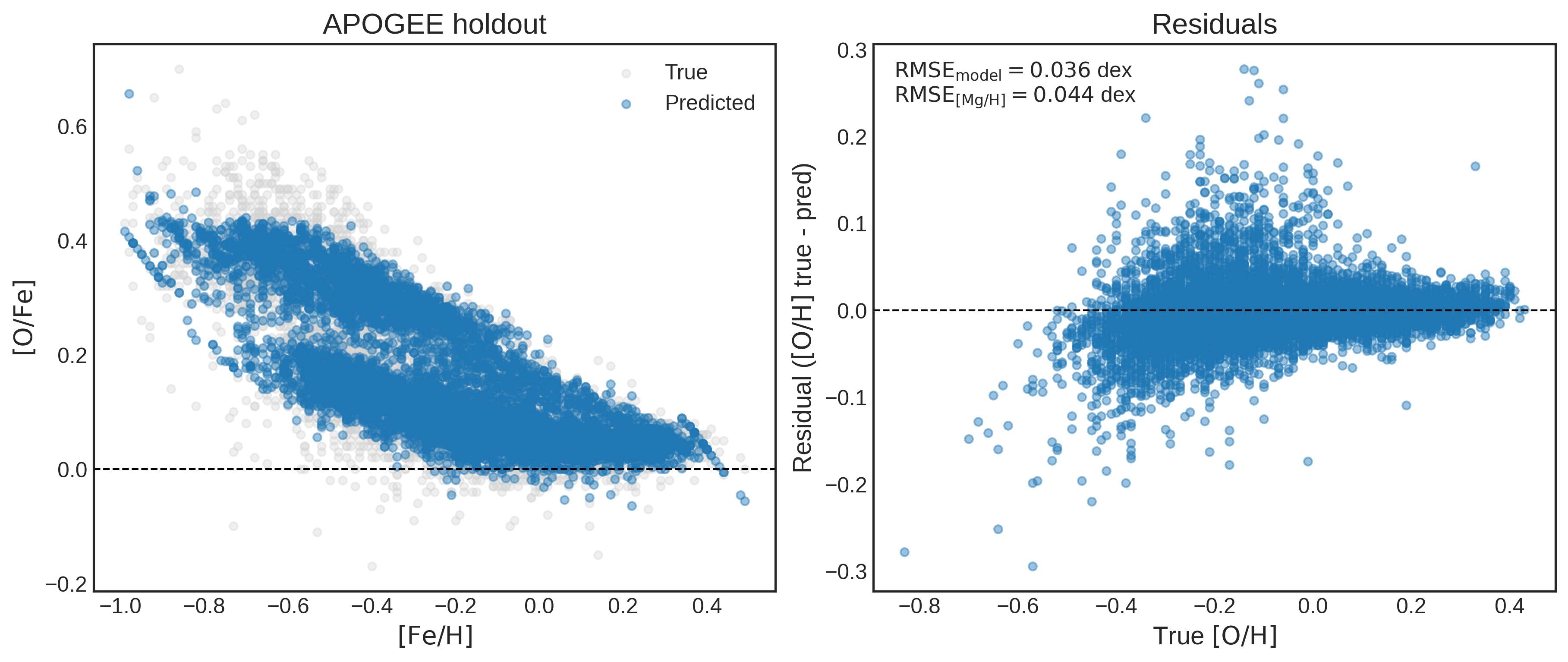}
\hfill
\includegraphics[width=0.48\textwidth]{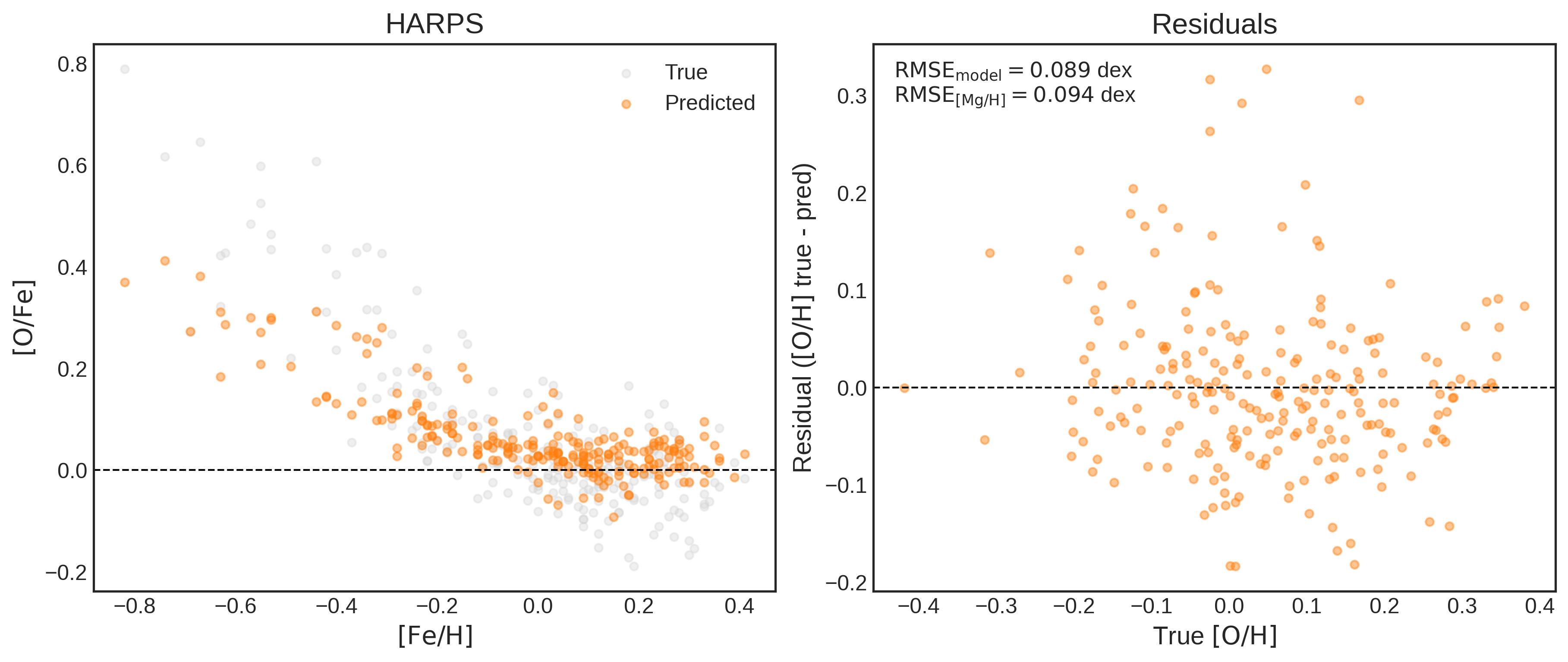}
\vspace{0.4cm}
\includegraphics[width=0.48\textwidth]{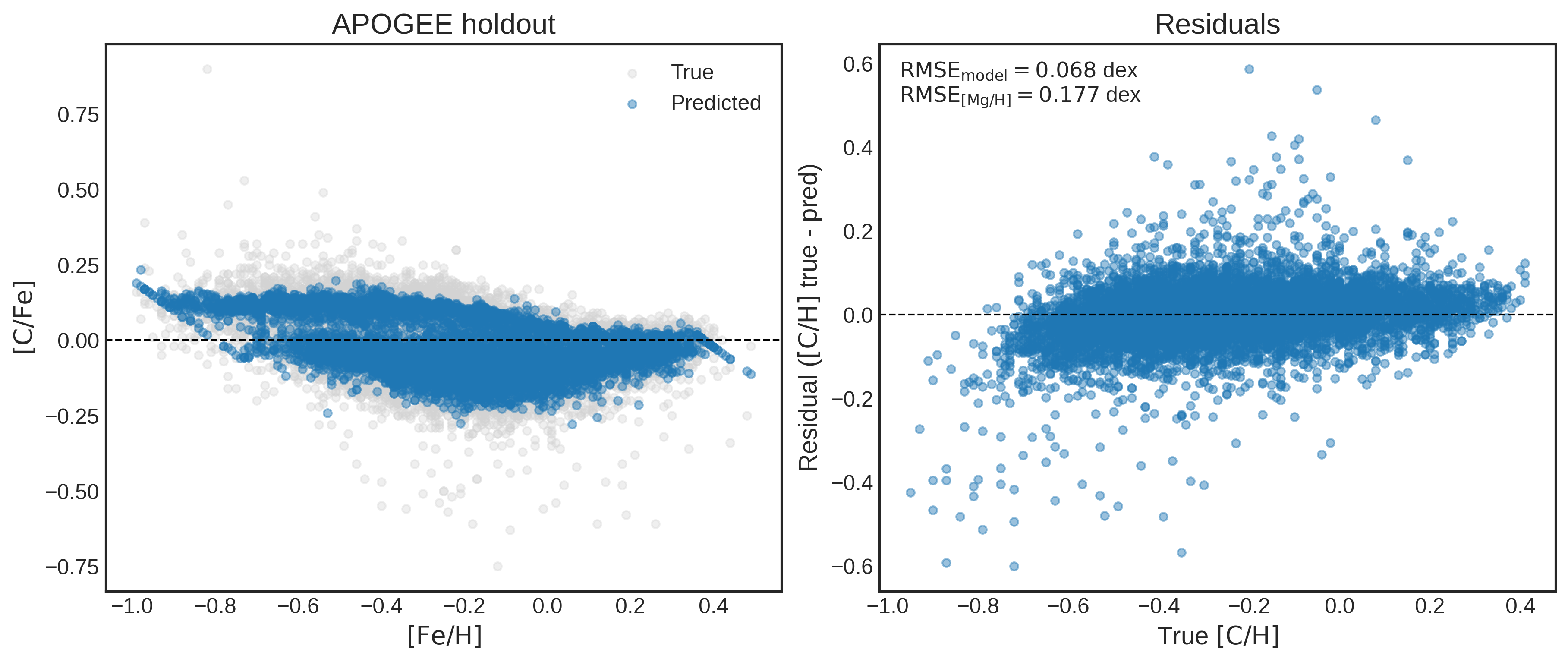}
\hfill
\includegraphics[width=0.48\textwidth]{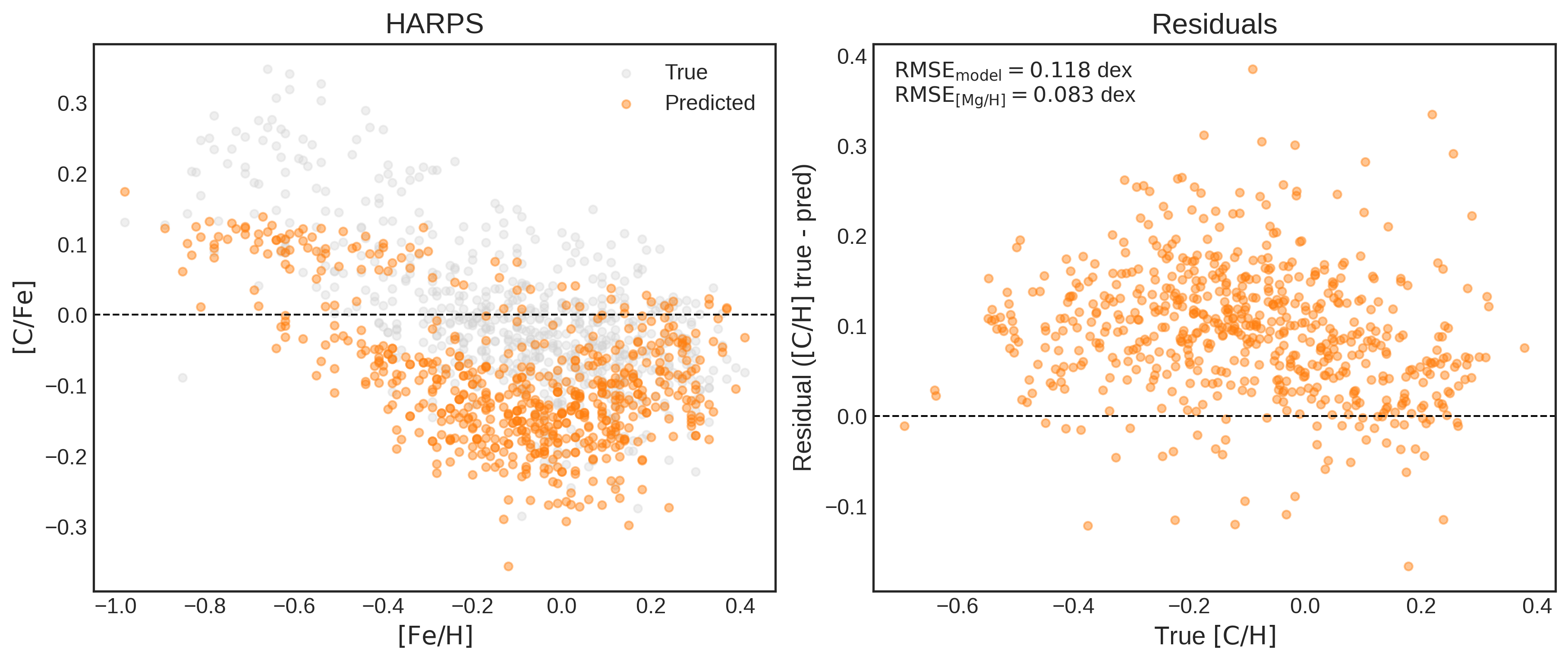}
\caption{Same as Fig.~\ref{fig:abundance_predictions}, but for the prediction of C and O abundances from Mg, Si, and [Fe/H].}
\label{fig:CO_predictions}
\end{figure*}

Figure~\ref{fig:CO_predictions} illustrates the performance of the tuned LightGBM models for predicting [O/H] and [C/H] using the APOGEE holdout sample and the independent HARPS dataset. The left panels show the predicted abundance ratios as a function of metallicity, while the right panels display the residuals as a function of the true abundance.

For the APOGEE sample, the model reproduces the overall abundance trends remarkably well, capturing both the overall slope and the intrinsic scatter of the [O/Fe]--[Fe/H] relation. The predicted values closely follow the true abundances across the full metallicity range, although a mild compression of the dynamic range is visible. This behaviour is consistent with the regression-to-the-mean effect, arising from incomplete information in the input features. The residuals exhibit a clear heteroscedastic pattern, with slightly larger scatter at lower abundances and a gradual tightening toward higher metallicities.

When applied to the independent HARPS sample, the model performance degrades. In particular, the [O/Fe] ratio at low metallicities is systematically underestimated relative to the HARPS values. Nevertheless, the achieved RMSE remains lower than that obtained from a simple baseline assuming [O/H] = [Mg/H], indicating that the model retains predictive power beyond this approximation.

The situation is less favourable for carbon. A clear offset is observed between the predicted and HARPS [C/H] abundances, with the model systematically underestimating the HARPS values. In fact, assuming [C/H] = [Mg/H] yields a lower RMSE than the direct model predictions. However, this discrepancy is largely driven by a systematic offset between the APOGEE and HARPS abundances as discussed in Sect.~\ref{abundances_from_feh}. After applying a constant correction of $\sim$0.1 dex to the predicted values, the RMSE decreases significantly to $\sim$0.07 dex.

\section{Abundance ratios relevant for exoplanet research} \label{sec:ratio_metallicity_trends}

In addition to the machine-learning experiments described above, we explored the empirical behaviour of several abundance ratios that are particularly relevant for exoplanet composition studies \citep{Dorn-15, Futo-25, Baumeister-25}. The purpose of this exercise is to provide a quick and simple way of obtaining these ratios directly from metallicity. Using the APOGEE and HARPS samples separately, we examined the abundance ratios Fe/Si, Mg/Si, C/O, and Fe/O as a function of metallicity, [Fe/H]. Each sample was also split into thin- and thick-disk populations based on their [Mg/Fe] abundances using the ad hoc relation
\[
[\mathrm{Mg/Fe}] = -0.15\,[\mathrm{Fe/H}] + 0.12,
\]
which visually separates the two populations very well. We opted against using a more rigorous population separation because the adopted relation provides a simple and practical way for users to classify stars and apply the empirical formulae derived here. We note, however, that we also tested a two-component Gaussian Mixture Model classification, and obtained practically indistinguishable results, especially for the APOGEE sample because of its large size. The uncertainties on the fitted parameters were estimated by bootstrap resampling. In addition to the slope and intercept uncertainties, which were very small, we also computed the rms scatter around each best-fit relation over the full metallicity range considered. In particular, the largest fractional offset between the two empirical relations at the whole range of metallicity is less than 17\%.

\begin{figure*}[ht]
    \centering
    \begin{subfigure}{0.49\textwidth}
        \centering
        \includegraphics[width=\linewidth]{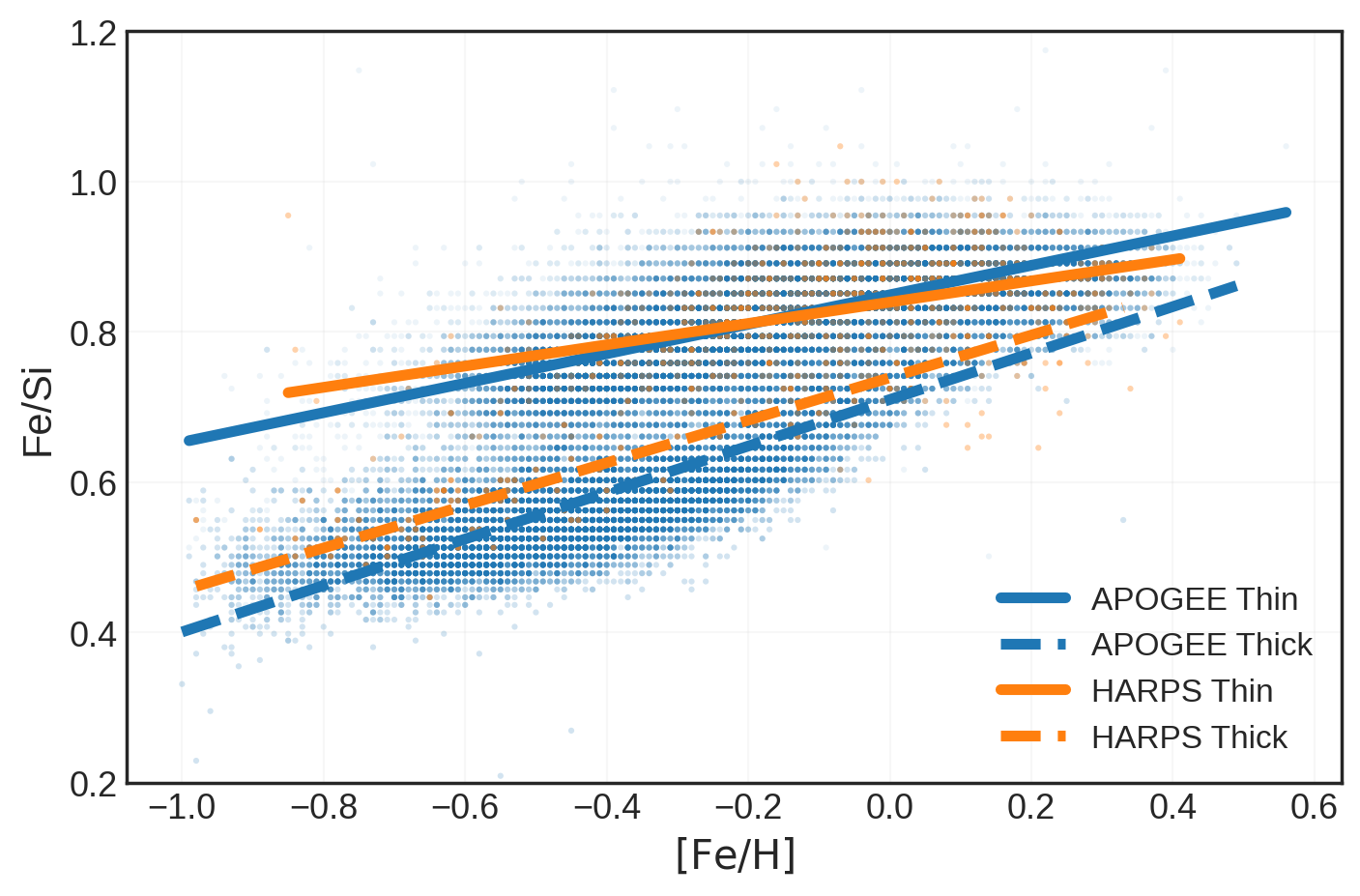}
    \end{subfigure}
    \hfill
    \begin{subfigure}{0.49\textwidth}
        \centering
        \includegraphics[width=\linewidth]{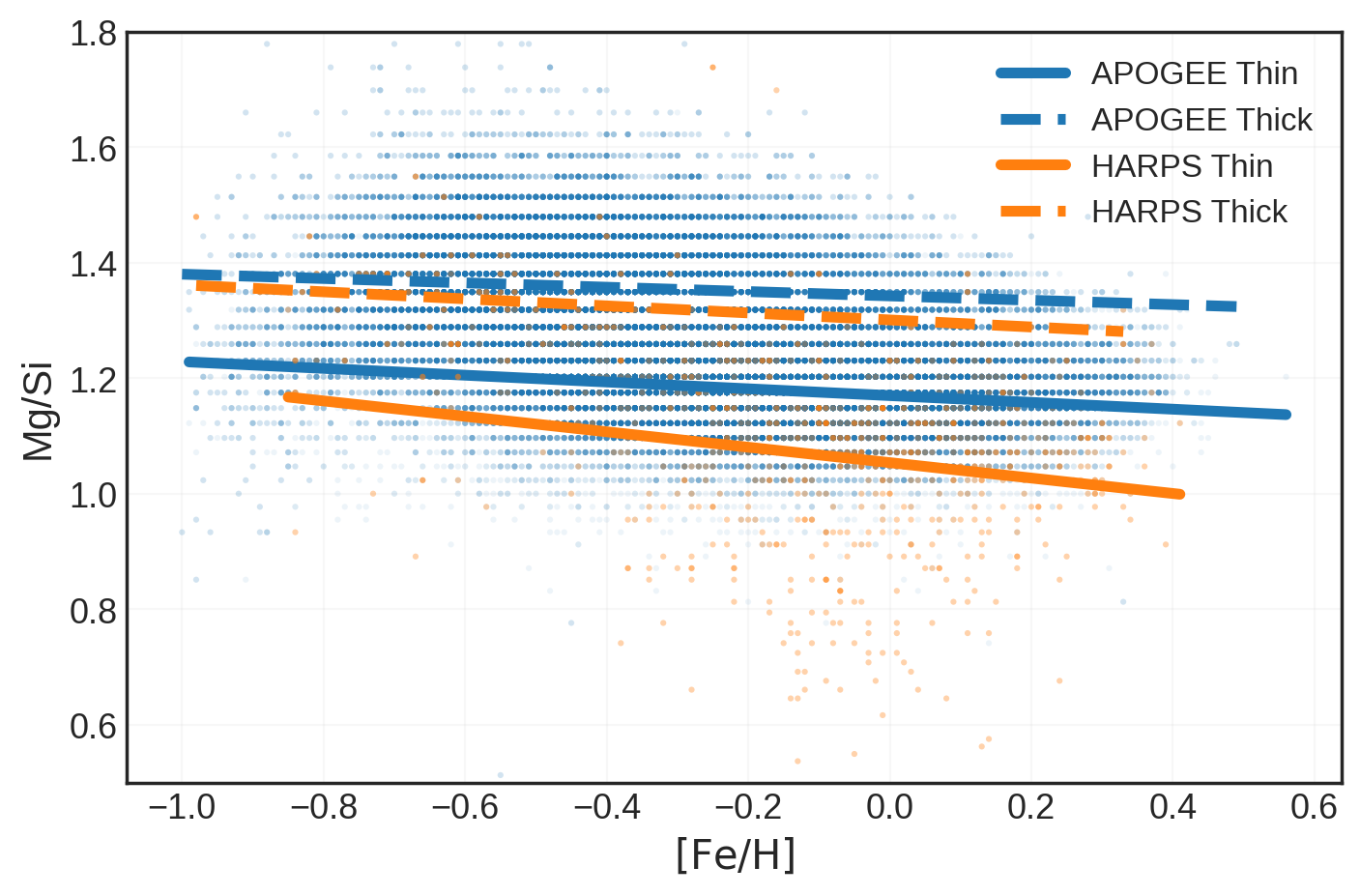}
    \end{subfigure}

    \vspace{0.5em}

    \begin{subfigure}{0.49\textwidth}
        \centering
        \includegraphics[width=\linewidth]{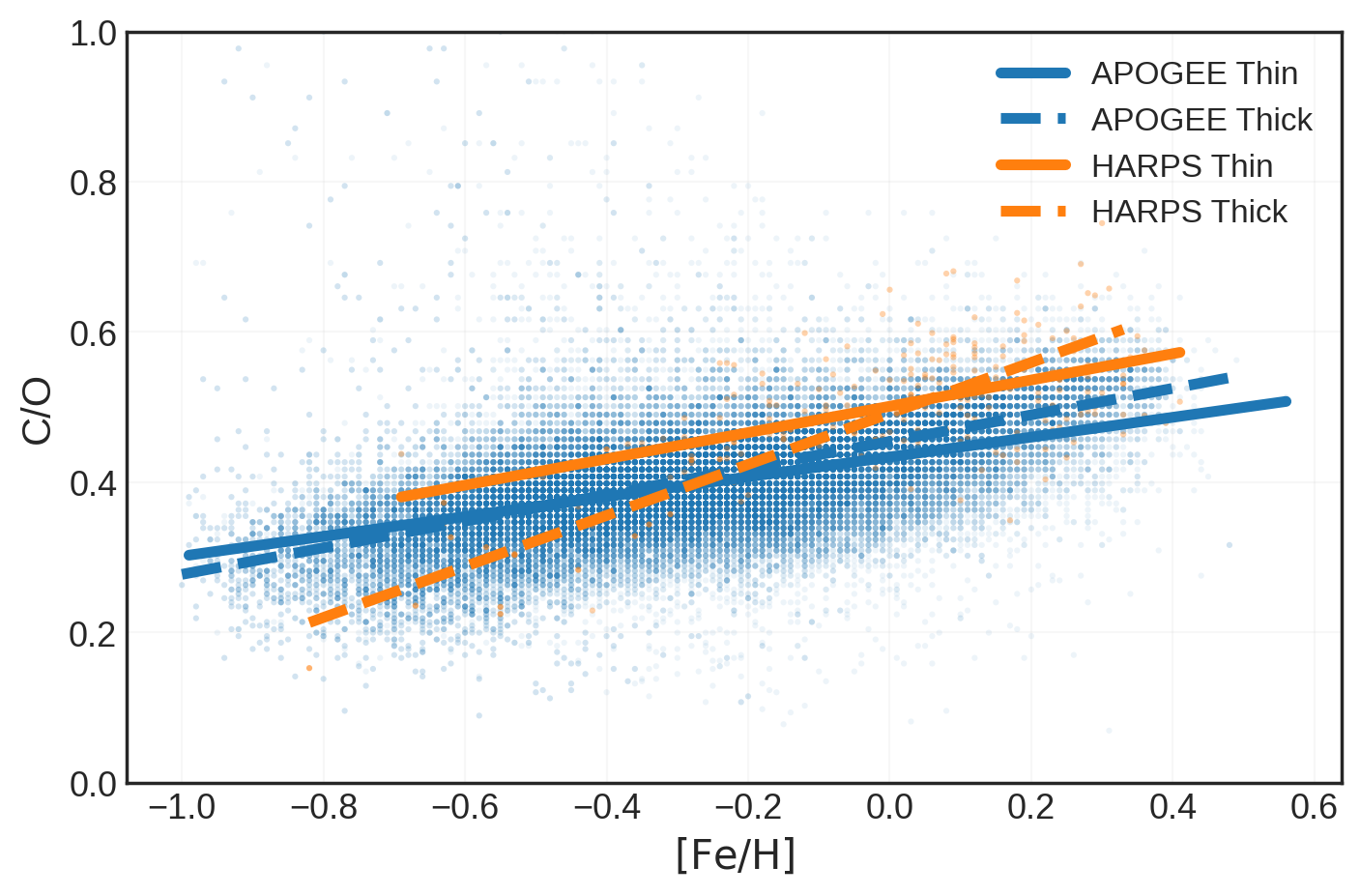}
    \end{subfigure}
    \hfill
    \begin{subfigure}{0.49\textwidth}
        \centering
        \includegraphics[width=\linewidth]{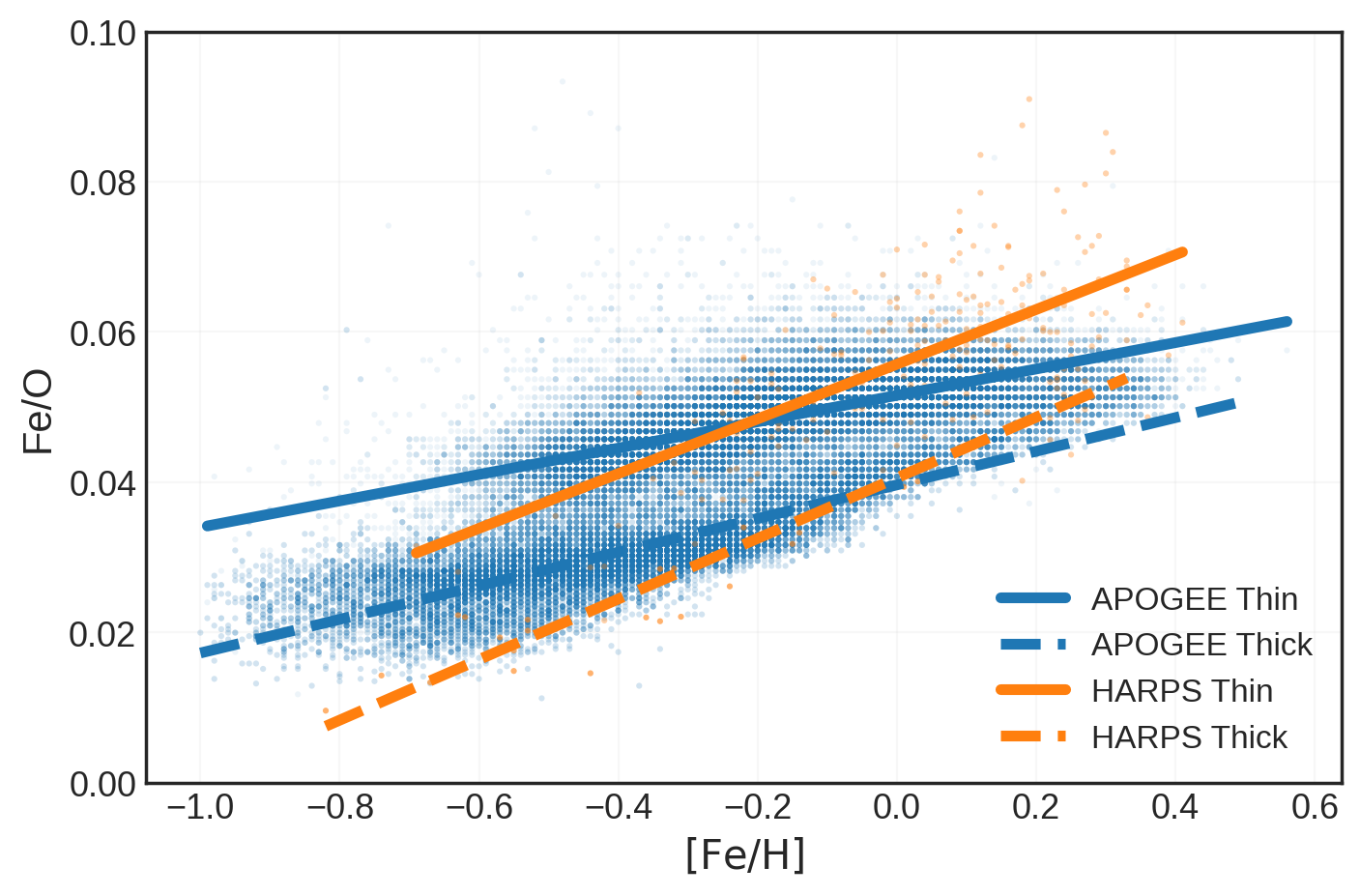}
    \end{subfigure}
    \caption{Linear abundance ratios as a function of metallicity for the APOGEE and HARPS thin- and thick-disk samples. In each panel, the left and right subplots show the APOGEE and HARPS data, respectively, together with the corresponding best-fit linear relation.}
    \label{fig:ratio_fits_feh}
\end{figure*}

The resulting trends are shown in Fig.~\ref{fig:ratio_fits_feh}, while the fitted slopes and rms scatter values are listed in Table~\ref{tab:ratio_slopes_rms}. Fe/Si and Fe/O increase with metallicity in both APOGEE and HARPS for both populations, whereas Mg/Si decreases with increasing [Fe/H]. The C/O ratio also increases with metallicity in both datasets and for both populations. The qualitative behaviour of the fits for the two samples is similar, although the fitted slopes are not identical. However, the differences are within the rms scatter of the samples.

These abundance ratios are relevant for exoplanet interior and composition studies. In particular, Fe/Si and Fe/O are linked to the amount of iron rich material and the inferred size of planetary cores, while Mg/Si affects mantle mineralogy and silicate composition \citep{Bond-10, Dorn-15, Wang-19, Adibekyan-21}. The C/O ratio may influence the chemistry of planet forming material, condensation pathways, volatile content, and the composition of planetary atmospheres \citep{Bond-10, Madhusudhan-12, Schneider-21}. Previous studies have shown that stellar abundance constraints can substantially improve the characterisation of rocky exoplanets, helping to reduce some of the degeneracies present in interior models based only on planetary mass and radius \citep{Dorn-15, Dorn-17, Hinkel-24}. Since uncertainties in planetary masses, radii, and interior models often dominate the final uncertainty budget \citep{Dorn-17, Plotnykov-24}, the rms scatter of our empirical relations appears sufficiently small for broad compositional characterisation of planetary systems.

\begin{table*}
\caption{Linear fits of abundance ratios as a function of metallicity for the APOGEE and HARPS thin- and thick-disk samples.}
\label{tab:ratio_slopes_rms}
\centering
\small
\begin{tabular}{llllc}
\hline\hline
Survey & Population & Ratio & Equation & RMS scatter \\
\hline
APOGEE & Thin  & Fe/Si & $y = (0.8493 \pm 0.0003) + (0.1961 \pm 0.0012)\,[\mathrm{Fe/H}]$ & 0.05 \\
APOGEE & Thick & Fe/Si & $y = (0.7099 \pm 0.0008) + (0.3094 \pm 0.0018)\,[\mathrm{Fe/H}]$ & 0.05 \\
HARPS  & Thin  & Fe/Si & $y = (0.8395 \pm 0.0027) + (0.1417 \pm 0.0125)\,[\mathrm{Fe/H}]$ & 0.07 \\
HARPS  & Thick & Fe/Si & $y = (0.7390 \pm 0.0091) + (0.2834 \pm 0.0151)\,[\mathrm{Fe/H}]$ & 0.05 \\
APOGEE & Thin  & Mg/Si & $y = (1.1697 \pm 0.0005) + (-0.0588 \pm 0.0018)\,[\mathrm{Fe/H}]$ & 0.08 \\
APOGEE & Thick & Mg/Si & $y = (1.3423 \pm 0.0015) + (-0.0376 \pm 0.0035)\,[\mathrm{Fe/H}]$ & 0.10 \\
HARPS  & Thin  & Mg/Si & $y = (1.0537 \pm 0.0055) + (-0.1335 \pm 0.0223)\,[\mathrm{Fe/H}]$ & 0.14 \\
HARPS  & Thick & Mg/Si & $y = (1.3006 \pm 0.0160) + (-0.0608 \pm 0.0288)\,[\mathrm{Fe/H}]$ & 0.09 \\
APOGEE & Thin  & C/O   & $y = (0.4334 \pm 0.0005) + (0.1321 \pm 0.0022)\,[\mathrm{Fe/H}]$ & 0.07 \\
APOGEE & Thick & C/O   & $y = (0.4541 \pm 0.0010) + (0.1769 \pm 0.0028)\,[\mathrm{Fe/H}]$ & 0.07 \\
HARPS  & Thin  & C/O   & $y = (0.5009 \pm 0.0046) + (0.1750 \pm 0.0233)\,[\mathrm{Fe/H}]$ & 0.06 \\
HARPS  & Thick & C/O   & $y = (0.4914 \pm 0.0192) + (0.3395 \pm 0.0448)\,[\mathrm{Fe/H}]$ & 0.05 \\
APOGEE & Thin  & Fe/O  & $y = (0.0515 \pm 0.0000) + (0.0176 \pm 0.0001)\,[\mathrm{Fe/H}]$ & 0.005 \\
APOGEE & Thick & Fe/O  & $y = (0.0396 \pm 0.0001) + (0.0224 \pm 0.0001)\,[\mathrm{Fe/H}]$ & 0.004 \\
HARPS  & Thin  & Fe/O  & $y = (0.0557 \pm 0.0006) + (0.0364 \pm 0.0032)\,[\mathrm{Fe/H}]$ & 0.009 \\
HARPS  & Thick & Fe/O  & $y = (0.0406 \pm 0.0027) + (0.0403 \pm 0.0058)\,[\mathrm{Fe/H}]$ & 0.004 \\
\hline
\end{tabular}
\end{table*}

\section{Conclusions and summary} \label{conclusion}

We have investigated the extent to which stellar metallicity and individual elemental abundances can be inferred from kinematics, orbital properties, and a limited set of chemical inputs. Using the APOGEE sample for training and the HARPS sample for external validation, we draw the following main conclusions.

First, kinematic and orbital information alone contains very limited, predictive signal for stellar metallicity. Gradient-boosting models achieve an RMSE of $\sim0.20$ dex for APOGEE and $\sim0.22$ dex for HARPS. The dominant role of $Z_{\mathrm{max}}$ across all feature-importance diagnostics demonstrates that the vertical structure of stellar orbits is the primary dynamical tracer of metallicity in our analysis.

Second, once [Fe/H] is included, the abundances of C, O, Mg, and Si can be predicted with significantly higher precision than the baseline assumption [X/H] = [Fe/H]. In these models, metallicity is by far the dominant predictor, while orbital parameters such as $R_{\mathrm{peri}}$ and $Z_{\mathrm{max}}$ provide secondary but non-negligible contributions.

Third, the abundances of carbon and oxygen can be inferred from Mg, Si, and [Fe/H] with useful accuracy. The inclusion of kinematic information does not lead to a measurable improvement, indicating that most of the predictive information is already encoded in the elemental abundances themselves, particularly Mg. The main limitation when transferring the models to independent datasets arises from systematic differences in abundance scales rather than model performance.

Finally, we provide simple empirical relations for key abundance ratios as a function of metallicity, offering a complementary and computationally inexpensive approach for estimating quantities relevant to exoplanet composition.

Overall, our results demonstrate that ML methods provide a practical framework for estimating stellar chemical properties when direct spectroscopic determinations are not possible. At the same time, the convergence of multiple models and extensive tests indicates that predictive performance is primarily limited by the intrinsic information content of the available features. Future improvements will therefore depend more on the quality, homogeneity, and physical richness of the input data than on further increases in model complexity.

%%%%%%%%%%%%%%%%%%%%%%%%%%%%%%%%%%%%%%%%%%%%%%%%%%%%%%%%%%%%%%
\begin{acknowledgements}
This work was funded by the European Union (ERC, FIERCE, 101052347). Views and opinions expressed are however those of the author(s) only and do not necessarily reflect those of the European Union or the European Research Council. Neither the European Union nor the granting authority can be held responsible for them. The computational part of this work was supported by Funda\c{c}\~ao para a Ci\^encia e Tecnologia (FCT) through the advanced computing projects 2024.18352.CPCA and 2024.13145.CPCA. This work was supported by FCT through national funds under the research grant UID/04434/2025 (DOI 10.54499/UID/04434/2025) V.~A. acknowledges support from FCT through a work contract funded by the FCT Scientific Employment Stimulus program (reference 2023.06055.CEECIND/CP2839/CT0005, DOI: 10.54499/2023.06055.CEECIND/CP2839/CT0005). R.~C. acknowledges support by UID/00147 -- Research Center for Systems and Technologies (SYSTEC), and the Associate Laboratory Advanced Production and Intelligent Systems (ARISE, 10.54499/LA/P/0112/2020) both funded by Fundação para a Ciência e a Tecnologia, I.P./MCTES through the national funds. V.~A., Zh.~M., and A.~A.~H. acknowledge support from the Higher Education and Science Committee of MESCS RA (Research project \textnumero~24LCG--1C021). EDM acknowledges the support by the Ramón y Cajal contract RyC2022-035854-I funded by the Spanish MICIU/AEI/10.13039/501100011033 and by ESF+. B.~M.~T.~B.~S. acknowledges support from FCT, I.P.  fellowship, under the grant number 2022.11805.BD (DOI: 10.54499/2022.11805.BD).
\end{acknowledgements}

%%%%%%%%%%%%%%%%%%%%%%%%%%%%%%%%%%%%%%%%%%%%%%%%%%%%%%%%%%%%%%
% Please note that we have included the references below in
% order to compile the document, but we ask you to:
\bibliographystyle{aa} % style aa.bst
\bibliography{references} % your references Yourfile.bib

%%%%%%%%%%%%%%%%%%%%%%%%%%%%%%%%%%%%%%%%%%%%%%%%%%%%%%%%%%%%%%%
% Appendices must be placed after   \end{thebibliography}
% They will be placed automatically on a new page.
%%%%%%%%%%%%%%%%%%%%%%%%%%%%%%%%%%%%%%%%%%%%%%%%%%%%%%%%%%%%%%%
\begin{appendix}

\section{Construction of the samples} \label{target_selection}

In this section, we describe in detail the selection criteria adopted for the different samples and the determination of the relevant parameters.

\subsection{APOGEE DR17}

For the APOGEE sample, we performed a careful quality selection to ensure reliable stellar parameters and elemental abundances. We excluded stars with problematic \texttt{APOGEE\_STARFLAG} entries, removing objects affected by high persistence (\texttt{PERSIST\_HIGH}, bit~9), low signal-to-noise ratio (\texttt{LOW\_SNR}, bit~4), contamination from a very bright neighbor (\texttt{VERY\_BRIGHT\_NEIGHBOR}, bit~3), and spectra showing significant positive or negative persistence jumps in the blue chip (\texttt{PERSIST\_JUMP\_POS}, bit~12; \texttt{PERSIST\_JUMP\_NEG}, bit~13). We also excluded stars flagged as having inconsistent radial velocity determinations (\texttt{SUSPECT\_RV\_COMBINATION}, bit~16).

To ensure a homogeneous and representative dataset, we selected only stars belonging to the main APOGEE survey by requiring \texttt{EXTRATARG}~=~0, thereby excluding ancillary and calibration targets. Using the \texttt{PROGRAMNAME} flag, we further removed stars that were part of minor programs (those containing fewer than 1500 stars) targeting specific objects.

To guarantee robust stellar parameters, elemental abundances, and kinematic measurements, we applied the following additional selection criteria \citep[see e.g.,][]{Feltzing-23}: 
we retained only red giant stars with effective temperatures $4000 < T_{\mathrm{eff}} < 6000$~K, surface gravities $1.0 < \log g < 2.5$~dex, and spectra with signal-to-noise ratios $\mathrm{S/N} > 100$. We required that \texttt{FE\_H\_FLAG}~=~0 and \texttt{X\_FE\_FLAG}~=~0 for the key elements C, O, Mg, Si, and Fe, and that the uncertainties in the corresponding abundances be smaller than 0.05~dex. We further required relative parallax uncertainties $\sigma_\pi / \pi < 0.1$ to ensure reliable distances. 

Because the main motivation of the present work is to predict metallicities and abundances of exoplanet host stars, we restricted the sample to stars within the metallicity range $-1.0 < [\mathrm{Fe/H}] < 0.6$~dex. The upper limit was imposed to avoid unrealistically high values or chemically peculiar objects. We also verified that all selected stars possess measured radial velocities, \textit{Gaia} parallaxes, and proper motions, which were used to compute the kinematic properties of the sample.

After applying all the above selection criteria, the final APOGEE sample consists of about 46\,000 red giant stars.

\subsection{HARPS}

The HARPS sample consists of 798 FGK-type dwarf stars with homogeneously derived atmospheric parameters and elemental abundances \citep{Adibekyan-12, BertrandeLis-15, DelgadoMena-21}. The astrometric data for these stars were taken from \textit{Gaia}~DR3, and as for the other samples, we required relative parallax uncertainties $\sigma_\pi / \pi < 0.1$ to ensure reliable distances.

The metallicities of the HARPS stars span the range $-1.0 < [\mathrm{Fe/H}] < 0.4$~dex. This sample, composed primarily of bright, nearby stars observed in the context of exoplanet searches, provides high-precision spectroscopic abundances and serves as an excellent benchmark for testing the machine learning models.

\subsection{Kinematics and dynamics of the stars}

Using the formulation of \citet{Johnson-87}, and combining the astrometric parameters and radial velocities of the stars, we computed the Galactic space velocity components ($U_{\mathrm{LSR}}$, $V_{\mathrm{LSR}}$, and $W_{\mathrm{LSR}}$) relative to the Local Standard of Rest (LSR). The solar motion with respect to the LSR was adopted from \citet{Schonrich-10}. 

We then estimated the probability of each star belonging to one of the three main Galactic stellar populations (the thin disk, thick disk, or halo) following the prescription of \citet{Reddy-06}. This method assumes that stars in the solar neighborhood can be described as a mixture of these three kinematic populations, each characterised by Gaussian velocity distributions in the $U$, $V$, and $W$ components \citep{Schwarzschild-1907}.

To further characterise the stellar orbits, we used the \texttt{galpy} package \citep{Bovy-15} to compute orbital parameters within a realistic Milky Way gravitational potential. For each star, we derived the pericentric ($R_{\mathrm{peri}}$) and apocentric ($R_{\mathrm{apo}}$) distances, orbital eccentricity ($e$), and the maximum vertical distance from the Galactic plane ($Z_{\mathrm{max}}$). The orbit integration was carried out assuming the \texttt{MWPotential2014} model provided by \texttt{galpy}, integrating each orbit over several Galactic rotations.

These quantities provide insight into the dynamical behaviour and population membership of the stars, and the methodology has been extensively applied in previous works \citep[e.g.,][]{Adibekyan-12}.

\section{Predicting metallicity from kinematics}
\label{appendix:metallicity}

This appendix presents supplementary material related to the prediction of stellar metallicity from kinematic and orbital parameters. In particular, we provide the full set of correlation analyses and additional figures used to explore the relationships between metallicity and the input features considered in this work.

\subsection{Exploratory data analysis for metallicity determination}

In this section, we provide the detailed results of the exploratory data analysis (EDA) performed to characterise the statistical relationships between the selected parameters and stellar metallicity.

\subsubsection{Spearman rank correlation analysis}

We computed Spearman’s rank correlation coefficient $\rho$ between each kinematic feature and [Fe/H] separately for the APOGEE and HARPS samples. Spearman’s coefficient measures monotonic associations without assuming linearity.

\begin{figure*}
    \centering
    \includegraphics[width=0.8\textwidth]{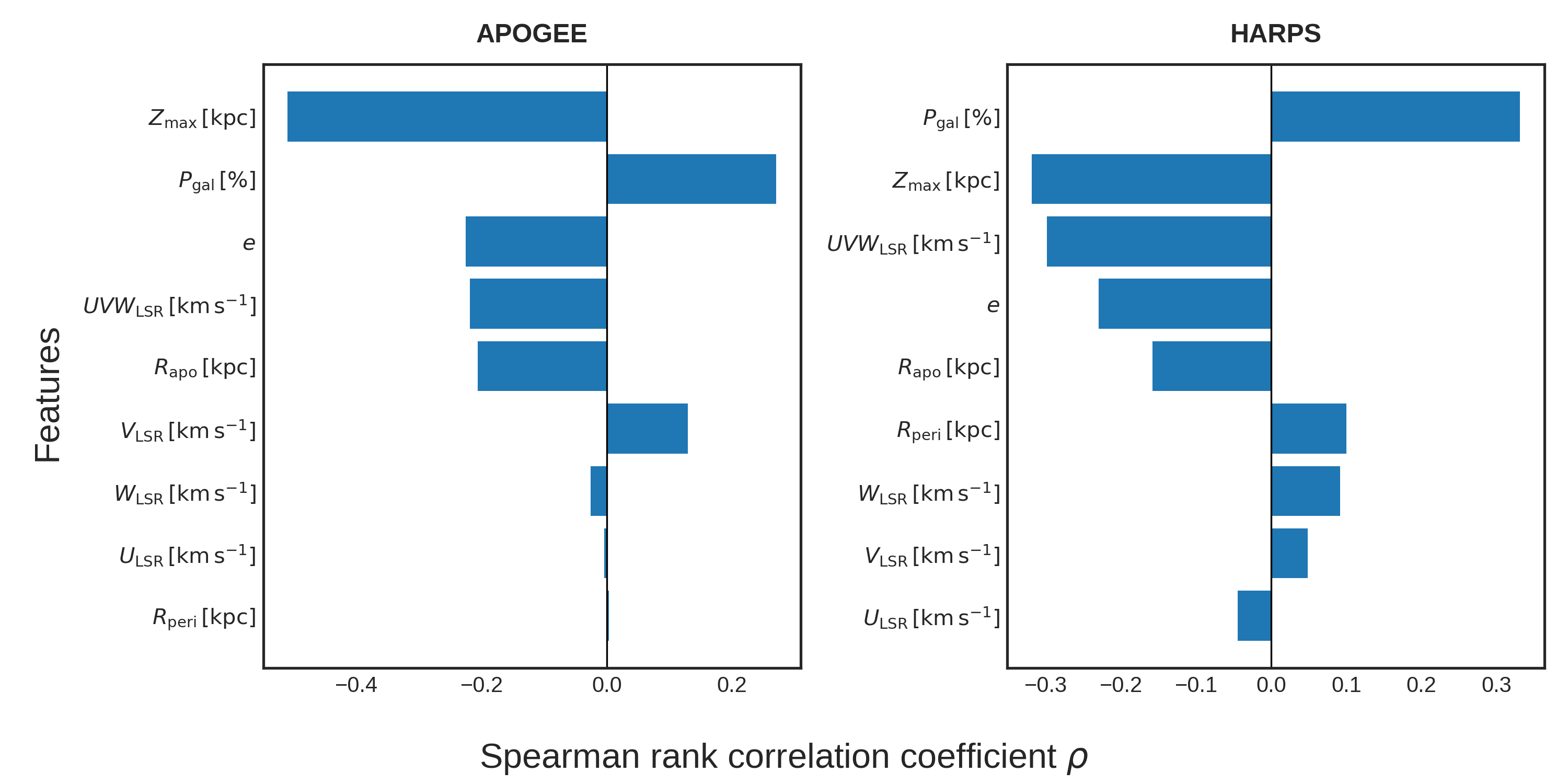}
    \caption{Spearman rank correlation coefficients between the parameters and [Fe/H] for the APOGEE (left) and HARPS (right) samples.}
    \label{fig:spearman_appendix}
\end{figure*}

Figure~\ref{fig:spearman_appendix} presents the full set of correlation coefficients. In both samples, $Z_{\max}$ exhibits the strongest negative correlation with metallicity, while $P_{\rm gal}$ shows the strongest positive correlation. The next most correlated parameters are the orbital eccentricity $e$ and the total space velocity $\mathrm{UVW_{LSR}}$.

Although the absolute values of the coefficients differ slightly between APOGEE and HARPS—likely due to differences in stellar populations and sample size—the overall ranking of highly correlated features remains consistent.

\subsubsection{Mutual information analysis}

To quantify nonlinear dependencies between the kinematic parameters and metallicity,
we computed the mutual information (MI) between each feature and [Fe/H].
Mutual information measures the general statistical dependence between two variables. Unlike correlation coefficients, MI does not assume linearity or monotonicity and is sensitive to arbitrary nonlinear relationships.

An MI value of zero indicates statistical independence, while larger values imply stronger dependence. In this work, MI was estimated using a non-parametric k-nearest-neighbour approach (with $k$ = 3), which is well suited for continuous variables \citep{kraskov2004}.

The MI values for both samples are shown in Figure~\ref{fig:mi_vif_appendix}. Consistent with the correlation analysis, $Z_{\max}$ and $P_{\rm gal}$ (also eccentricity) contain the largest amount of predictive information.

In addition, parameters such as $R_{\rm peri}$ show enhanced importance in the MI analysis compared to its rank correlation, suggesting the presence of nonlinear contributions to the metallicity-kinematics relation.

\subsubsection{Multicollinearity diagnostics}

We evaluated multicollinearity among the predictors using the variance inflation factor (VIF), defined as
\begin{equation}
\mathrm{VIF}_j = \frac{1}{1 - R_j^2},
\end{equation}
where $R_j^2$ is obtained by regressing feature $j$ against all remaining predictors. The VIF quantifies how much the variance of a regression coefficient is inflated due to linear dependence among input variables.

A VIF value of 1 indicates no multicollinearity, values between 1 and 5 are generally considered mild, values above 5 indicate moderate collinearity, and values exceeding 10 are typically interpreted as substantial multicollinearity.

High multicollinearity can lead to unstable coefficient estimates and inflated variance in linear models. However, its impact depends on the nature of the predictive algorithm. Tree-based ensemble methods, which perform recursive partitioning rather than coefficient estimation, are generally more robust to correlated predictors.

\begin{figure}
    \centering
    \includegraphics[width=0.49\textwidth]{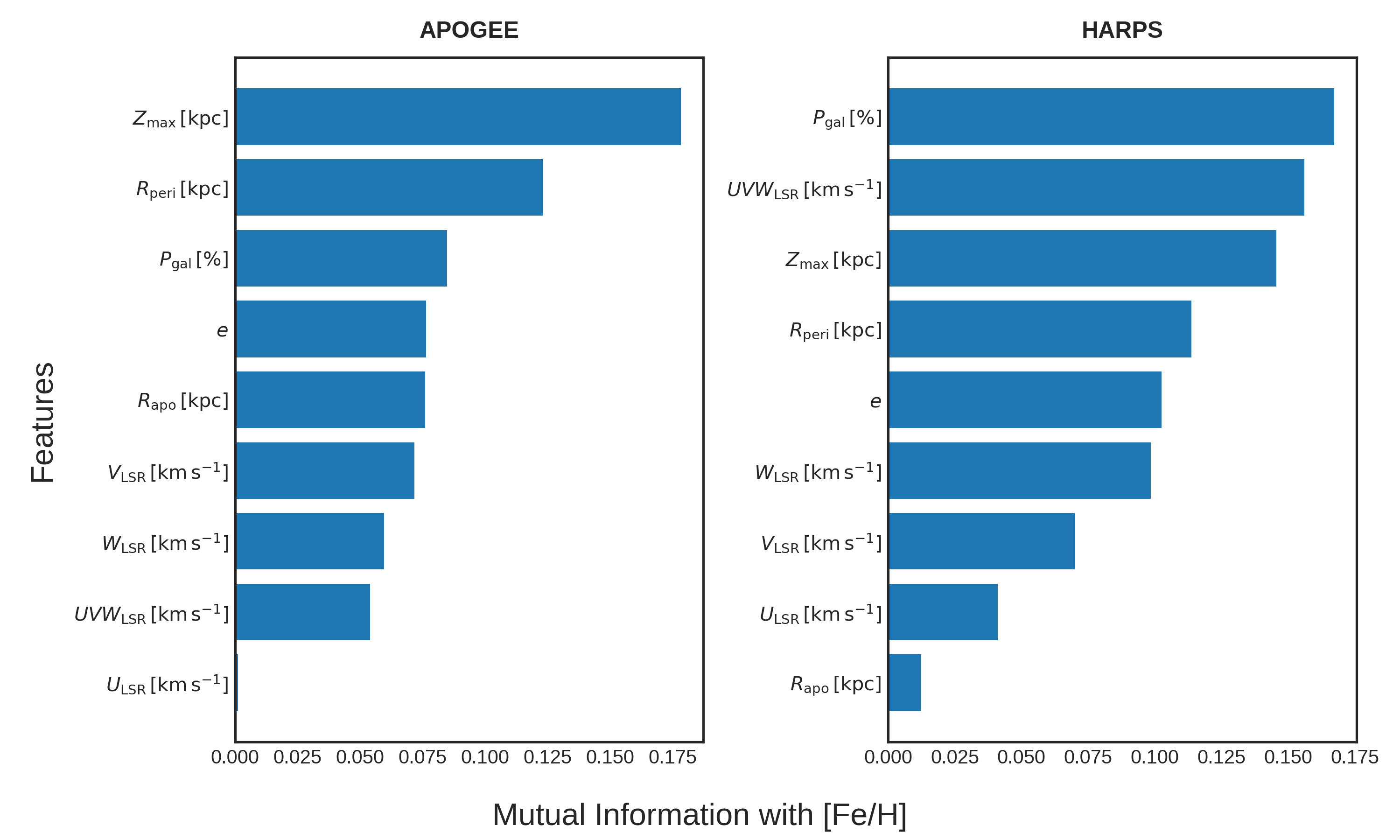}
    \vspace{0.4cm}
    \includegraphics[width=0.49\textwidth]{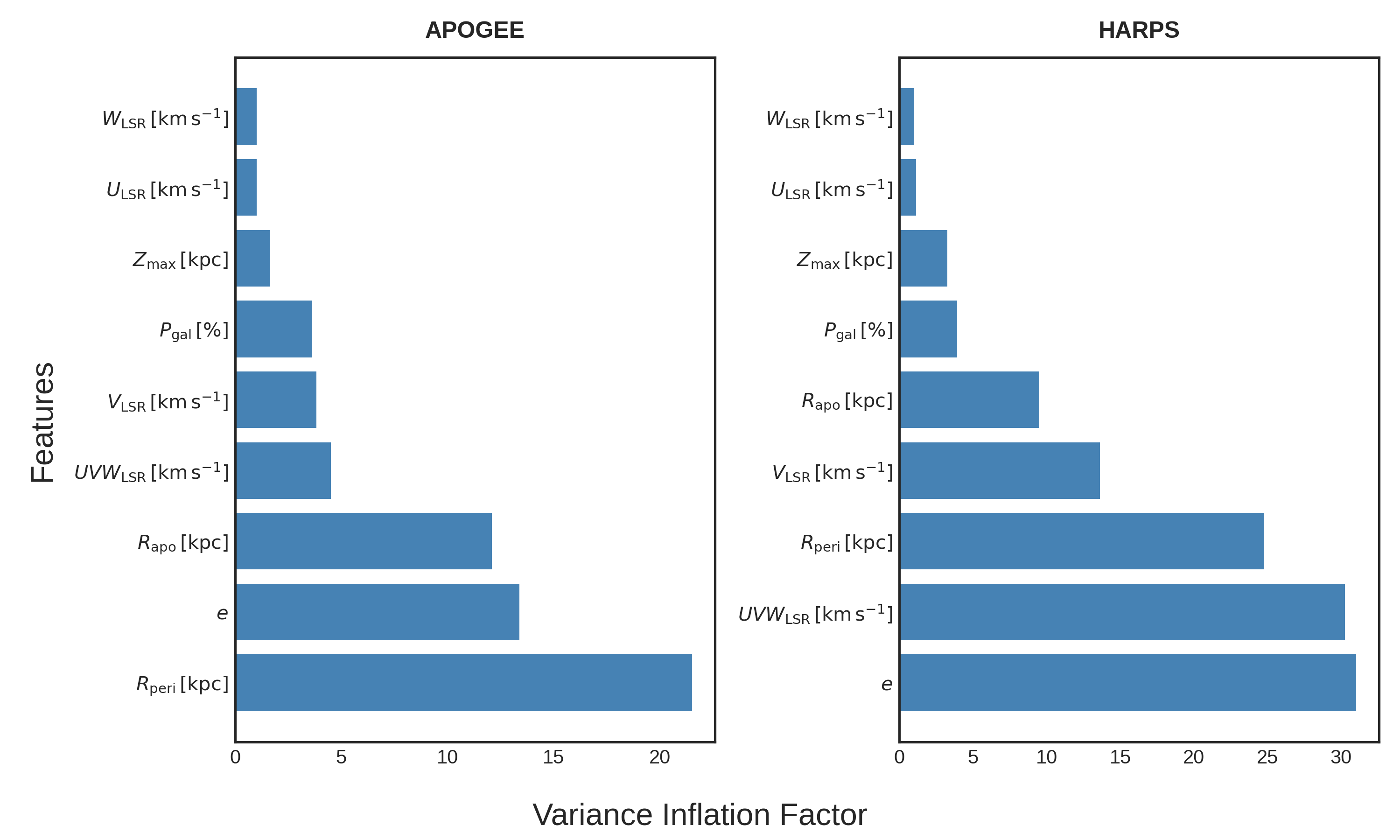}
    \caption{Mutual information between kinematic parameters and [Fe/H] (top), and Variance inflation factors (VIF) for the kinematic parameters (bottom) in the APOGEE and HARPS samples.}
    \label{fig:mi_vif_appendix}
\end{figure}

The resulting VIF values are shown in Figure~\ref{fig:mi_vif_appendix}. As expected, several parameters exhibit large VIF values due to their deterministic dependence on the other parameters. $Z_{\max}$ and $P_{\rm gal}$ display comparatively lower VIF values, indicating that they are not strongly linearly redundant with the remaining parameters.

\subsubsection{Comparison of parameter distributions}

\begin{figure*}
\centering
\begin{subfigure}{0.25\textwidth}
    \includegraphics[width=\linewidth]{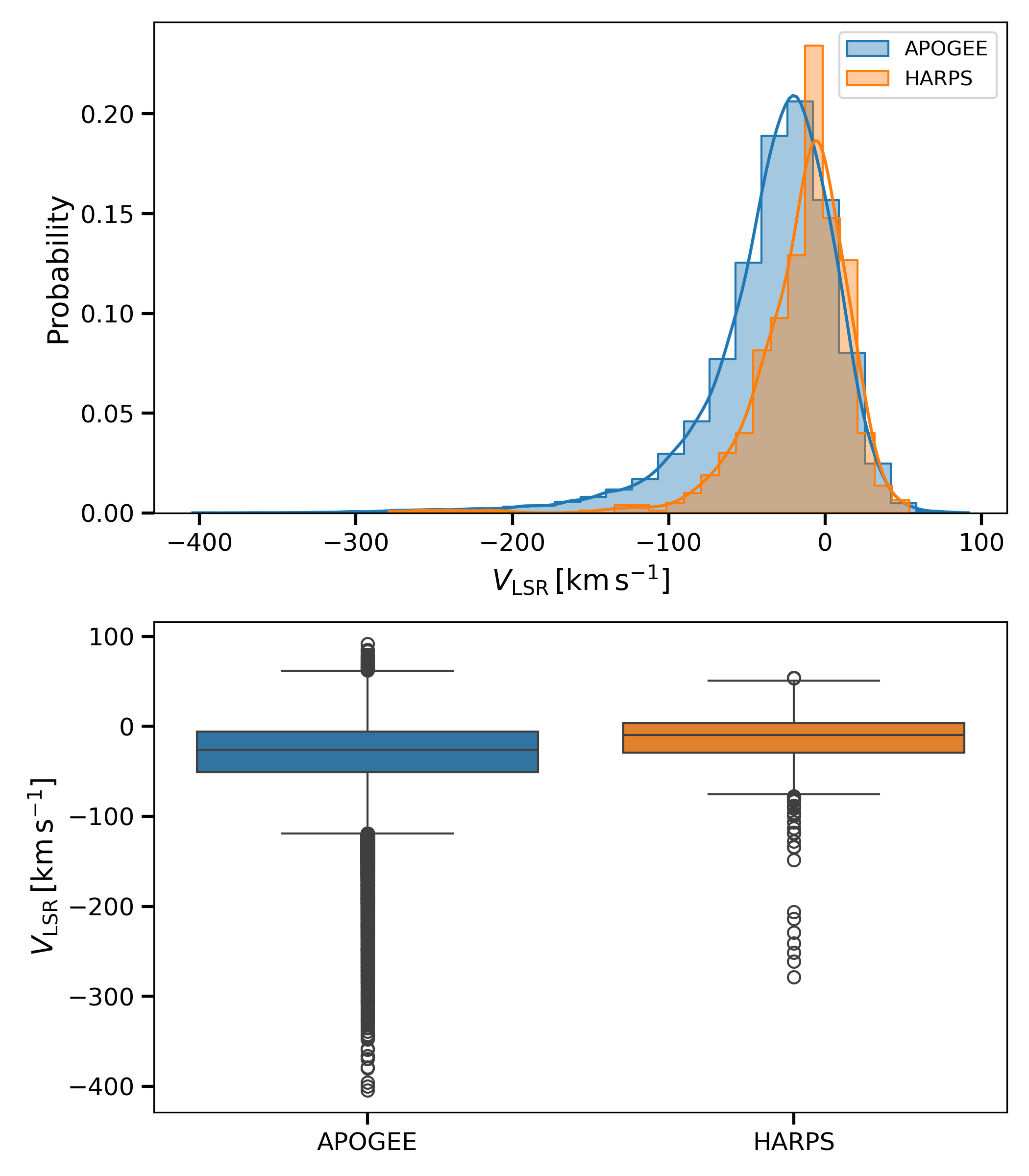}
    \caption{$V_{\mathrm{LSR}}\, [km/s]$}
\end{subfigure}
\begin{subfigure}{0.25\textwidth}
    \includegraphics[width=\linewidth]{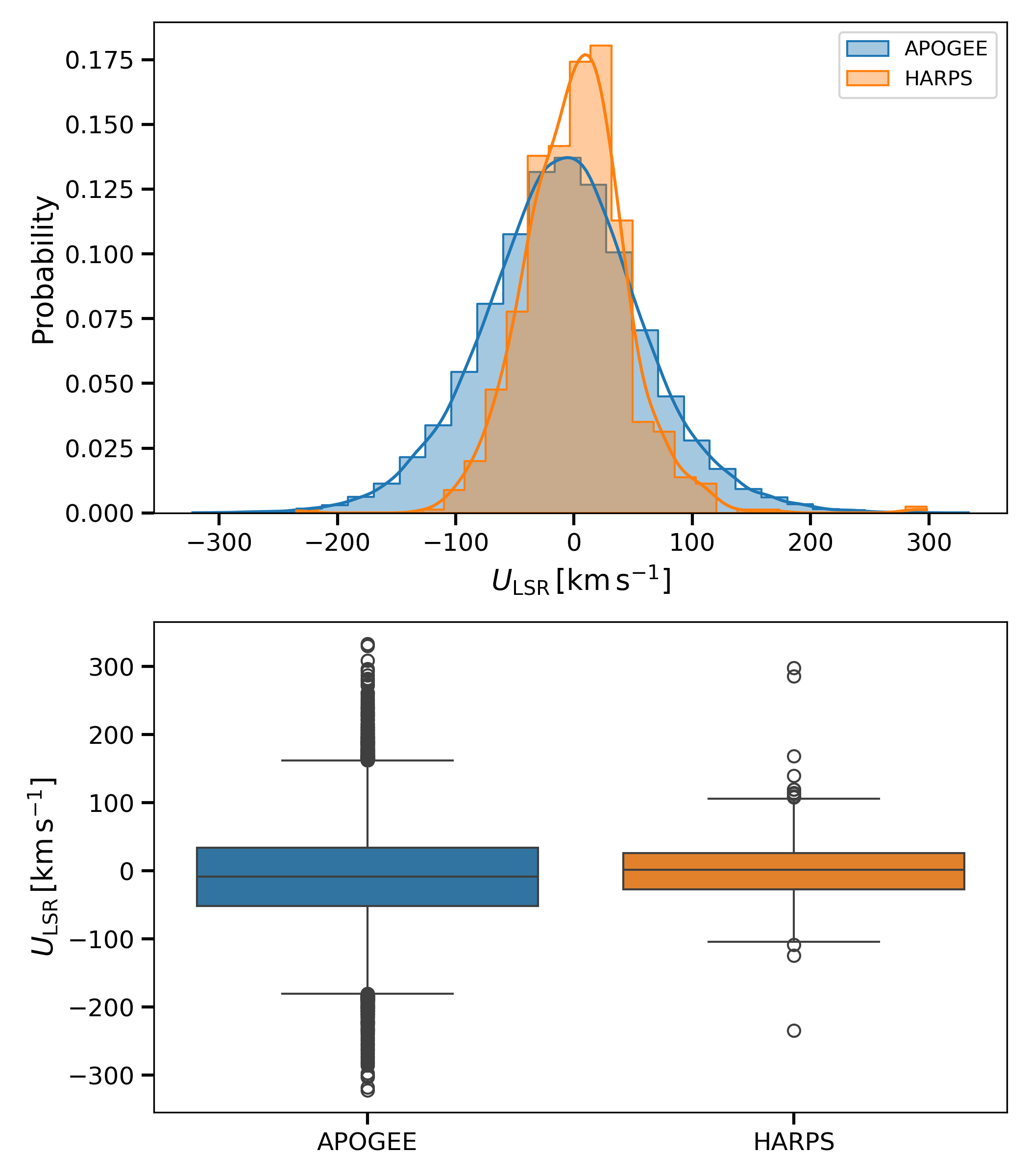}
    \caption{$U_{\mathrm{LSR}}\, [km/s]$}
\end{subfigure}
\begin{subfigure}{0.25\textwidth}
    \includegraphics[width=\linewidth]{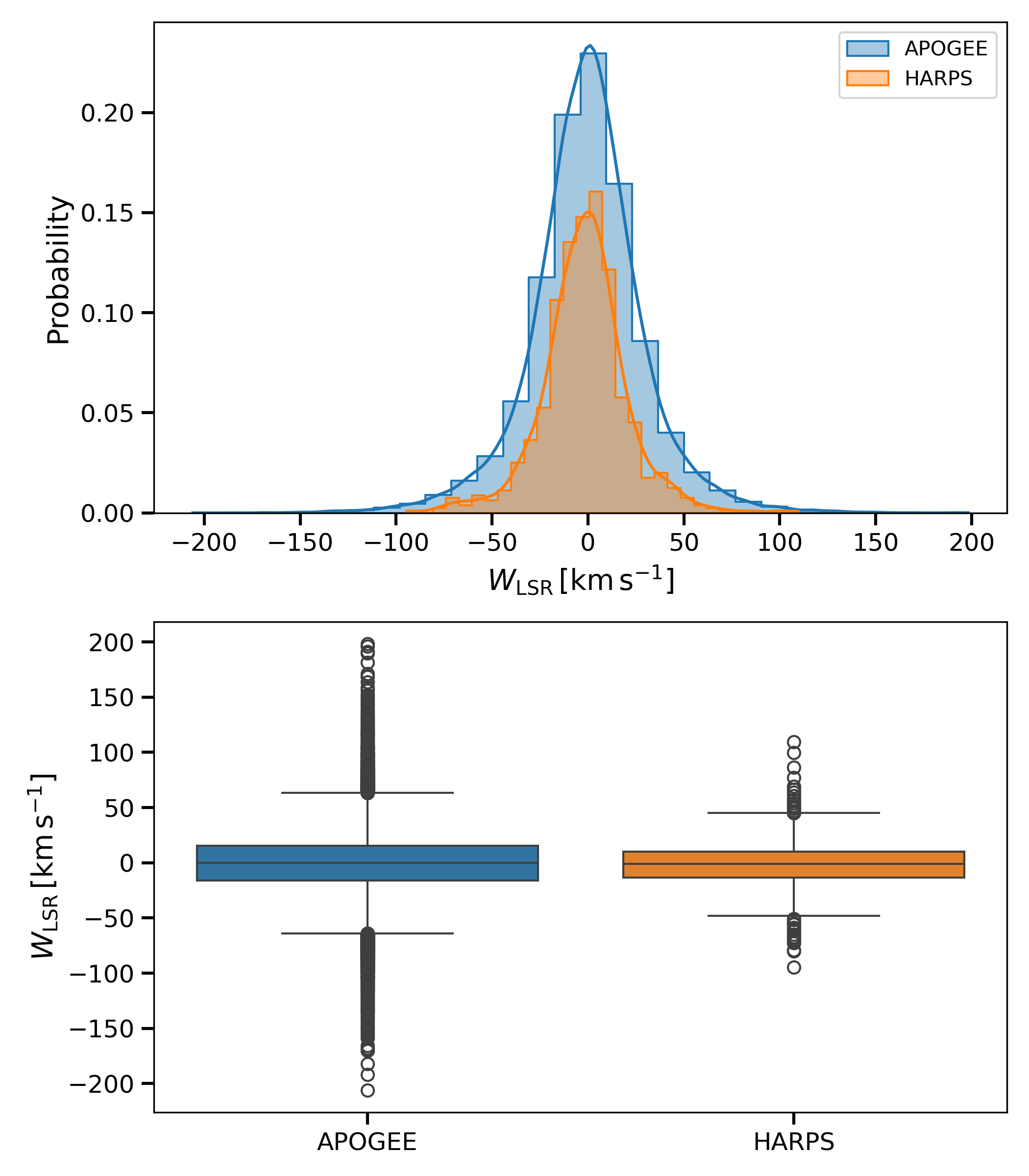}
    \caption{$W_{\mathrm{LSR}}\, [km/s]$}
\end{subfigure}

\vspace{0.4cm}

\begin{subfigure}{0.25\textwidth}
    \includegraphics[width=\linewidth]{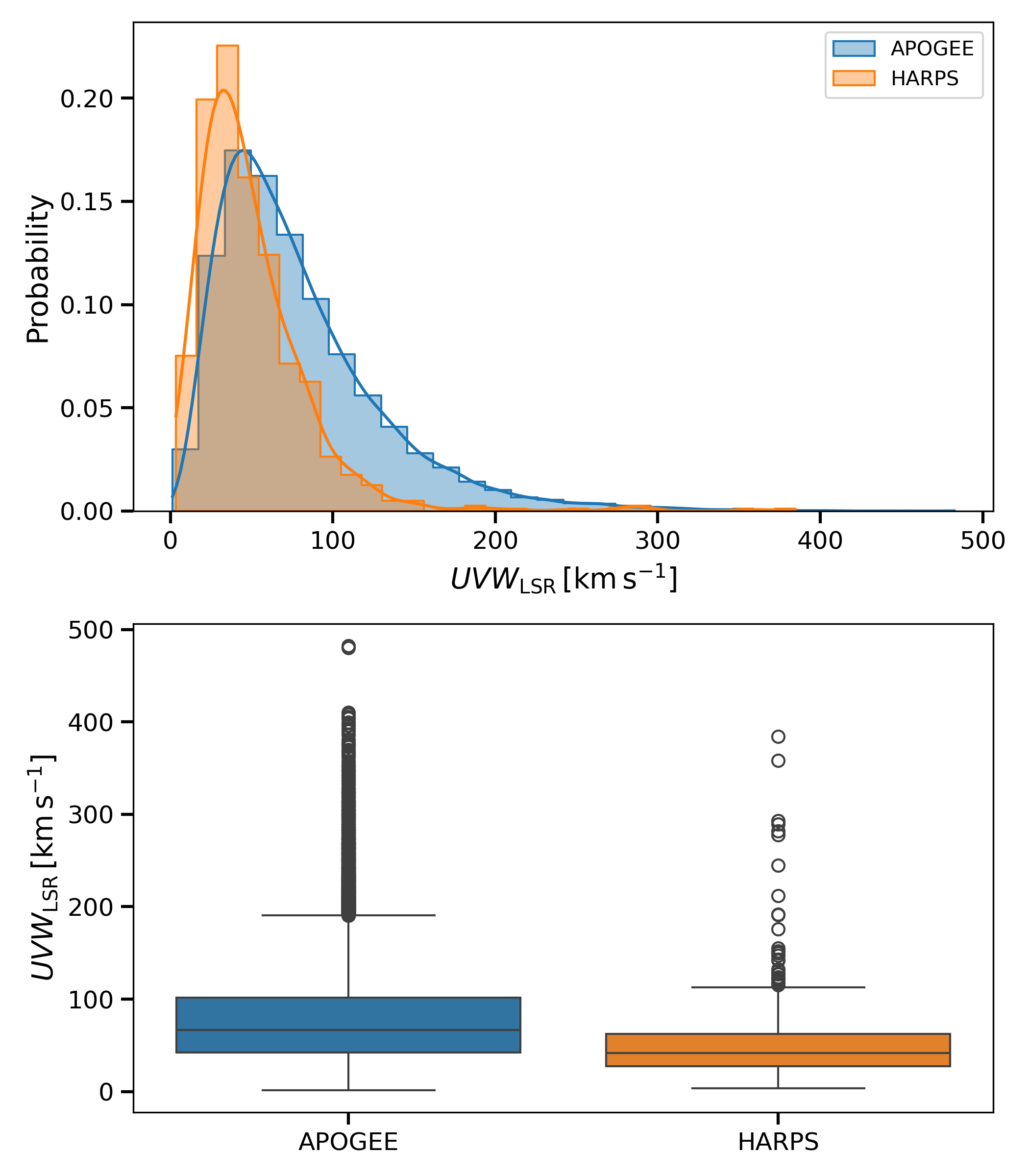}
    \caption{$UVW_{\mathrm{LSR}}\, [km/s]$}
\end{subfigure}
\begin{subfigure}{0.25\textwidth}
    \includegraphics[width=\linewidth]{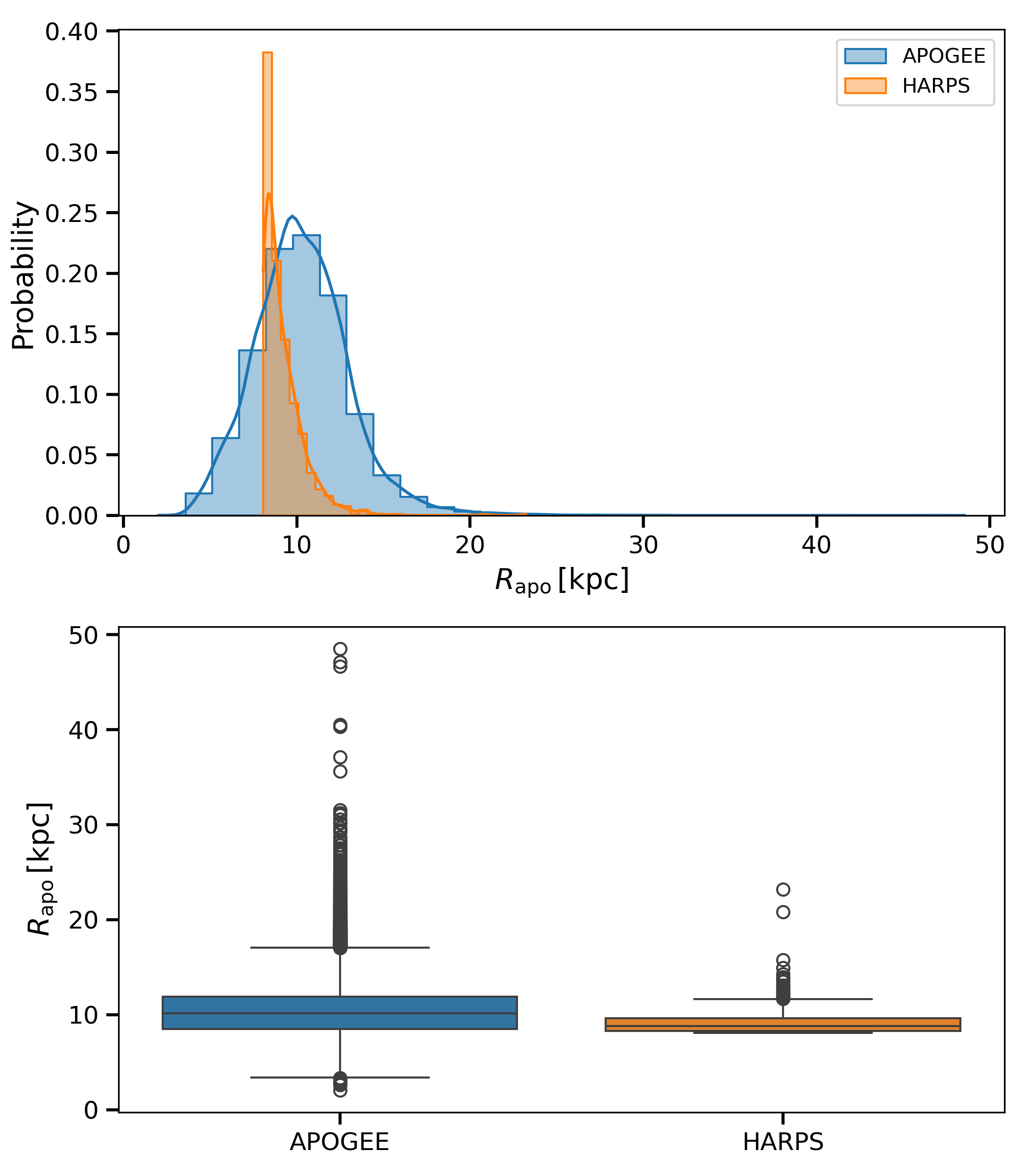}
    \caption{$R_{\mathrm{apo}}\, [Kpc]$}
\end{subfigure}
\begin{subfigure}{0.25\textwidth}
    \includegraphics[width=\linewidth]{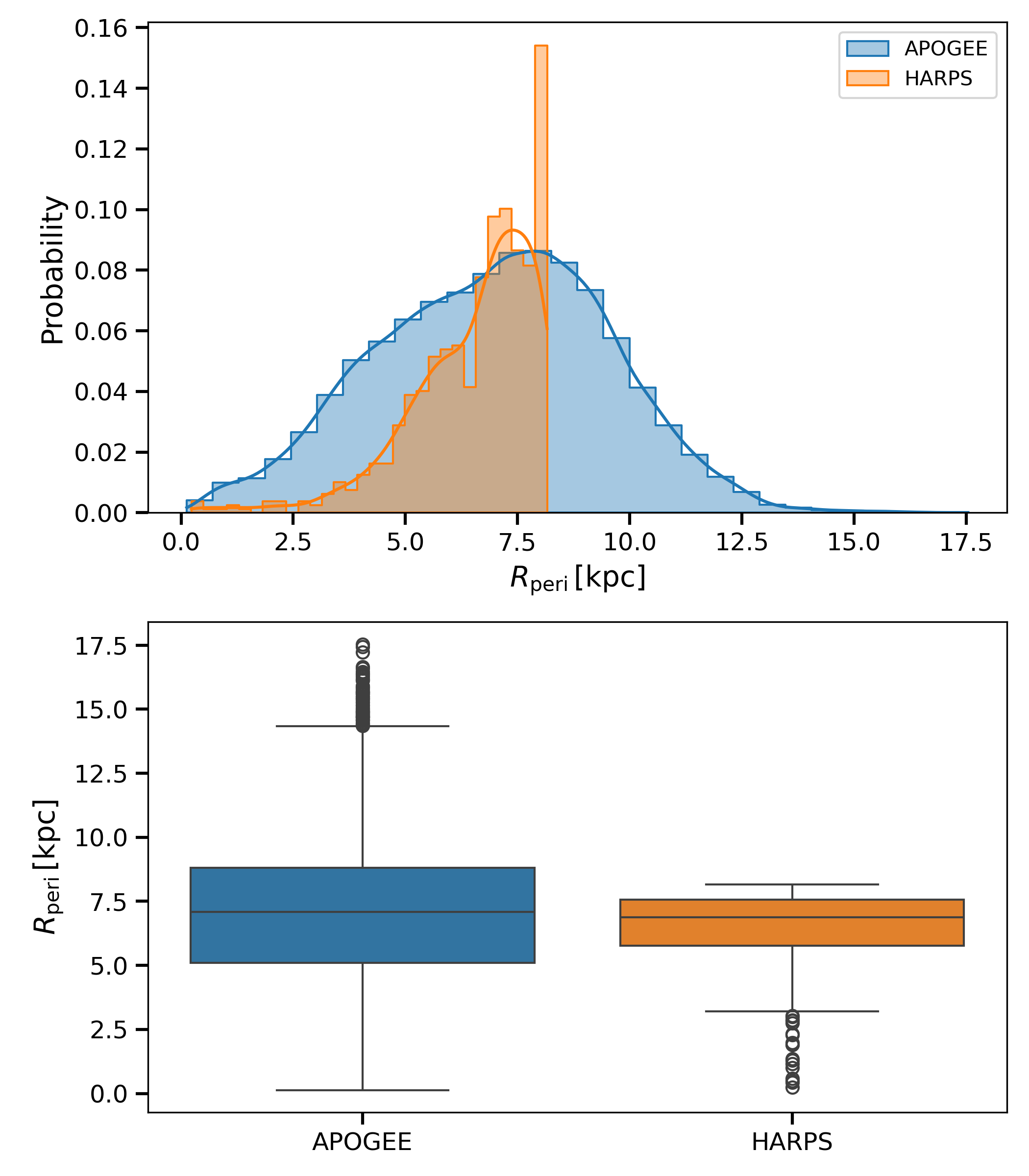}
    \caption{$R_{\mathrm{peri}}\, [Kpc]$}
\end{subfigure}

\vspace{0.4cm}

\begin{subfigure}{0.25\textwidth}
    \includegraphics[width=\linewidth]{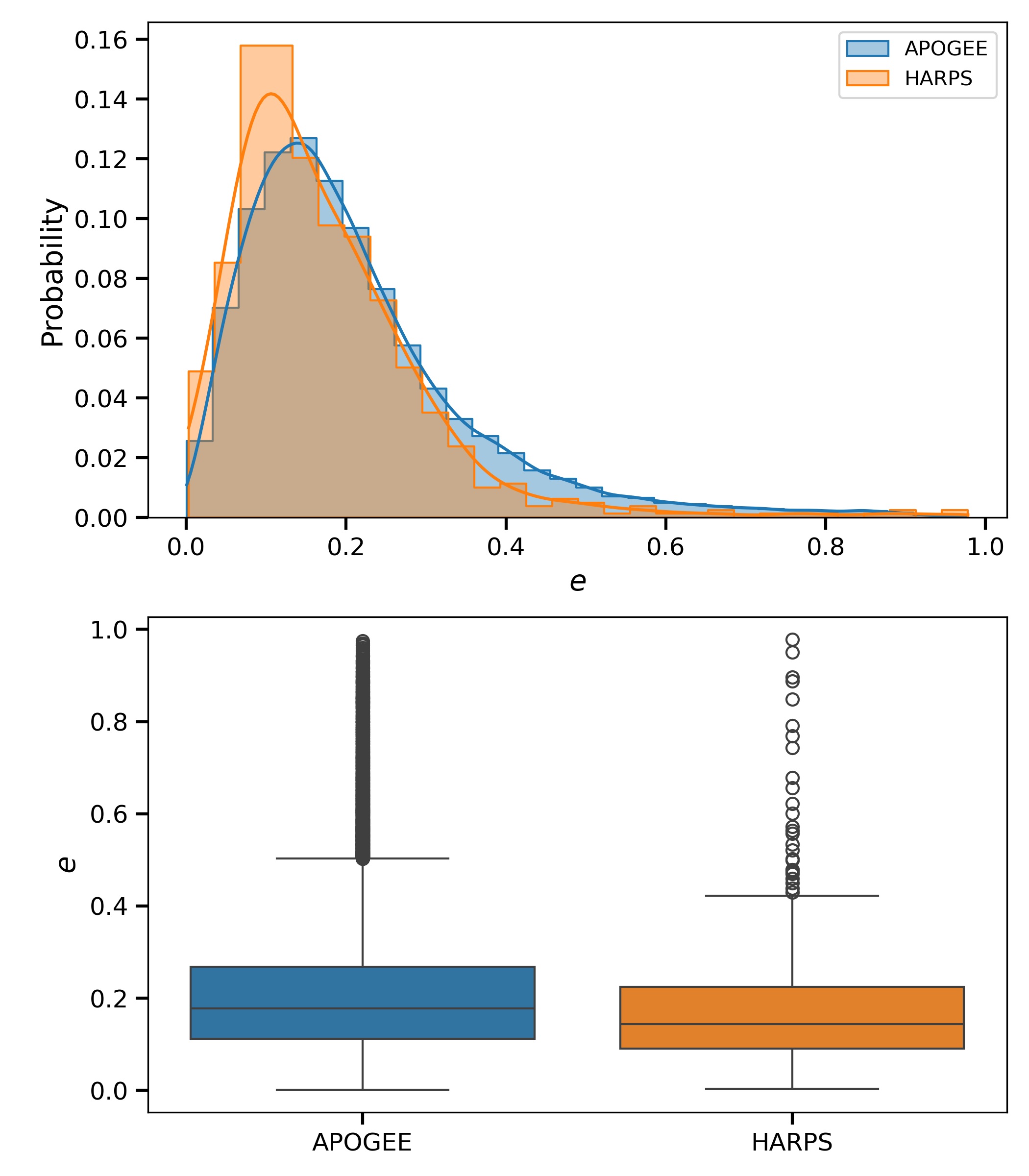}
    \caption{$e$}
\end{subfigure}
\begin{subfigure}{0.25\textwidth}
    \includegraphics[width=\linewidth]{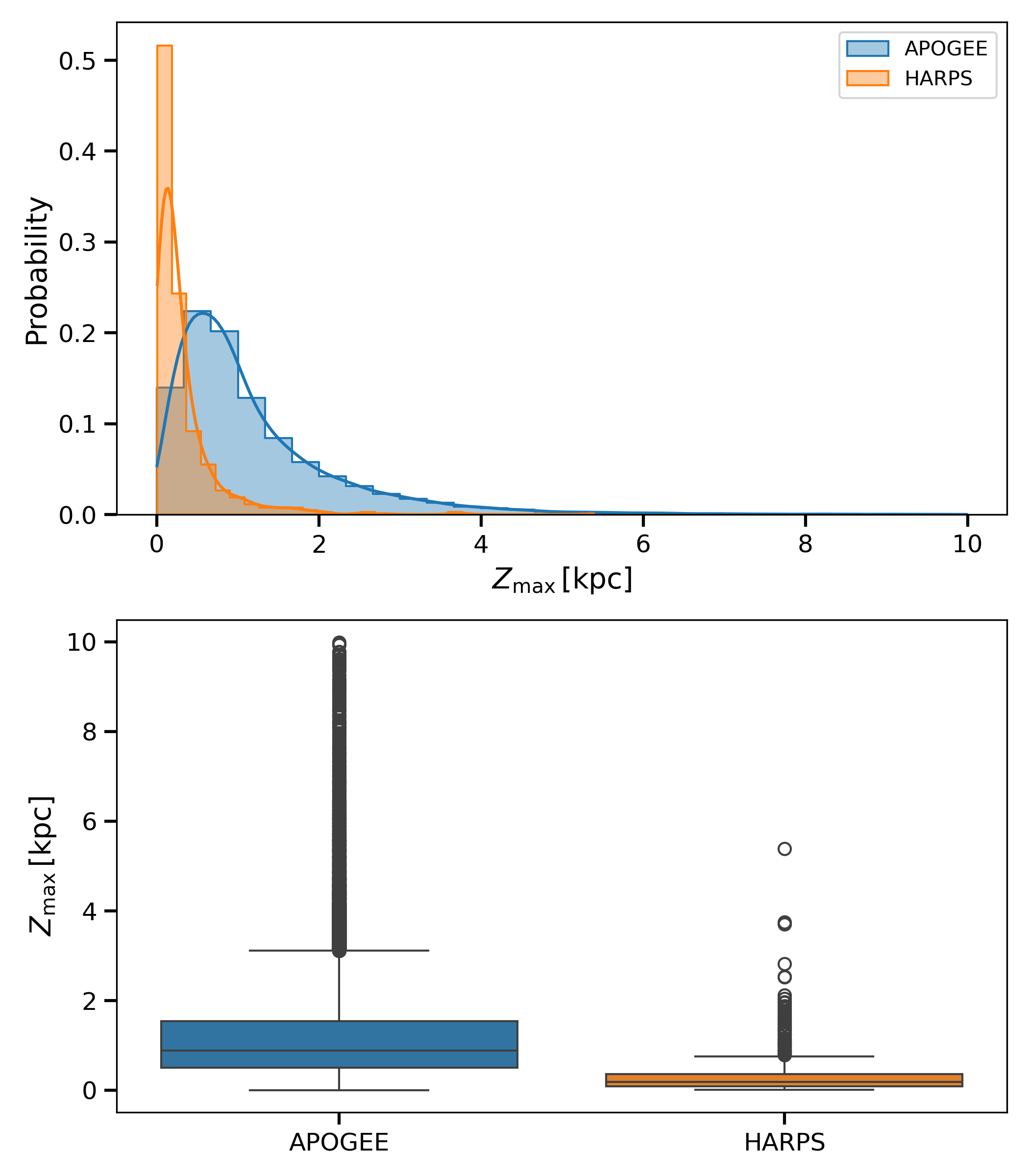}
    \caption{$Z_{\max}\, [Kpc]$}
\end{subfigure}
\begin{subfigure}{0.25\textwidth}
    \includegraphics[width=\linewidth]{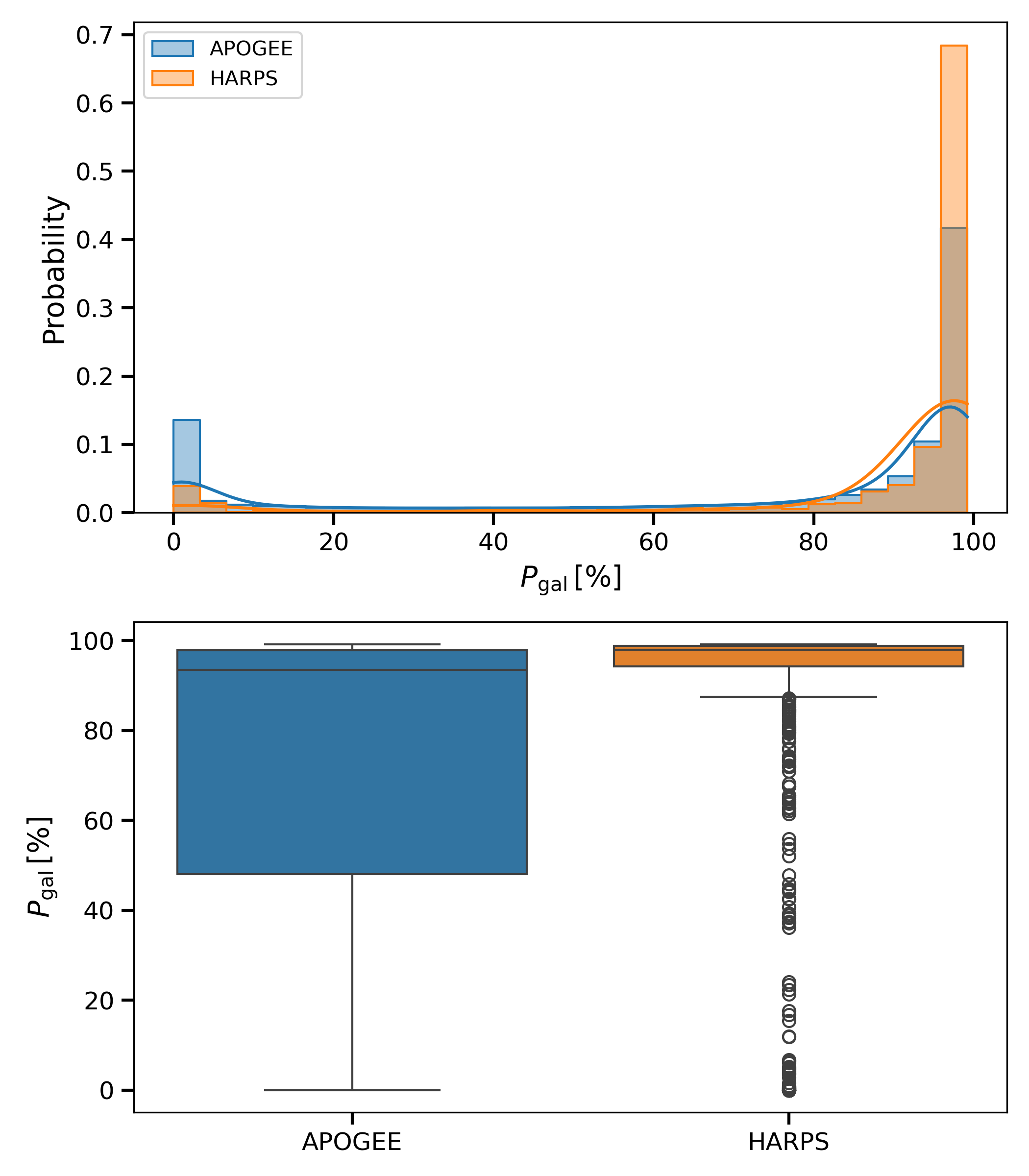}
    \caption{$P_{\mathrm{gal}}\, [\%]$}
\end{subfigure}

\caption{Comparison of kinematic and orbital parameters distributions between the APOGEE and HARPS samples. Upper panels show probability-normalised histograms; lower panels show boxplots.}
\label{fig:feature_distributions}
\end{figure*}

Figure~\ref{fig:feature_distributions} presents the distributions of the kinematic and orbital parameters for the APOGEE and HARPS samples. The upper panels show probability-normalised histograms, while the lower panels display boxplots summarizing the median and spread of each parameter.

Some differences are observed between the two samples. The HARPS dataset, composed primarily of nearby solar-type dwarf stars, is confined to a relatively narrow range of velocities and orbital parameters, reflecting its local nature and selection for exoplanet surveys.
In contrast, the APOGEE red giant sample spans a significantly broader range in velocity space and orbital properties, including larger values of $Z_{\max}$, eccentricity, and total space velocity $\mathrm{UVW_{LSR}}$, consistent with its wider spatial coverage across different Galactic environments.

Despite these differences in dispersion, the APOGEE sample largely encompasses the parameter range covered by HARPS. This overlap ensures that the machine-learning models trained on APOGEE data operate within the dynamical regime relevant for the local HARPS stars, while also capturing a broader diversity of Galactic populations.

\subsection{Regression benchmarks and ensemble models for [Fe/H]}

This section provides the detailed benchmarking results for the regression models used to predict stellar metallicity from kinematic parameters, together with the configuration of the ensemble methods. The goal of this analysis was to assess the intrinsic predictive power of the kinematic features under a consistent evaluation framework rather than to optimise individual algorithms.

All models were evaluated using 5-fold cross-validation with identical data splits. Performance was quantified using the root mean squared error (RMSE), mean absolute error (MAE), and coefficient of determination ($R^2$). Table~\ref{tab:cv_scores_full} summarizes the mean performance across folds, with the quoted RMSE uncertainties corresponding to the standard deviation over folds. The small standard deviations of the RMSE across folds indicate stable model performance.

\begin{table}
\caption{Performance of regression models evaluated using 5-fold cross-validation.
Reported values correspond to the mean across folds, with RMSE uncertainties
indicating the standard deviation.}
\centering
\begin{tabular}{lccc}
\hline
Model & RMSE (dex) & MAE (dex) & $R^2$ \\
\hline
Linear Regression & $0.213 \pm 0.001$ & 0.169 & 0.329 \\
Random Forest & $0.203 \pm 0.001$ & 0.159 & 0.391 \\
Extra Trees & $0.204 \pm 0.001$ & 0.160 & 0.384 \\
XGBoost & $0.204 \pm 0.001$ & 0.160 & 0.386 \\
XGBoost (QT) & $0.204 \pm 0.001$ & 0.160 & 0.386 \\
LightGBM & $0.199 \pm 0.001$ & 0.156 & 0.412 \\
LightGBM (QT) & $0.199 \pm 0.001$ & 0.156 & 0.412 \\
CatBoost & $0.200 \pm 0.001$ & 0.157 & 0.411 \\
CatBoost (QT) & $0.200 \pm 0.001$ & 0.156 & 0.408 \\
Voting Regressor & $0.200 \pm 0.001$ & 0.157 & 0.410 \\
Stacking Regressor & $0.199 \pm 0.001$ & 0.156 & 0.413 \\
\hline
\end{tabular}
\label{tab:cv_scores_full}
\end{table}

We also considered target ([Fe/H]) transformation using a quantile normalization (QT), which maps the metallicity distribution onto a Gaussian form via its empirical cumulative distribution function, but it did not improve predictive accuracy. This is consistent with the approximately symmetric metallicity distribution and the robustness of tree-based ensemble methods to monotonic target transformations.

For reference, the standard deviation of the metallicity distribution is $\sigma_{\rm [Fe/H]} \approx 0.26$ dex, corresponding to the RMSE of a trivial mean predictor. The fraction of explained variance can be expressed as $R^2 = 1 - \mathrm{RMSE}^2 / \sigma_{\rm [Fe/H]}^2$. Using the best-performing model (RMSE $\approx 0.199$ dex) yields $R^2 \approx 0.41$, indicating that approximately 40\% of the total metallicity variance is captured by the kinematic features.

We further explored ensemble strategies. The voting regressor combines multiple base learners through a weighted average of their predictions,
\begin{equation}
\hat{y}_{\rm vote} = \sum_{m=1}^{M} w_m \hat{y}_m ,
\end{equation}
where $w_m$ are non-negative weights satisfying $\sum w_m = 1$. The ensemble included LightGBM, CatBoost, Extra Trees, XGBoost, and Random Forest, with weights assigned inversely proportional to their cross-validated RMSE.

Stacking was implemented using the same set of base learners. To avoid information leakage, out-of-fold (OOF) predictions were generated for each base model using 5-fold cross-validation. These OOF predictions were used as input features to train a ridge regression meta-learner, yielding the final prediction
\begin{equation}
\hat{y}_{\rm stack} = f_{\rm meta}(\hat{y}_{\rm lgbr}^{\rm OOF},
\hat{y}_{\rm catbr}^{\rm OOF},
\hat{y}_{\rm extr}^{\rm OOF},
\hat{y}_{\rm xgbr}^{\rm OOF},
\hat{y}_{\rm rfr}^{\rm OOF}) .
\end{equation}

To assess ensemble diversity, we computed the Pearson correlation between the OOF predictions of the base models. The boosting models exhibit very high mutual correlations ($r \gtrsim 0.98$), while Extra Trees shows slightly lower correlation ($r \sim 0.95$). This strong similarity between base-model predictions explains the limited improvement obtained from voting and stacking.

\subsubsection{Hyperparameter optimisation of gradient boosting models}

We performed hyperparameter optimisation for two gradient boosting models: LightGBM, and XGBoost. Although XGBoost was not among the best-performing models in the default-configuration benchmark (Table~\ref{tab:cv_scores_full}), it was included in the tuning analysis because of its strong empirical performance in tabular-data machine learning tasks and its widespread use in competitive benchmarks (e.g., Kaggle competitions\footnote{https://www.kaggle.com/}).

Hyperparameter optimisation was carried out using the Optuna framework \citep{Akiba-19}, which implements efficient Bayesian optimisation strategies for hyperparameter search. The objective was to minimize the RMSE using the same 5-fold splits adopted in the baseline experiments.

The tuning focused on the most relevant hyperparameters controlling model complexity, regularization, and generalization performance. These include the number of boosting iterations, learning rate, tree depth (or number of leaves), subsampling fractions, and $L_1$/$L_2$ regularization terms. These parameters directly regulate the bias–variance trade-off and the effective model capacity.

The resulting best cross-validated RMSE values are:

\begin{itemize}
\item LightGBM: RMSE = 0.1988 dex
\item XGBoost: RMSE = 0.1987 dex
\end{itemize}

Compared to the default configurations reported in Table~\ref{tab:cv_scores_full}, the improvement is marginal for LightGBM and appears only at the third decimal level in RMSE. For XGBoost, hyperparameter optimisation brings its performance to essentially the same precision level as the other two boosting methods. This behaviour further confirms that predictive performance is primarily constrained by the intrinsic information content of the kinematic features rather than by hyperparameter configuration or the specific boosting architecture.

\subsubsection{Feature importance analysis} \label{appendix:feature_importance_feh}

This appendix presents the full set of feature-importance figures discussed in Sect.\ref{main_feat_imp_feh}. For completeness, we show the results of the built-in gain importance, permutation importance, SHAP global importance, and single-feature predictive performance for both LightGBM and XGBoost. These figures illustrate the consistency of the feature ranking across different importance metrics and model architectures.

\begin{figure}[ht]
\centering
\includegraphics[width=0.24\textwidth]{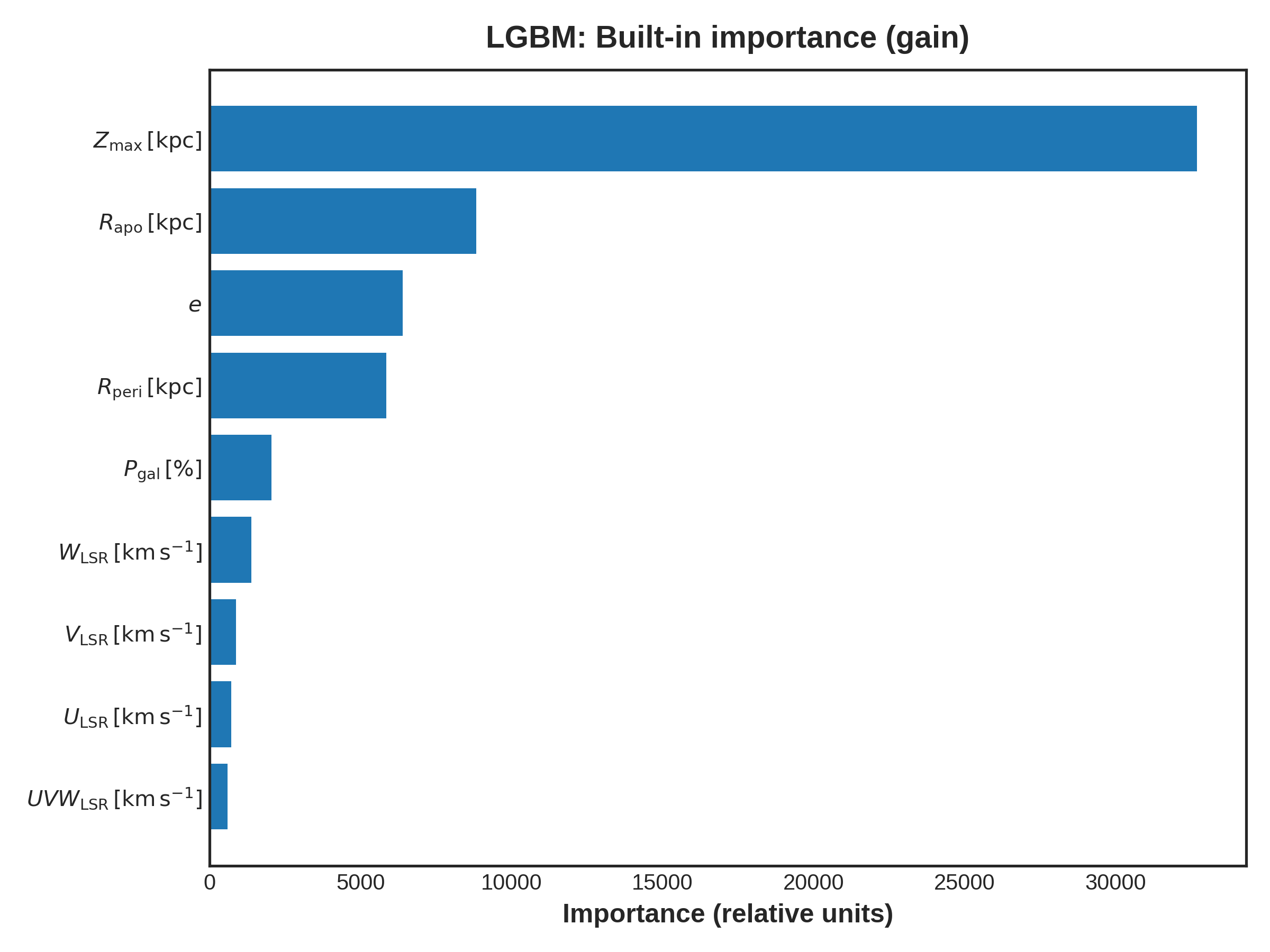}
\includegraphics[width=0.24\textwidth]{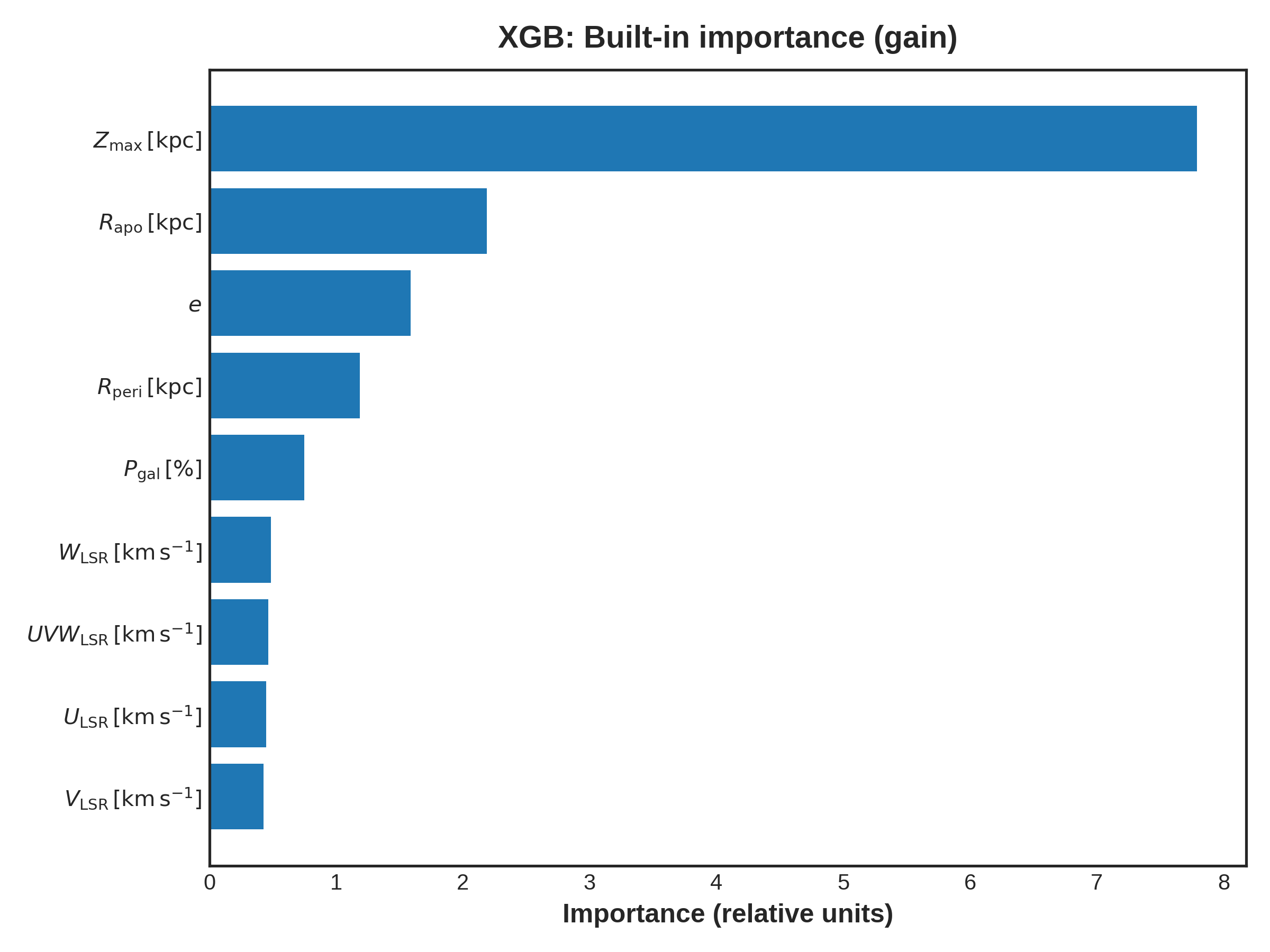}
\vspace{0.4cm}
\includegraphics[width=0.24\textwidth]{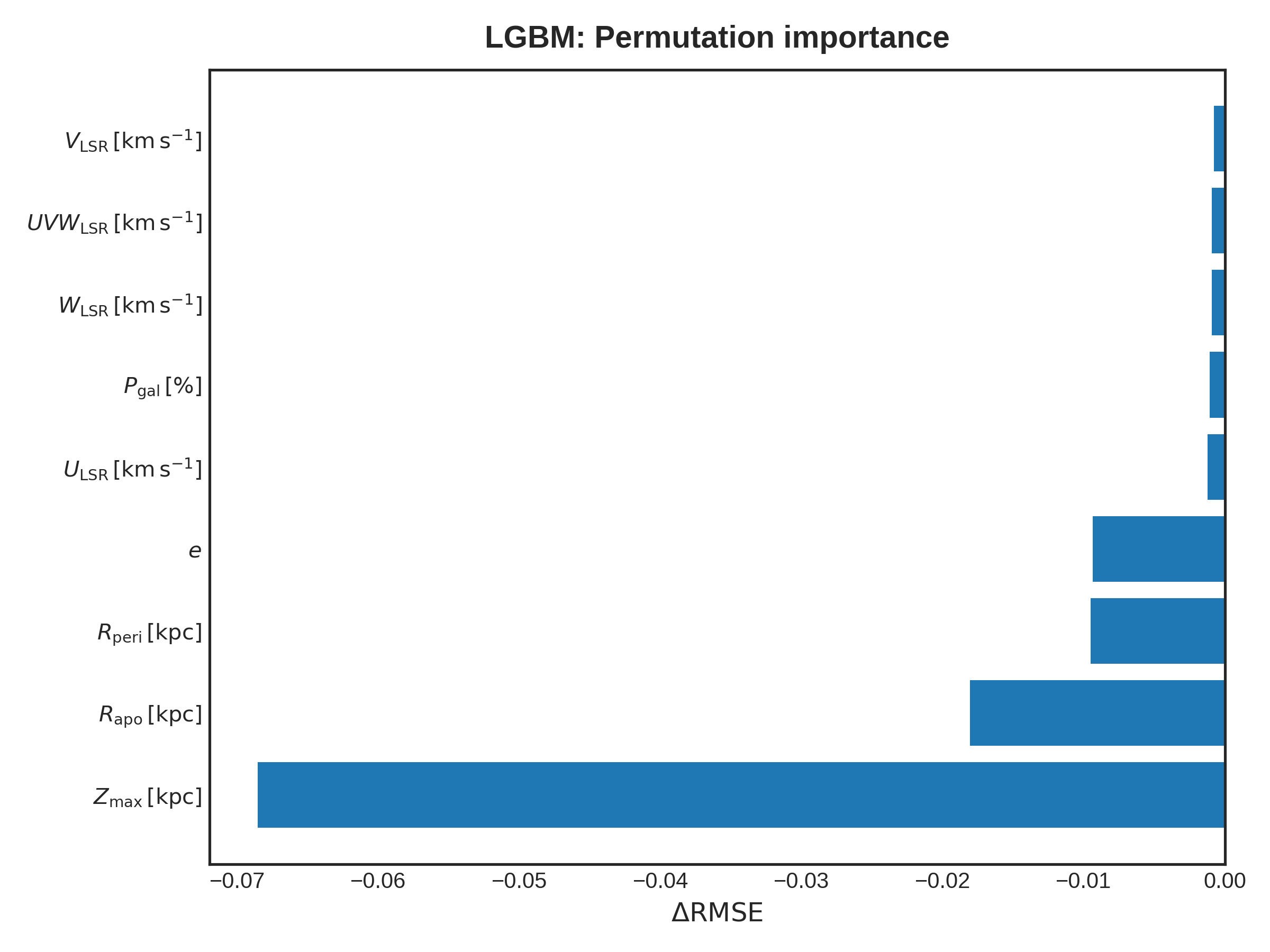}
\includegraphics[width=0.24\textwidth]{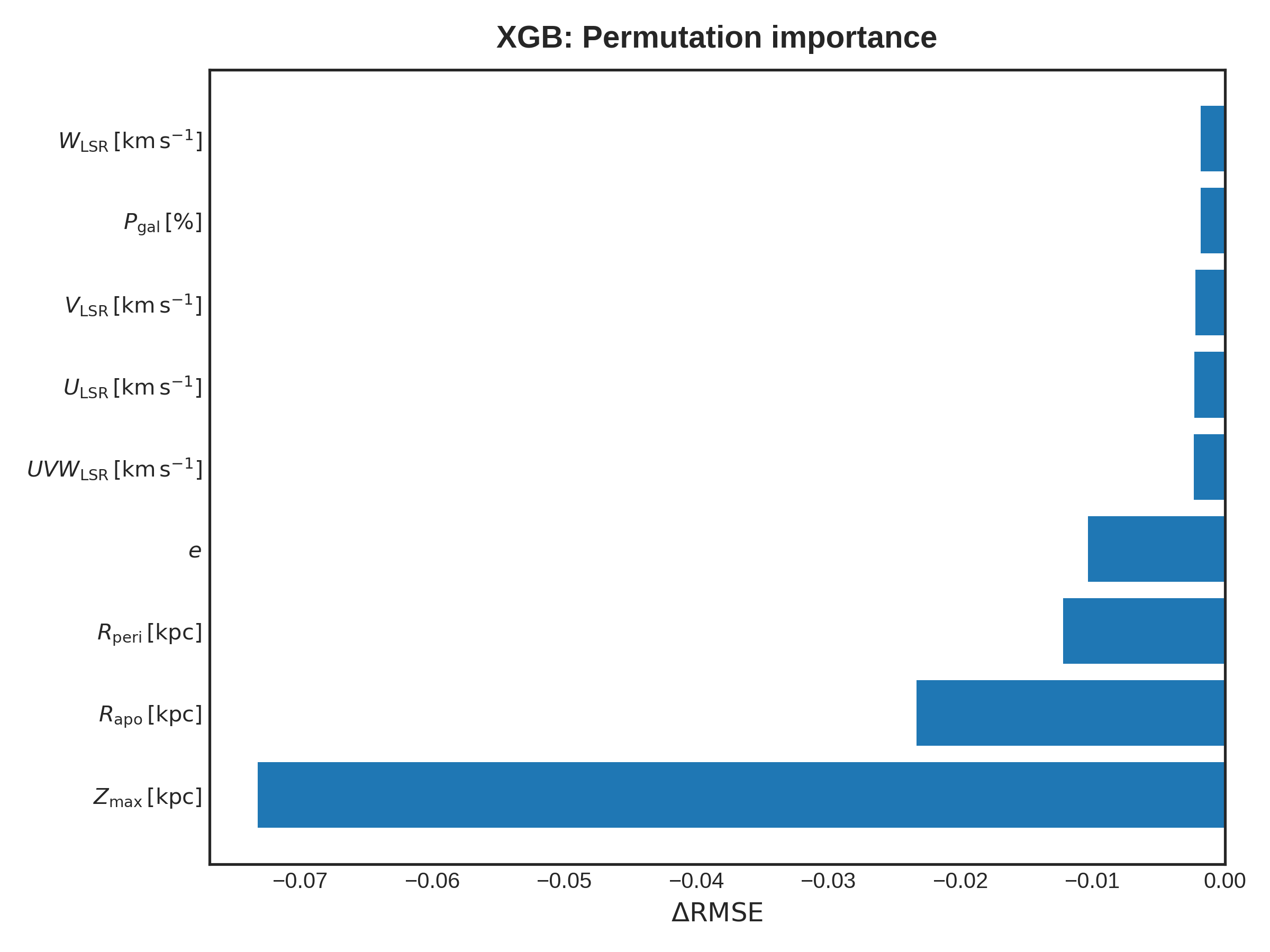}
\vspace{0.4cm}
\includegraphics[width=0.24\textwidth]{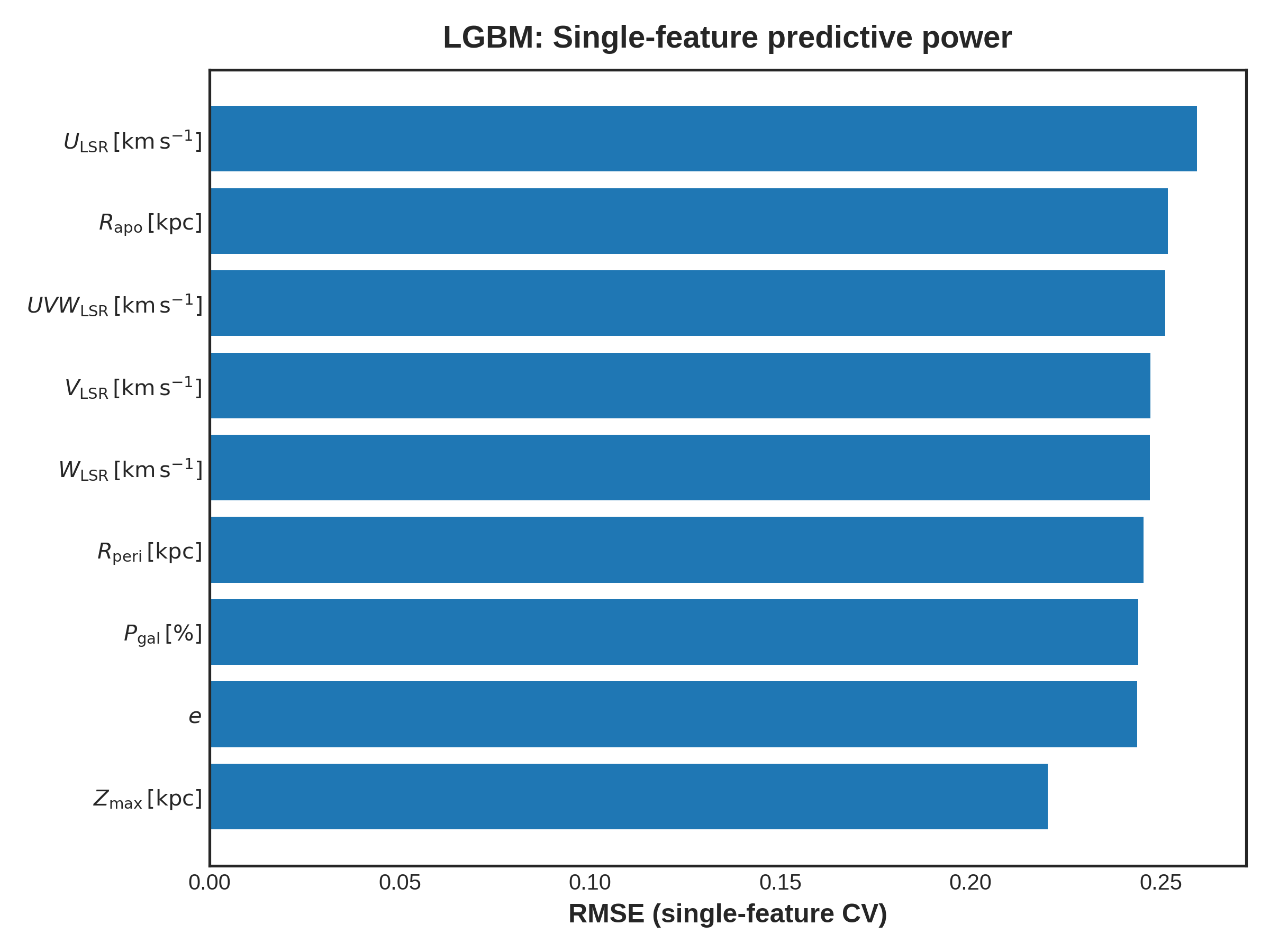}
\includegraphics[width=0.24\textwidth]{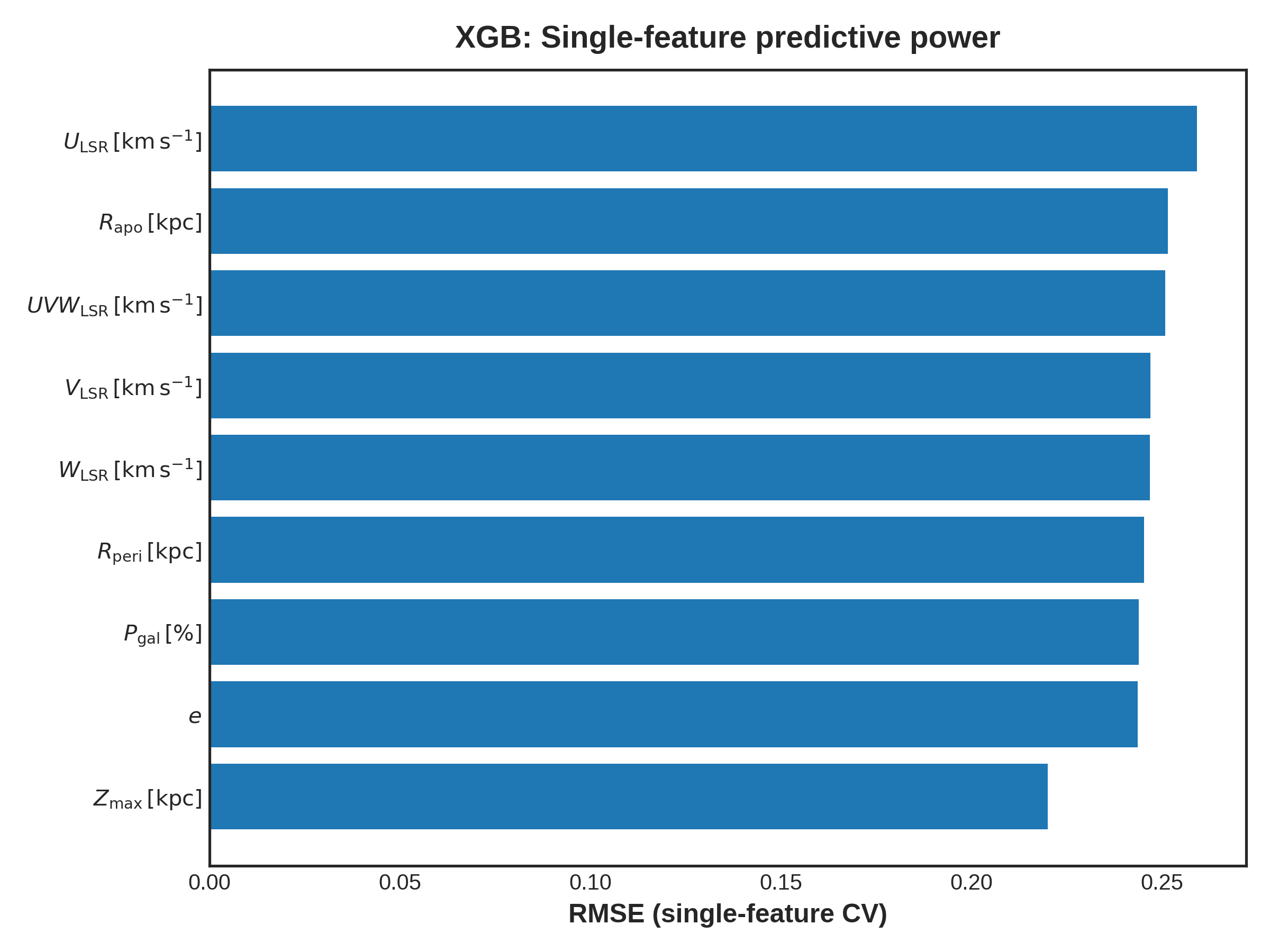}
\vspace{0.4cm}
\includegraphics[width=0.24\textwidth]{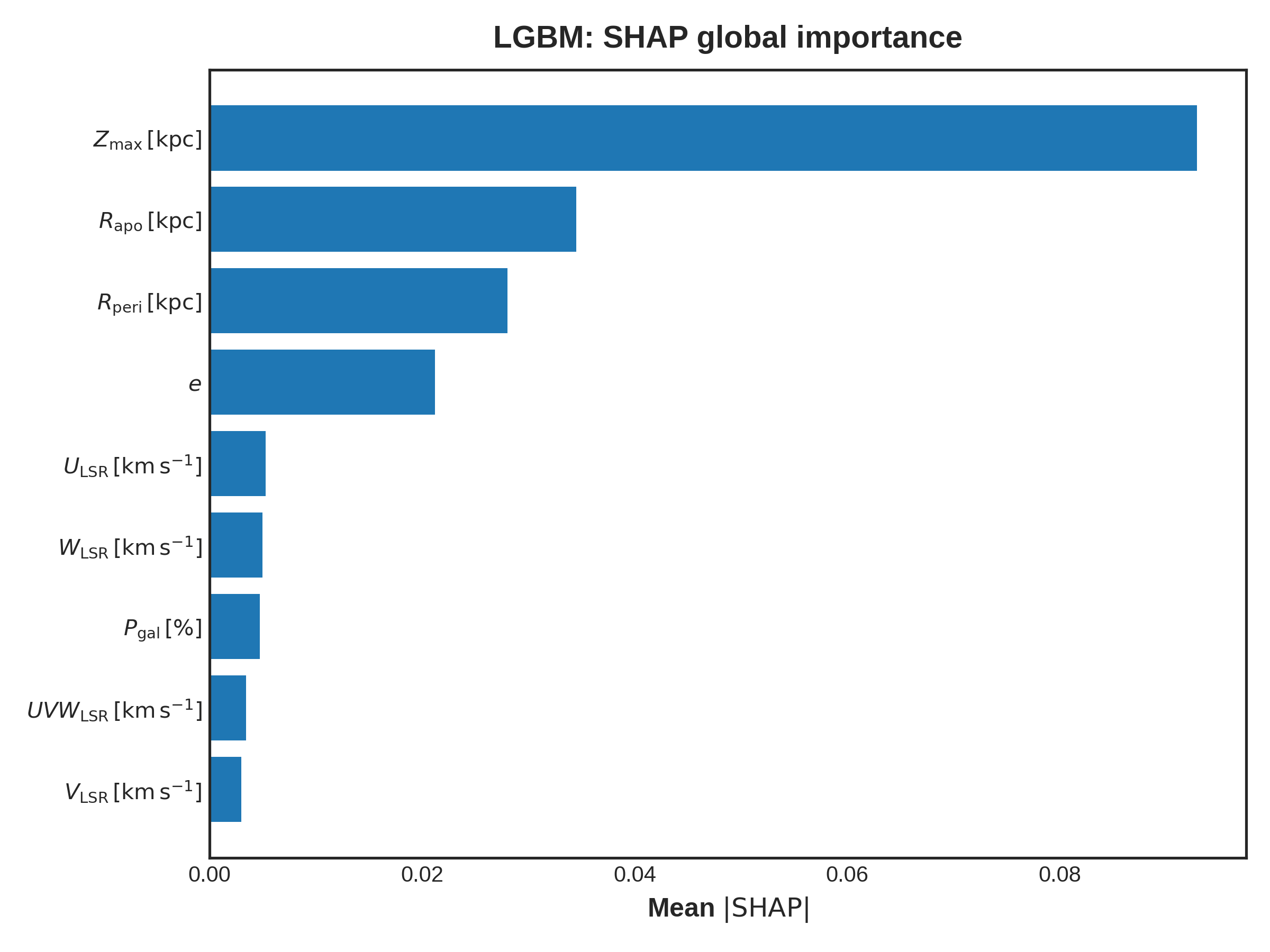}
\includegraphics[width=0.24\textwidth]{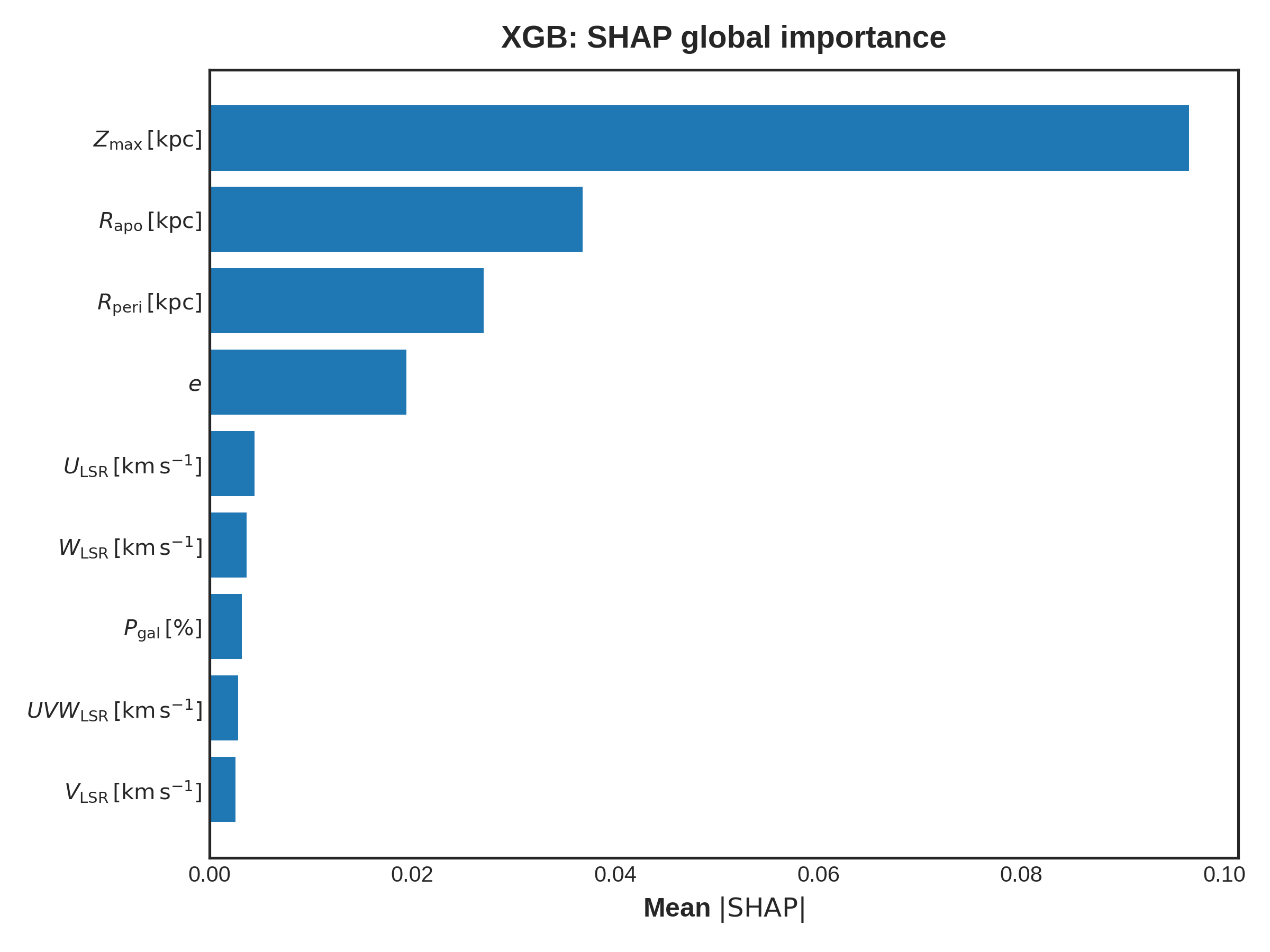}
\caption{Feature-importance diagnostics for the tuned LightGBM and XGBoost models predicting [Fe/H]. Built-in gain feature importance derived from the gradient boosting trees (top). Permutation feature importance, measured as the increase in prediction error ($\Delta$RMSE) after randomly shuffling each feature (second to top).  Predictive performance (RMSE from cross-validation) obtained when training the model using only a single feature at a time (second to bottom). Global SHAP importance, showing the mean absolute SHAP value for each feature (bottom).}
\label{fig:feature_importance_feh}
\end{figure}

\section{Predicting individual abundances from metallicity and kinematics}
\label{appendix:abundances}

This appendix provides additional material related to the prediction of individual elemental abundances using stellar metallicity together with kinematic and orbital parameters.

\subsection{Correlation and mutual information analysis}
\label{appendix:correlations_mi_abund}

This appendix presents the full set of Spearman rank correlation and mutual information diagrams between the elemental abundances and the input features used in this work. The figures are shown separately for the APOGEE and HARPS samples.

\begin{figure}[ht]
\centering
\includegraphics[width=0.48\textwidth]{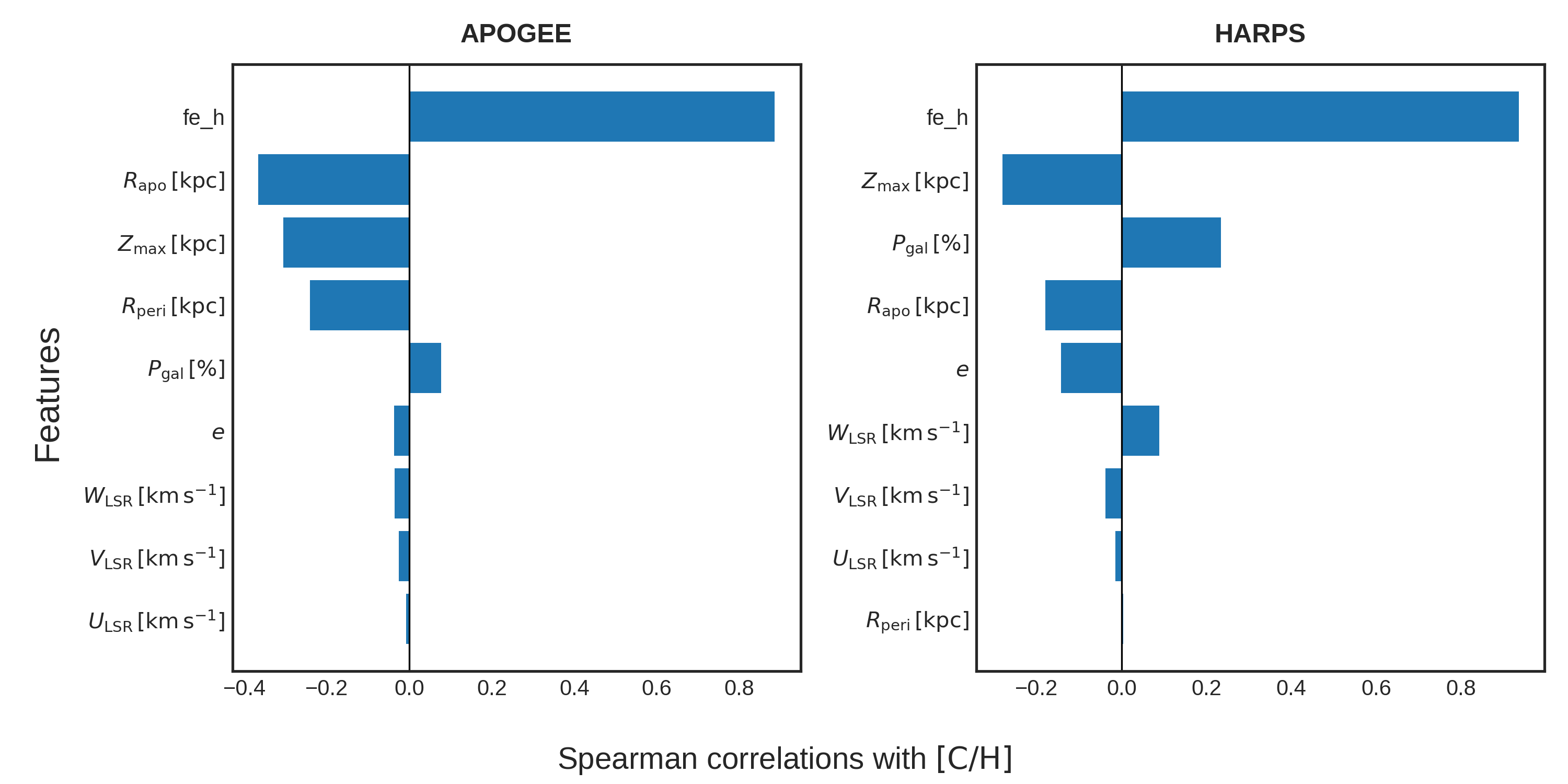}
\vspace{0.4cm}
\includegraphics[width=0.48\textwidth]{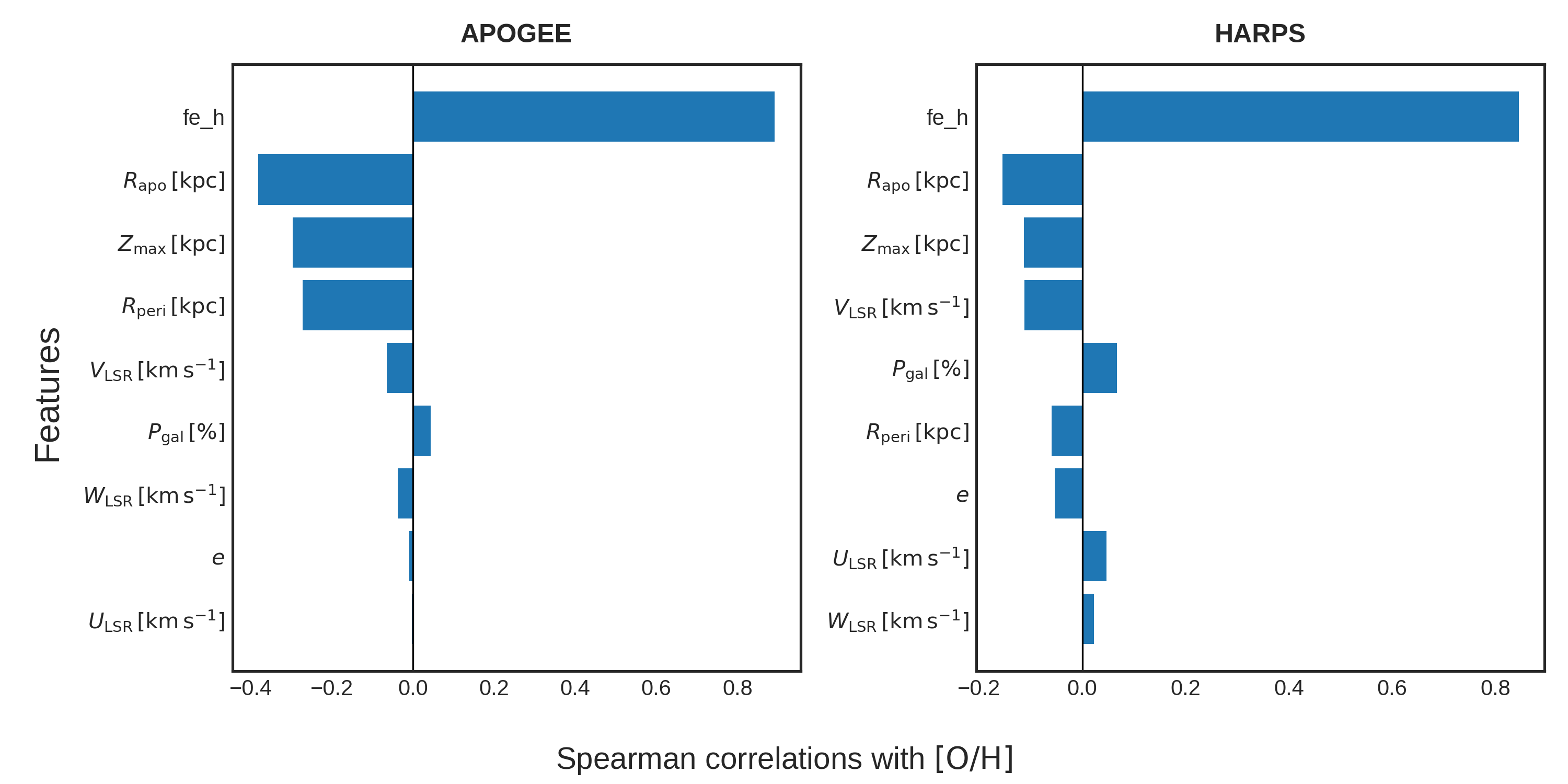}
\vspace{0.4cm}
\includegraphics[width=0.48\textwidth]{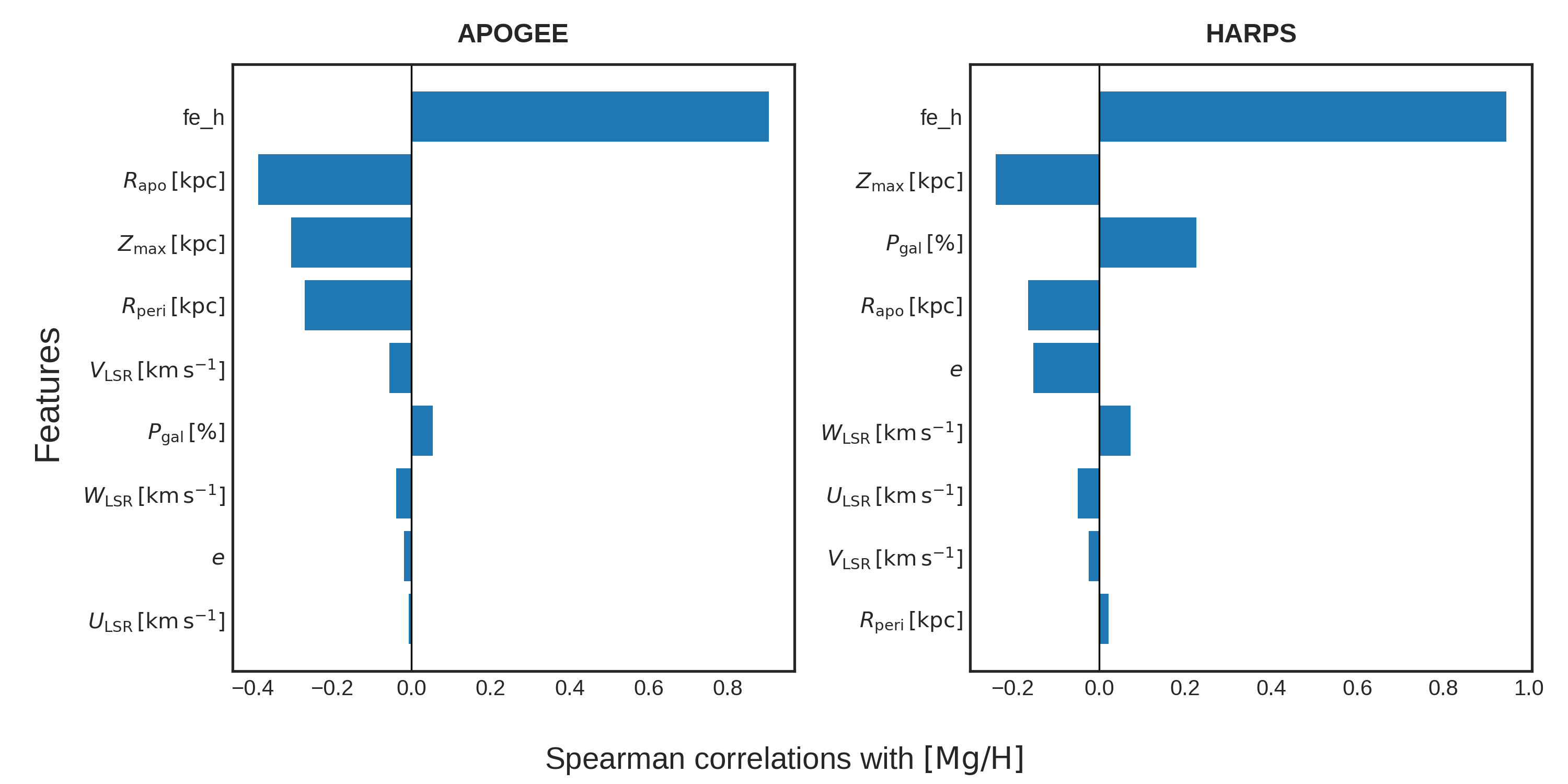}
\vspace{0.4cm}
\includegraphics[width=0.48\textwidth]{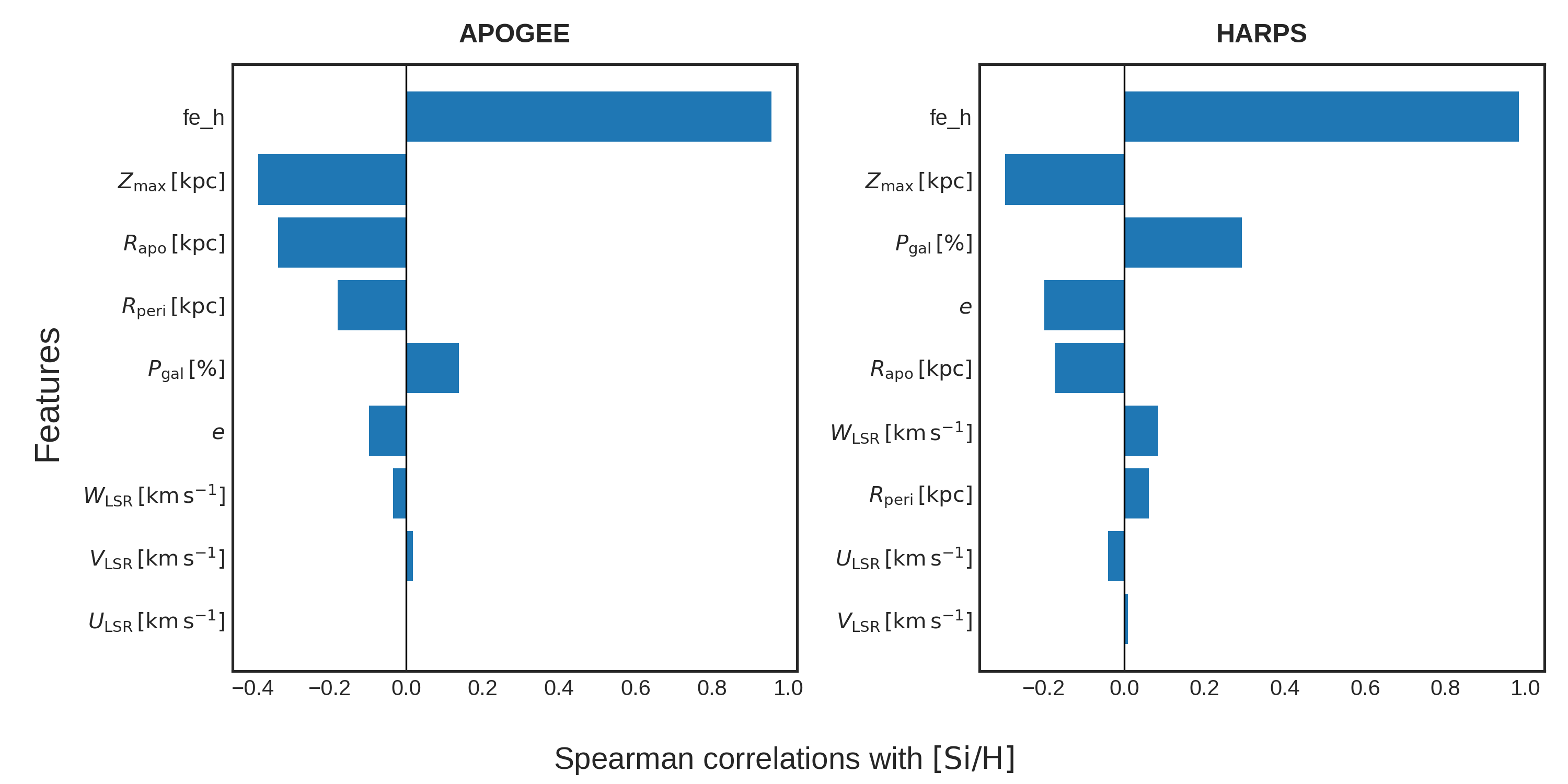}
\caption{Spearman rank correlation coefficients between the input features and the abundances of elements for the APOGEE (left) and HARPS (right) samples.}
\label{fig:cor_abund_fe}
\end{figure}

\begin{figure*}[ht]
\centering
\includegraphics[width=0.48\textwidth]{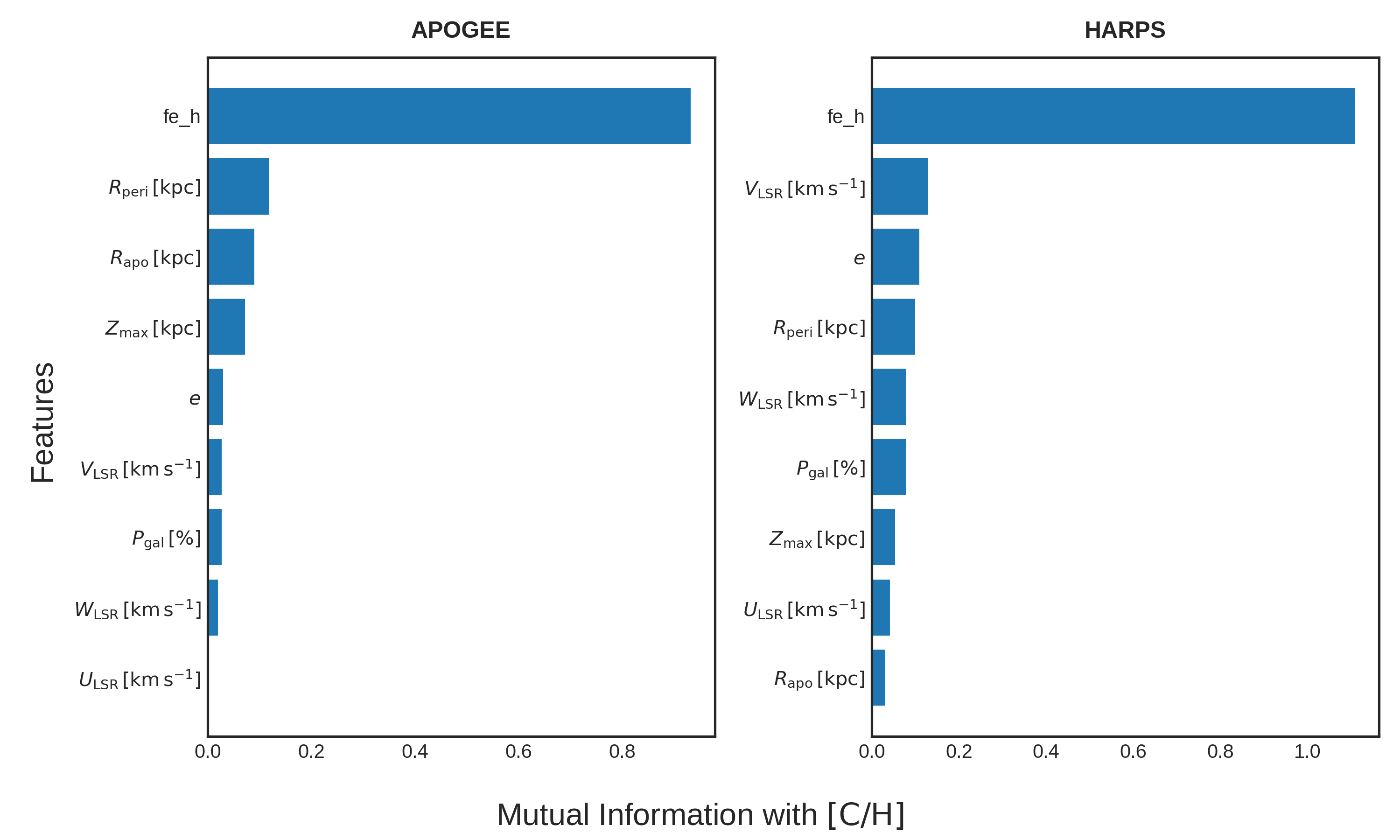}
\hfill
\includegraphics[width=0.48\textwidth]{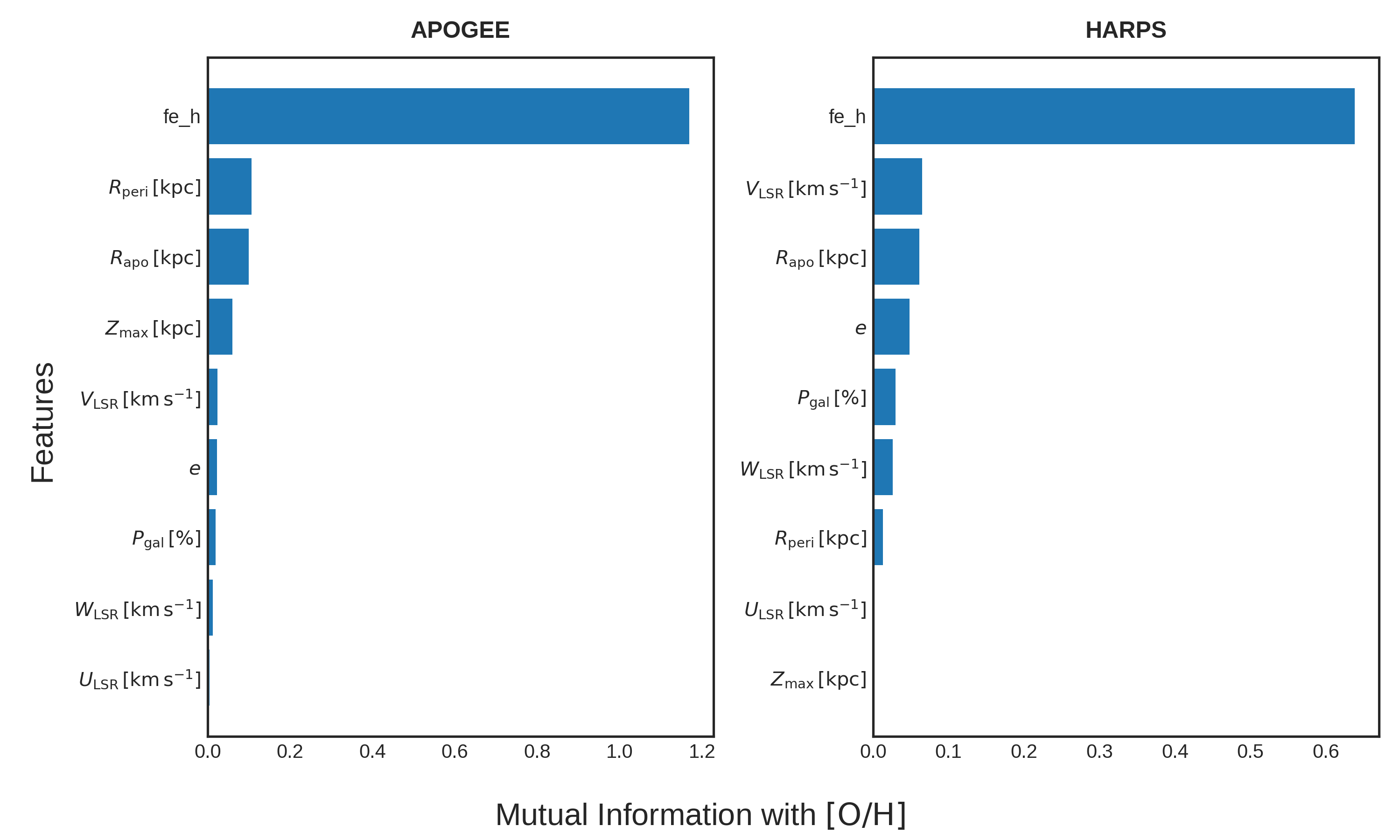}
\vspace{0.4cm}
\includegraphics[width=0.48\textwidth]{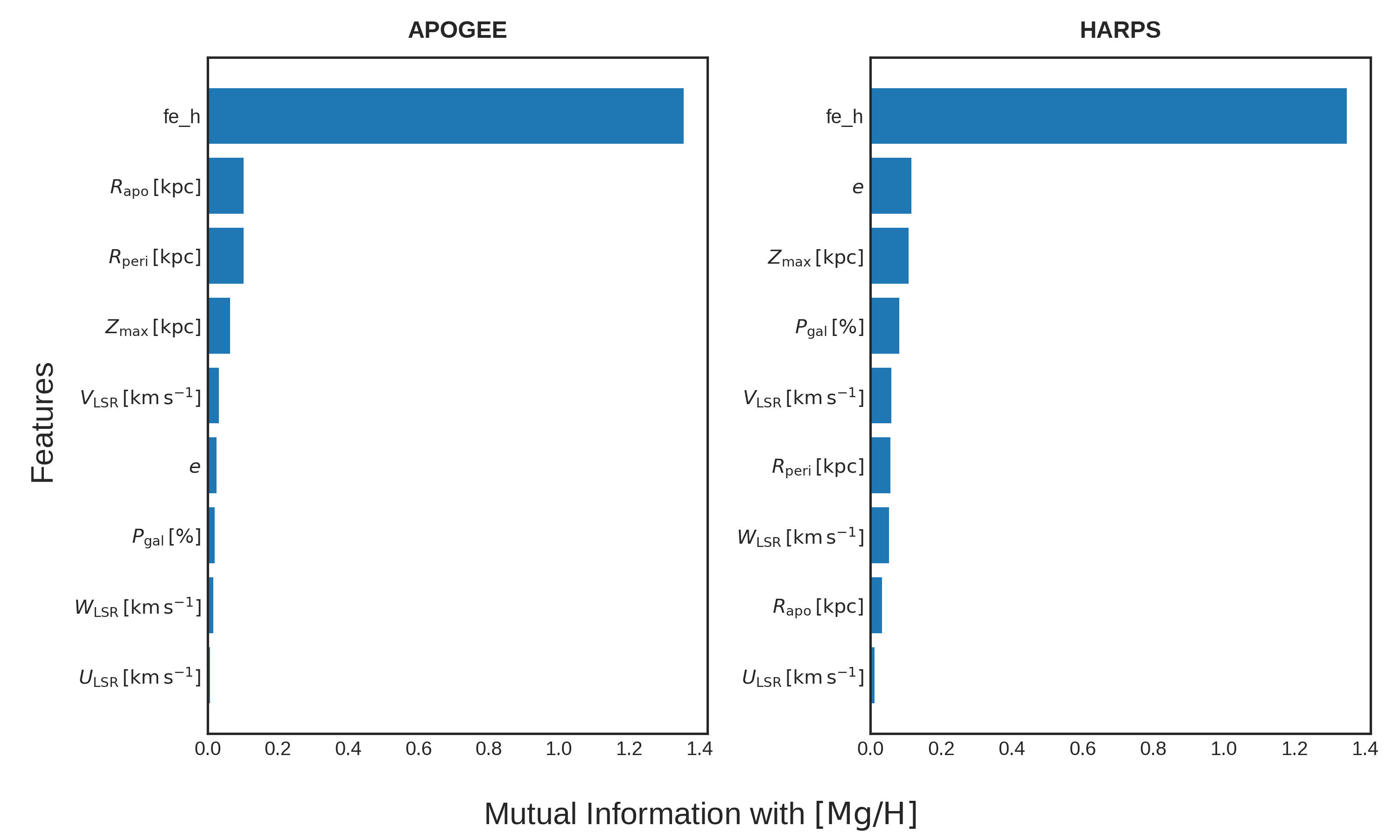}
\hfill
\includegraphics[width=0.48\textwidth]{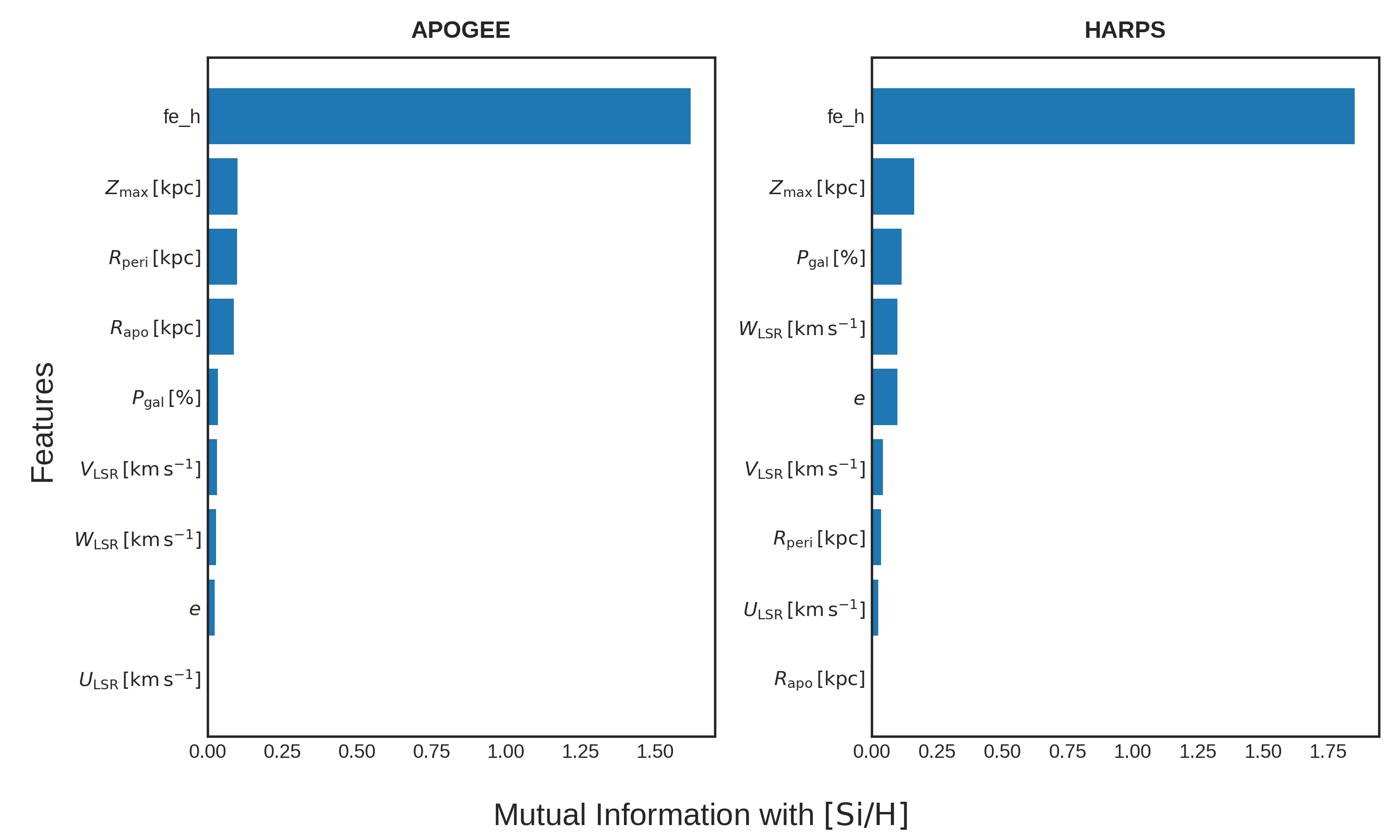}
\caption{Mutual information (MI) between the input features and the abundances of elements for the APOGEE (left) and HARPS (right) samples.}
\label{fig:mi_abund_fe}
\end{figure*}

\begin{table}
\caption{Performance of regression models for predicting elemental abundances
from [Fe/H] and kinematics using 5-fold cross-validation on the APOGEE
training sample. Reported values correspond to the mean across folds, with RMSE
uncertainties indicating the standard deviation.}
\centering
\begin{tabular}{lccc}
\hline
Model & RMSE (dex) & MAE (dex) & $R^2$ \\
\hline

\multicolumn{4}{l}{[O/H]} \\
\hline
Linear Regression & $0.061 \pm 0.000$ & 0.046 & 0.901 \\
Random Forest & $0.056 \pm 0.000$ & 0.040 & 0.916 \\
Extra Trees & $0.057 \pm 0.000$ & 0.041 & 0.914 \\
XGBoost & $0.056 \pm 0.000$ & 0.040 & 0.915 \\
XGBoost (QT) & $0.057 \pm 0.001$ & 0.040 & 0.914 \\
LightGBM & $0.055 \pm 0.000$ & 0.039 & 0.920 \\
LightGBM (QT) & $0.055 \pm 0.000$ & 0.039 & 0.919 \\
CatBoost & $0.055 \pm 0.001$ & 0.039 & 0.920 \\
CatBoost (QT) & $0.055 \pm 0.001$ & 0.039 & 0.919 \\
Voting Regressor & $0.055 \pm 0.000$ & 0.039 & 0.920 \\
\hline

\multicolumn{4}{l}{[Mg/H]} \\
\hline
Linear Regression & $0.057 \pm 0.000$ & 0.044 & 0.920 \\
Random Forest & $0.051 \pm 0.000$ & 0.037 & 0.937 \\
Extra Trees & $0.051 \pm 0.000$ & 0.037 & 0.937 \\
XGBoost & $0.051 \pm 0.000$ & 0.036 & 0.938 \\
XGBoost (QT) & $0.051 \pm 0.000$ & 0.036 & 0.937 \\
LightGBM & $0.049 \pm 0.001$ & 0.035 & 0.941 \\
LightGBM (QT) & $0.049 \pm 0.001$ & 0.035 & 0.941 \\
CatBoost & $0.049 \pm 0.000$ & 0.035 & 0.941 \\
CatBoost (QT) & $0.049 \pm 0.000$ & 0.035 & 0.941 \\
Voting Regressor & $0.049 \pm 0.000$ & 0.035 & 0.941 \\
\hline

\multicolumn{4}{l}{[Si/H]} \\
\hline
Linear Regression & $0.041 \pm 0.000$ & 0.032 & 0.961 \\
Random Forest & $0.038 \pm 0.000$ & 0.028 & 0.967 \\
Extra Trees & $0.038 \pm 0.000$ & 0.028 & 0.967 \\
XGBoost & $0.038 \pm 0.000$ & 0.028 & 0.968 \\
XGBoost (QT) & $0.038 \pm 0.000$ & 0.028 & 0.967 \\
LightGBM & $0.037 \pm 0.000$ & 0.027 & 0.969 \\
LightGBM (QT) & $0.037 \pm 0.000$ & 0.027 & 0.969 \\
CatBoost & $0.037 \pm 0.000$ & 0.027 & 0.969 \\
CatBoost (QT) & $0.037 \pm 0.000$ & 0.027 & 0.969 \\
Voting Regressor & $0.037 \pm 0.000$ & 0.027 & 0.969 \\
\hline

\multicolumn{4}{l}{[C/H]} \\
\hline
Linear Regression & $0.086 \pm 0.000$ & 0.065 & 0.872 \\
Random Forest & $0.081 \pm 0.001$ & 0.060 & 0.887 \\
Extra Trees & $0.082 \pm 0.000$ & 0.061 & 0.885 \\
XGBoost & $0.081 \pm 0.001$ & 0.060 & 0.888 \\
XGBoost (QT) & $0.081 \pm 0.001$ & 0.060 & 0.886 \\
LightGBM & $0.079 \pm 0.001$ & 0.058 & 0.893 \\
LightGBM (QT) & $0.079 \pm 0.001$ & 0.058 & 0.893 \\
CatBoost & $0.079 \pm 0.001$ & 0.058 & 0.893 \\
CatBoost (QT) & $0.079 \pm 0.001$ & 0.059 & 0.892 \\
Voting Regressor & $0.079 \pm 0.001$ & 0.059 & 0.892 \\
\hline

\end{tabular}
\label{tab:abundance_cv_scores}
\end{table}

\subsubsection{Feature importance analysis} \label{appendix:feature_importance_abundances_feh}

This appendix presents the full set of feature-importance figures discussed in Sect.\ref{main_feat_imp_abund_feh} For completeness, we show the results of the built-in gain importance, permutation importance, SHAP global importance, and single-feature predictive performance for the LightGBM model.

\begin{figure*}[ht]
\centering
\includegraphics[width=0.24\textwidth]{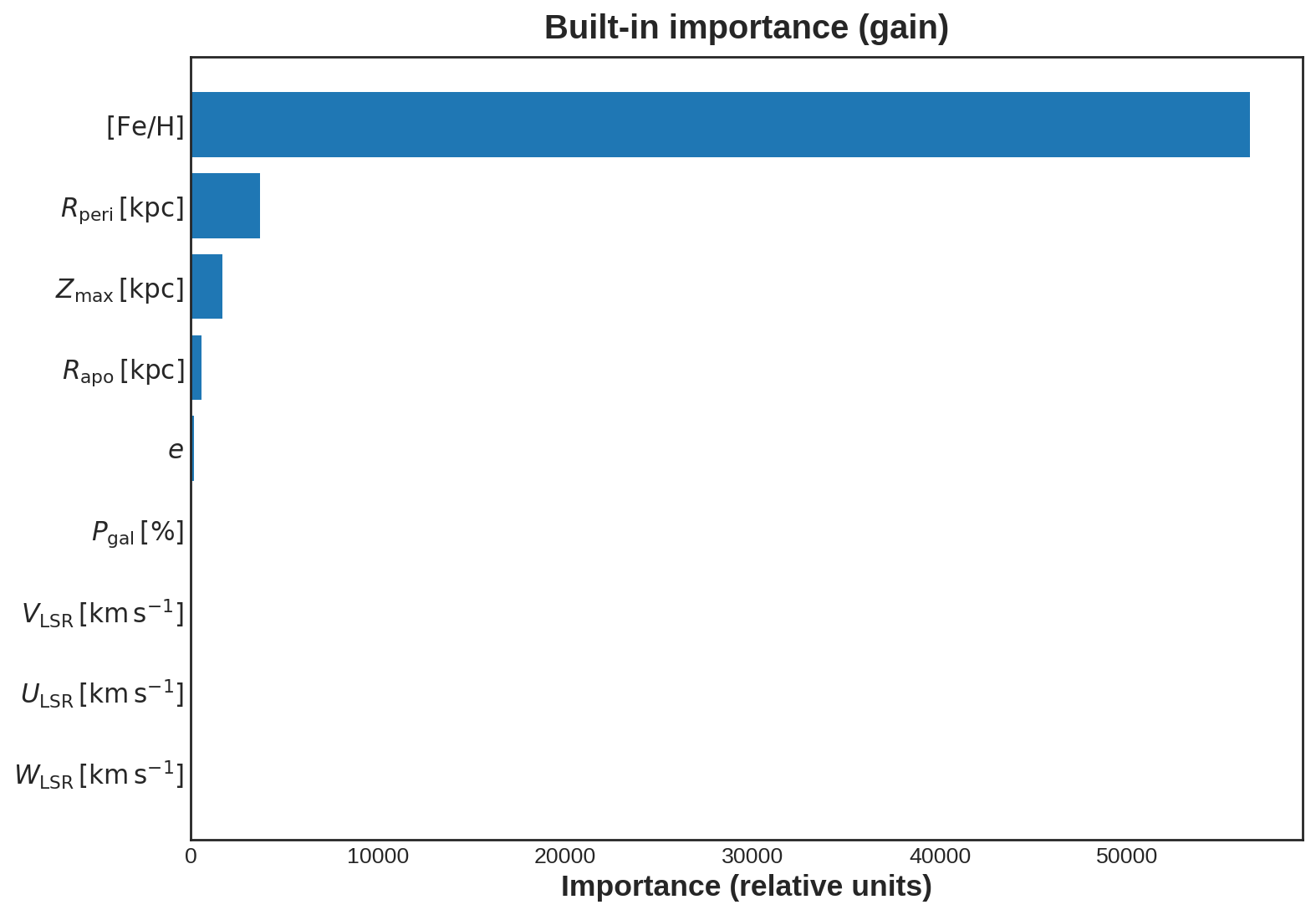}
\includegraphics[width=0.24\textwidth]{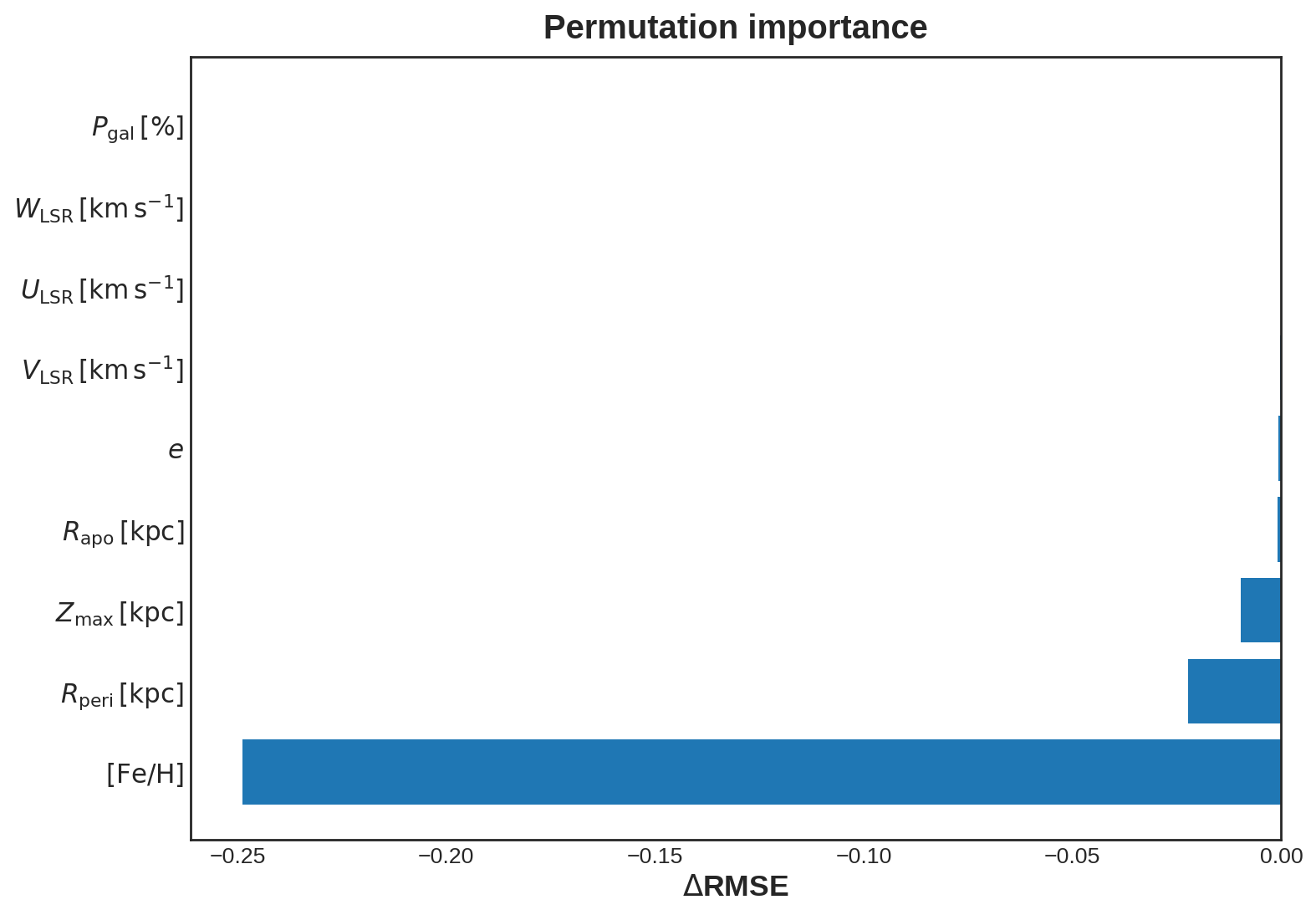}
\includegraphics[width=0.24\textwidth]{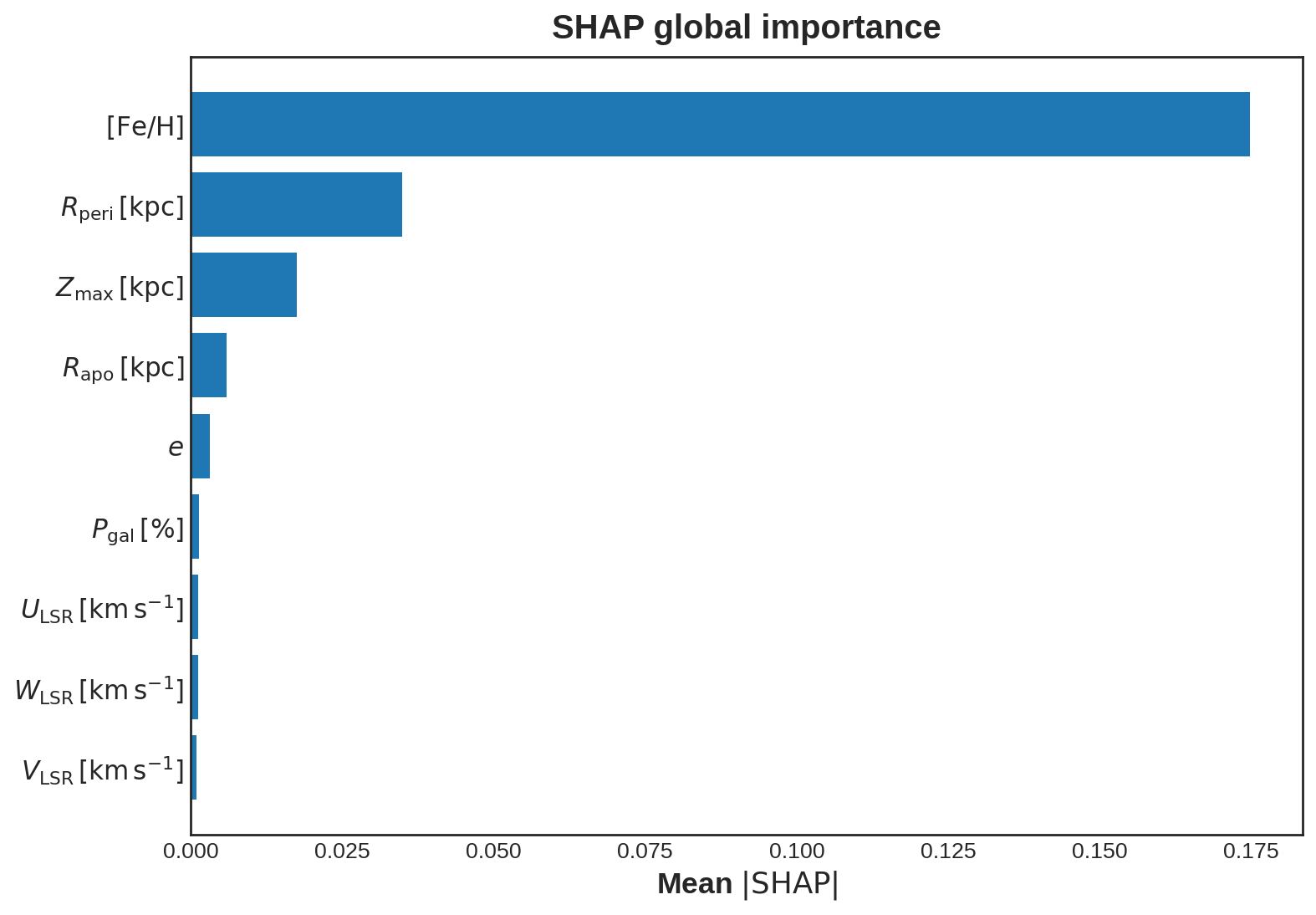}
\includegraphics[width=0.24\textwidth]{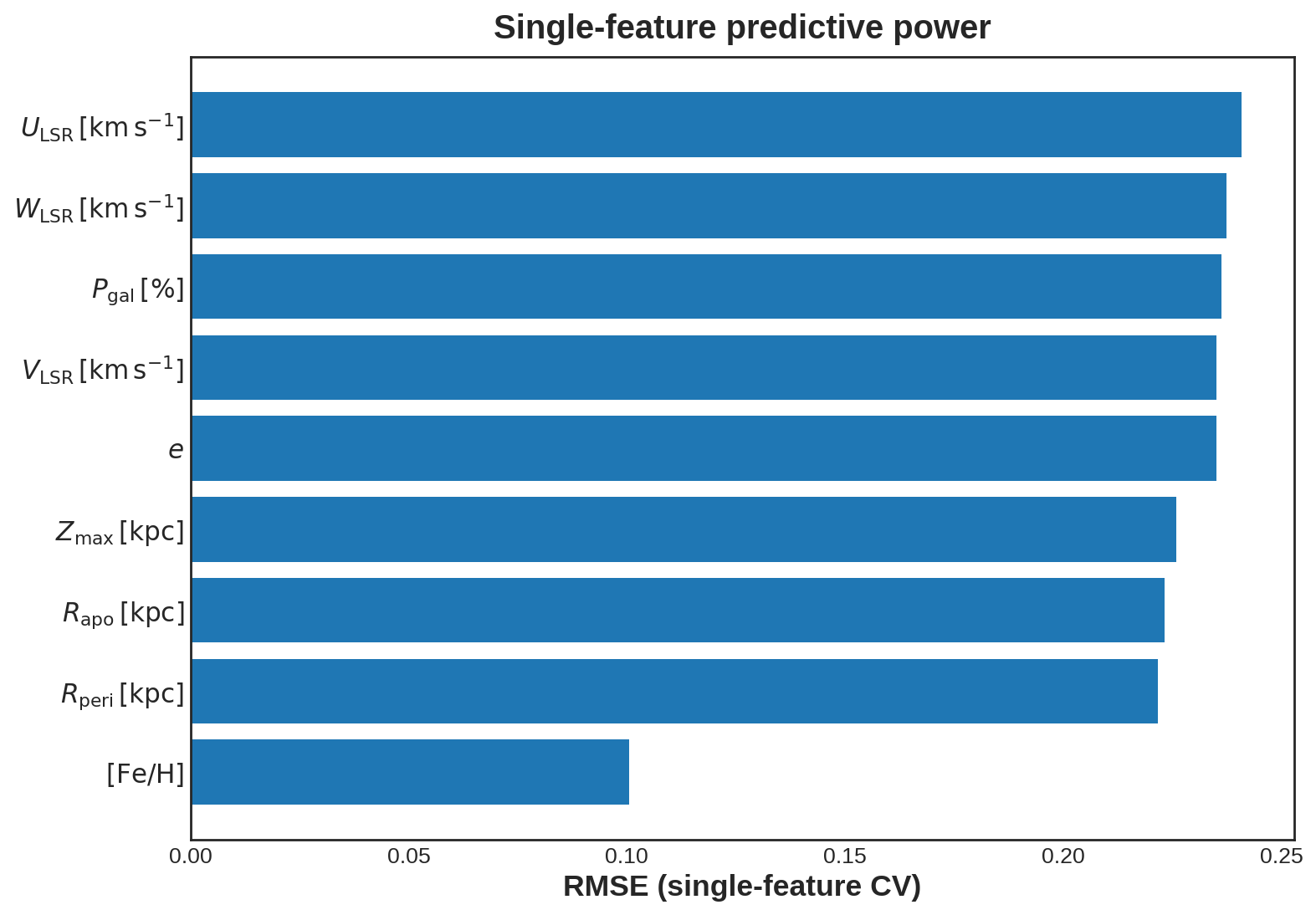}
\caption{
Feature-importance diagnostics for the tuned LightGBM model predicting $\mathrm{[C/H]}$. Built-in gain feature importance derived from the gradient boosting trees (top left). Permutation feature importance, measured as the increase in prediction error ($\Delta$RMSE) after randomly shuffling each feature (top right). Global SHAP importance, showing the mean absolute SHAP value for each feature (bottom left). Predictive performance (RMSE from cross-validation) obtained when training the model using only a single feature at a time (bottom right).}
\label{fig:feature_importance_ch}
\end{figure*}

\begin{figure*}[ht]
\centering
\includegraphics[width=0.24\textwidth]{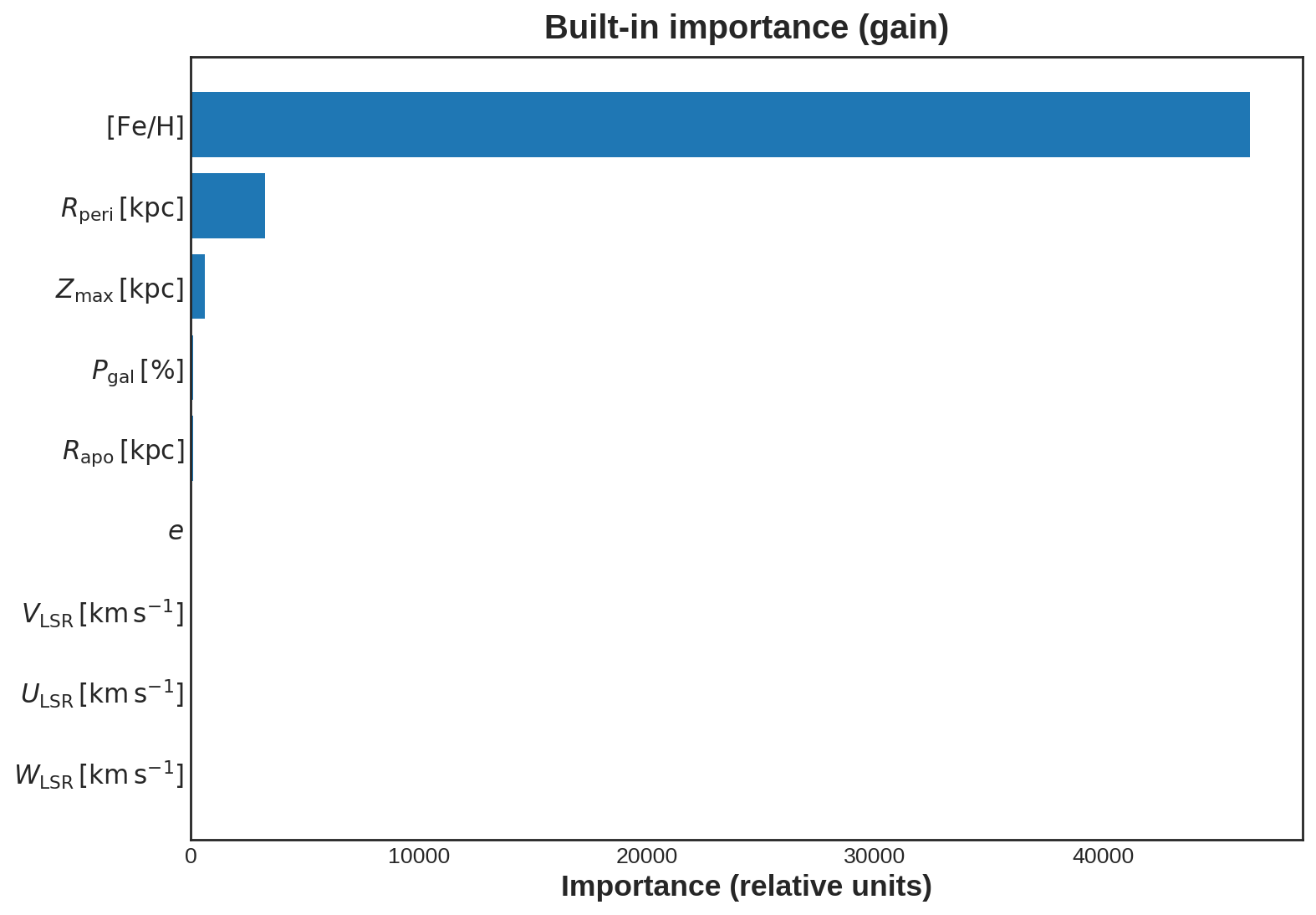}
\includegraphics[width=0.24\textwidth]{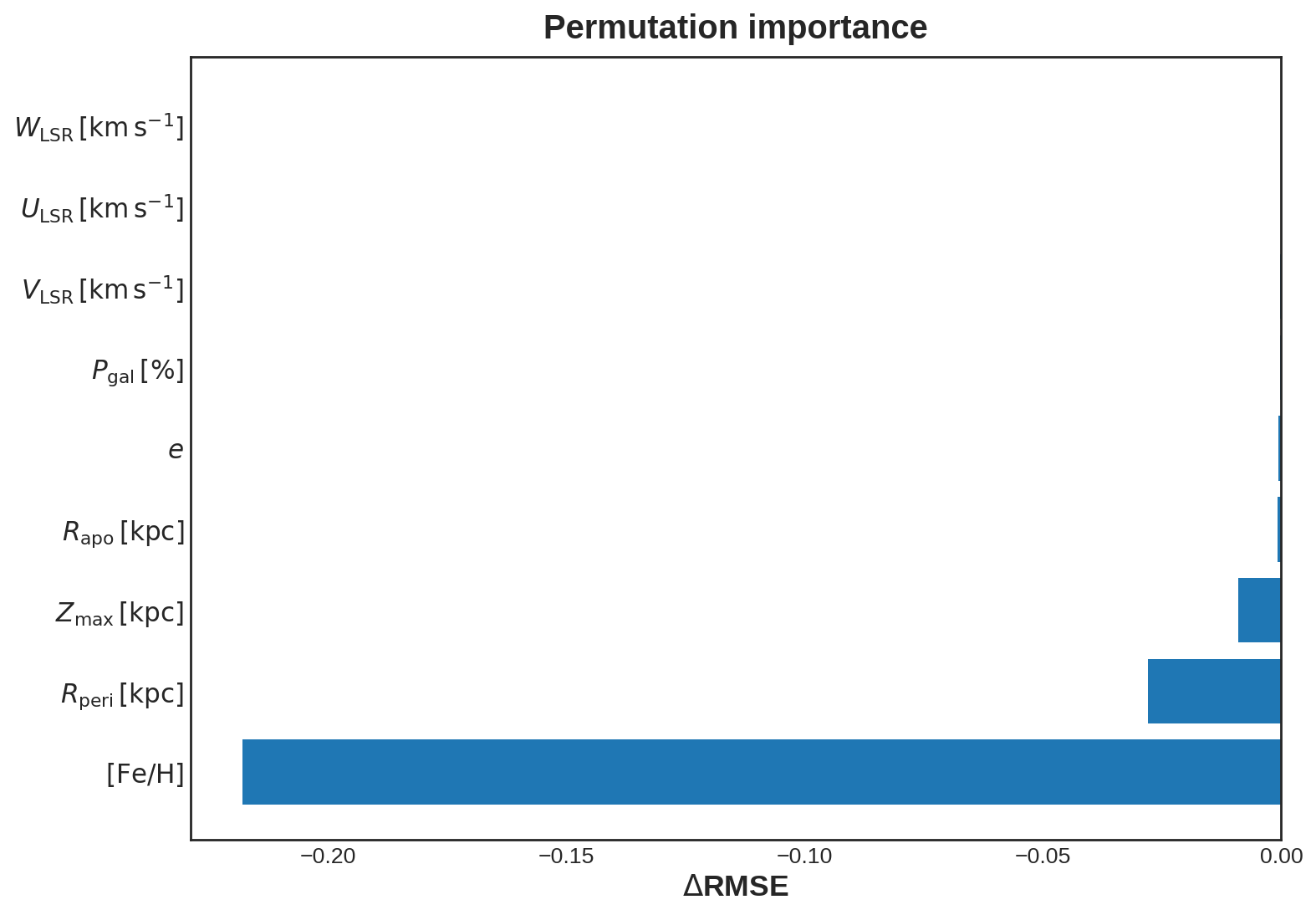}
\includegraphics[width=0.24\textwidth]{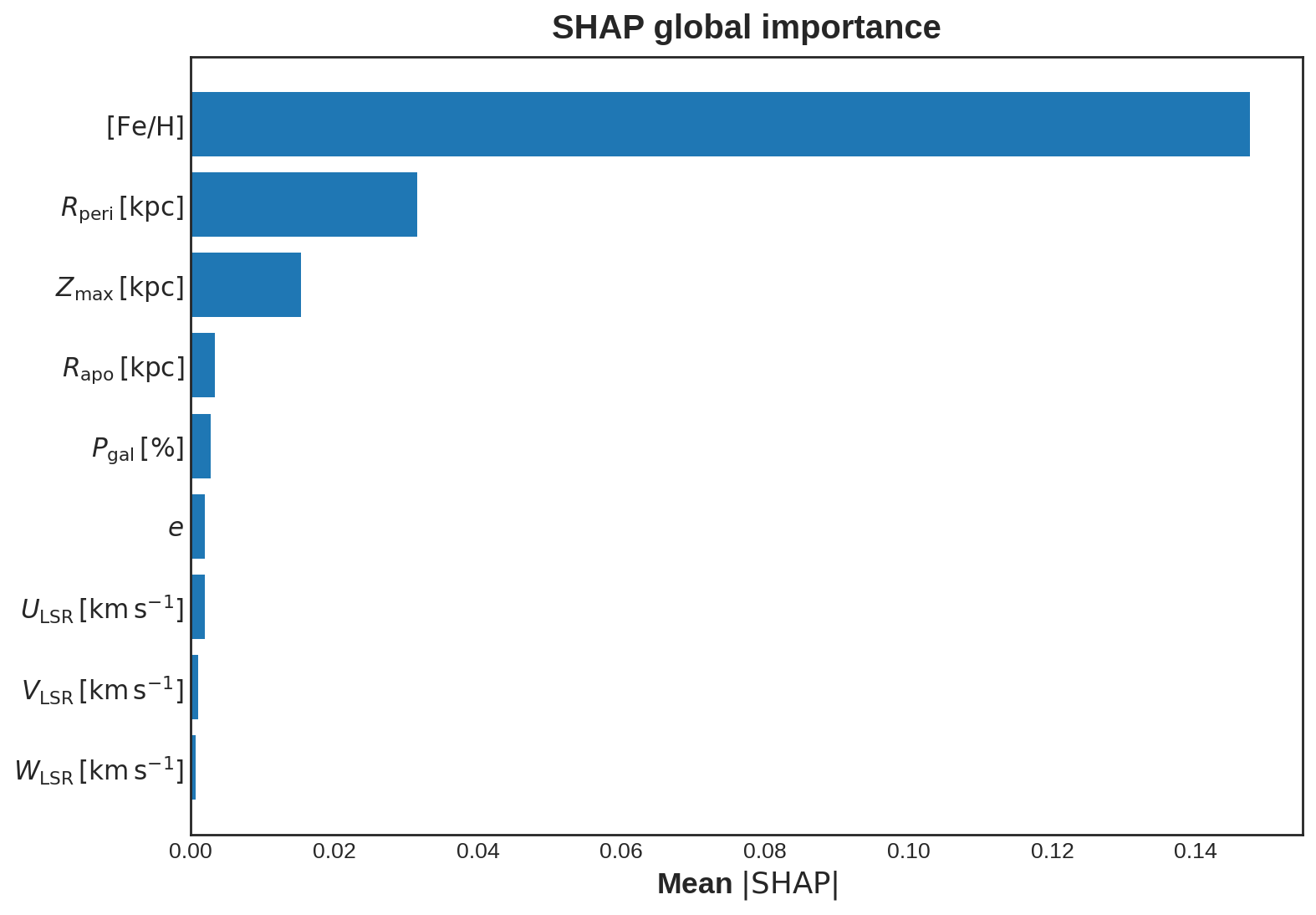}
\includegraphics[width=0.24\textwidth]{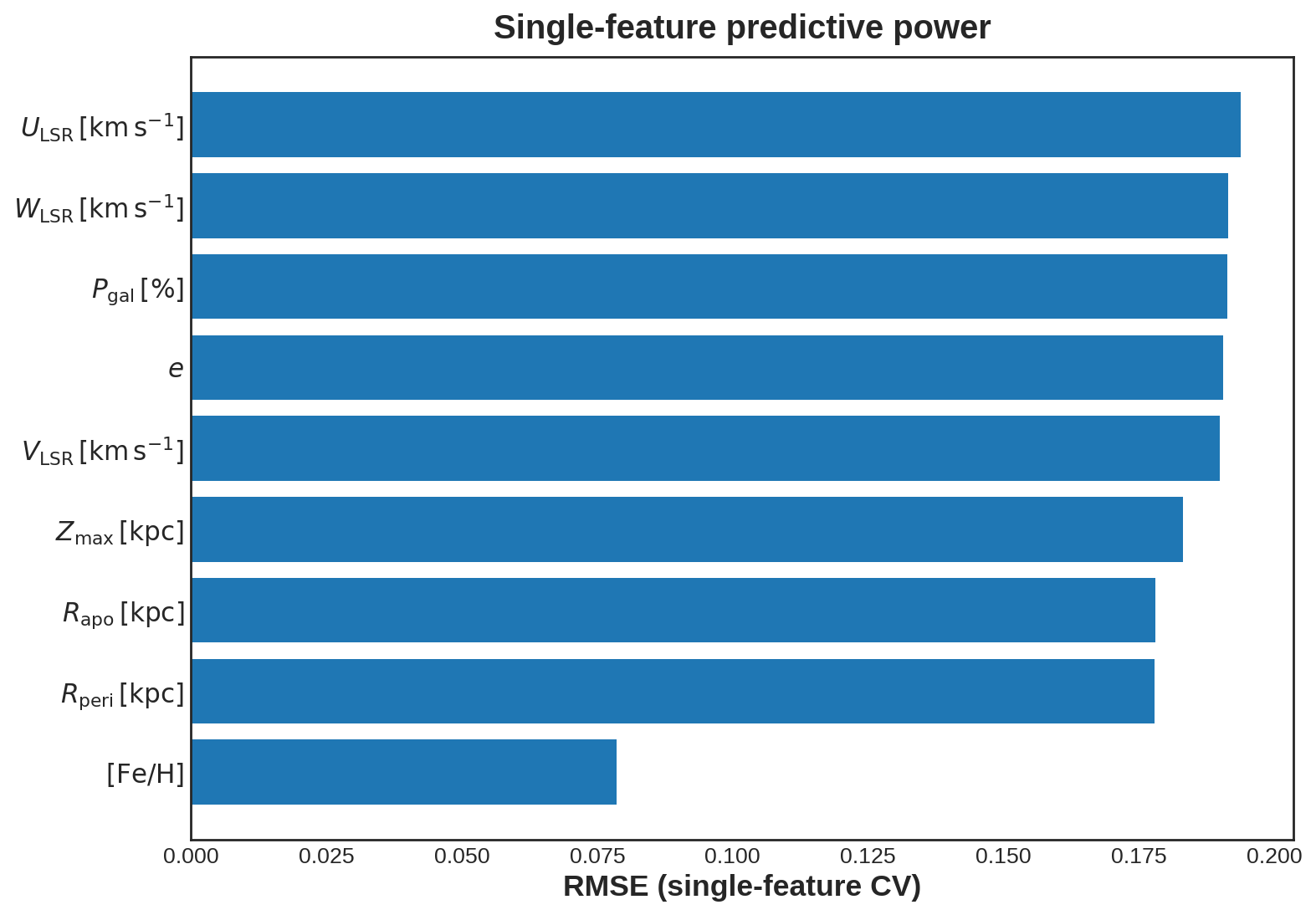}
\caption{Same as Fig.\ref{fig:feature_importance_ch} but for $\mathrm{[O/H]}$}
\label{fig:feature_importance_oh}
\end{figure*}

\begin{figure*}[ht]
\centering
\includegraphics[width=0.24\textwidth]{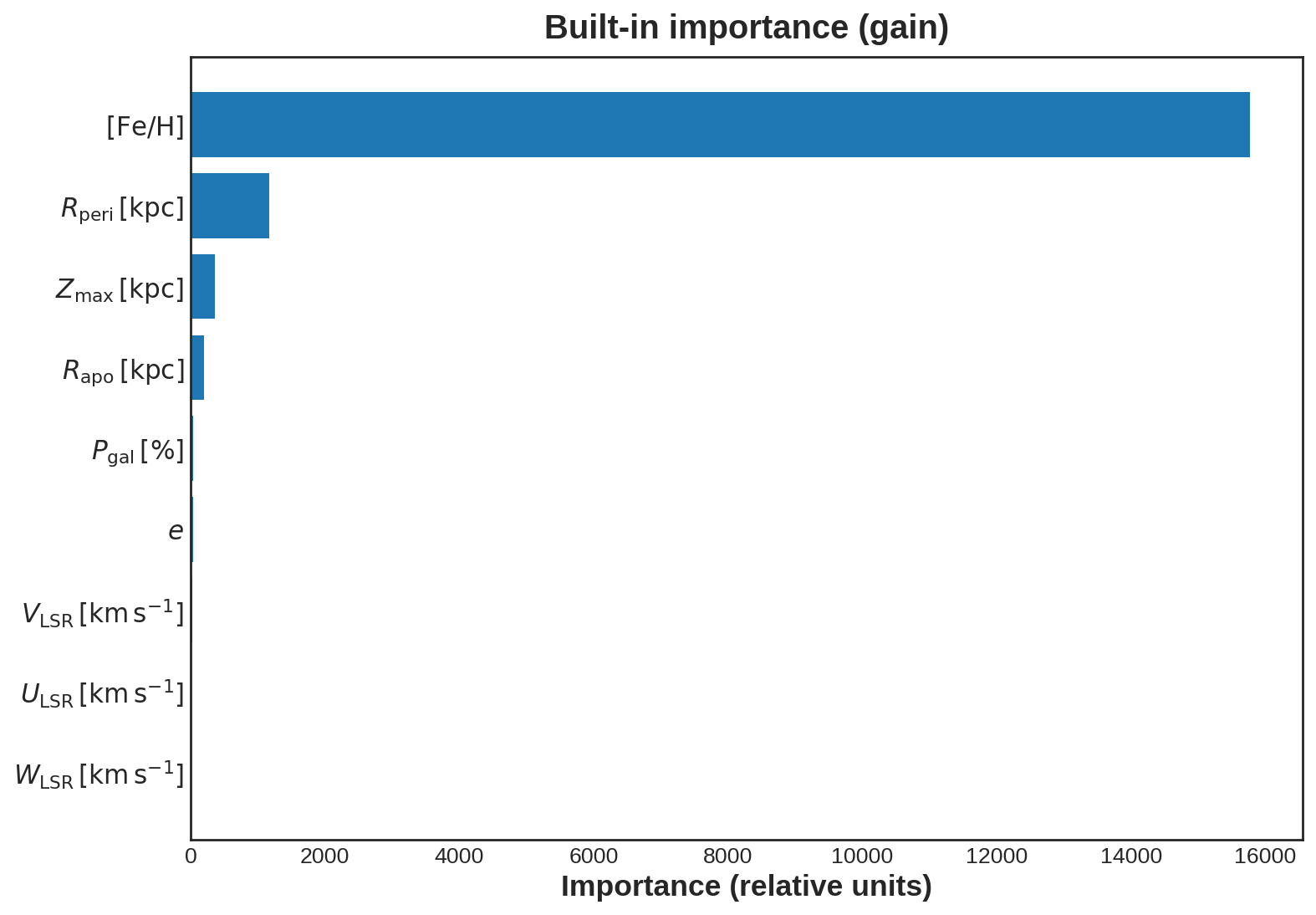}
\includegraphics[width=0.24\textwidth]{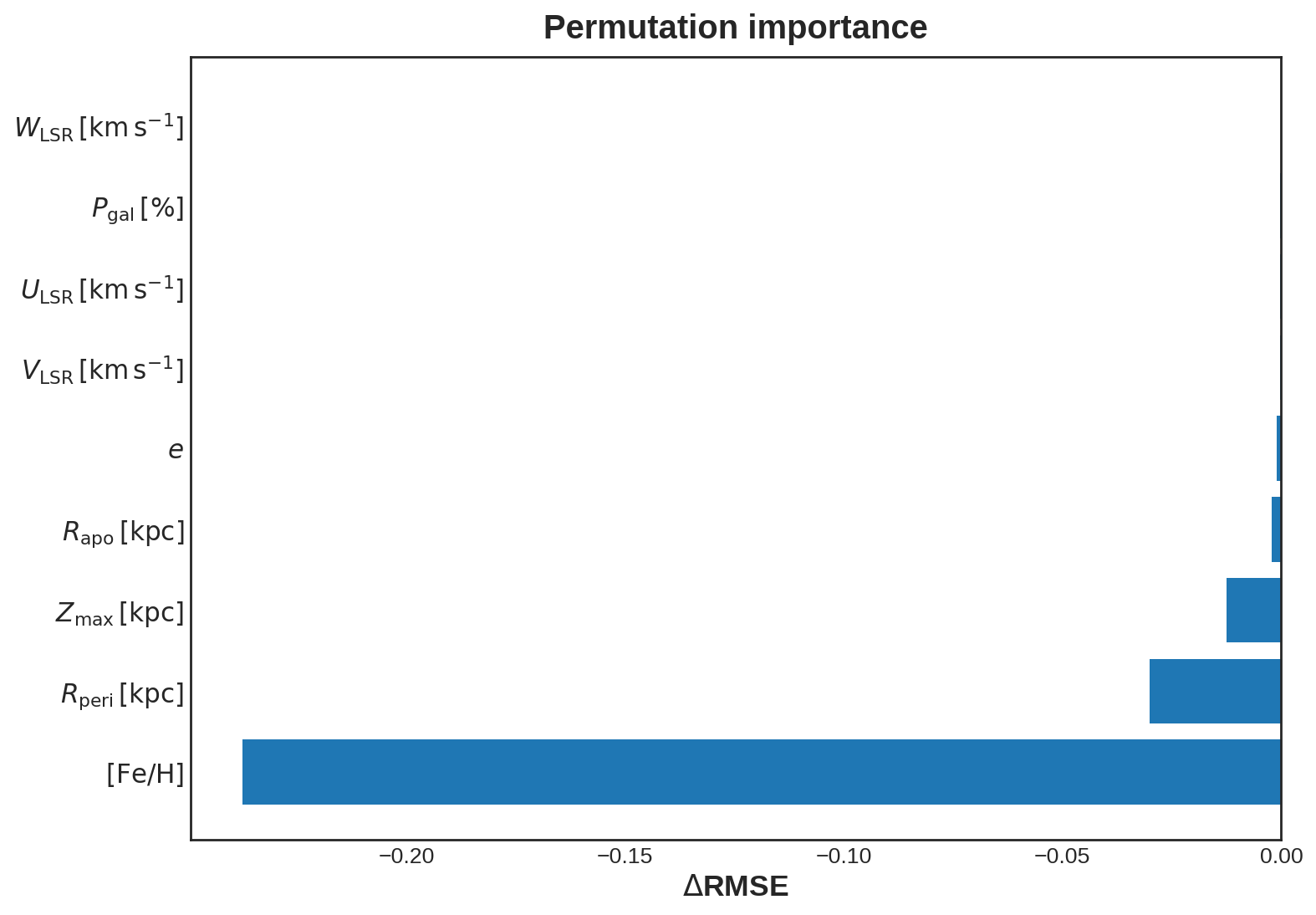}
\includegraphics[width=0.24\textwidth]{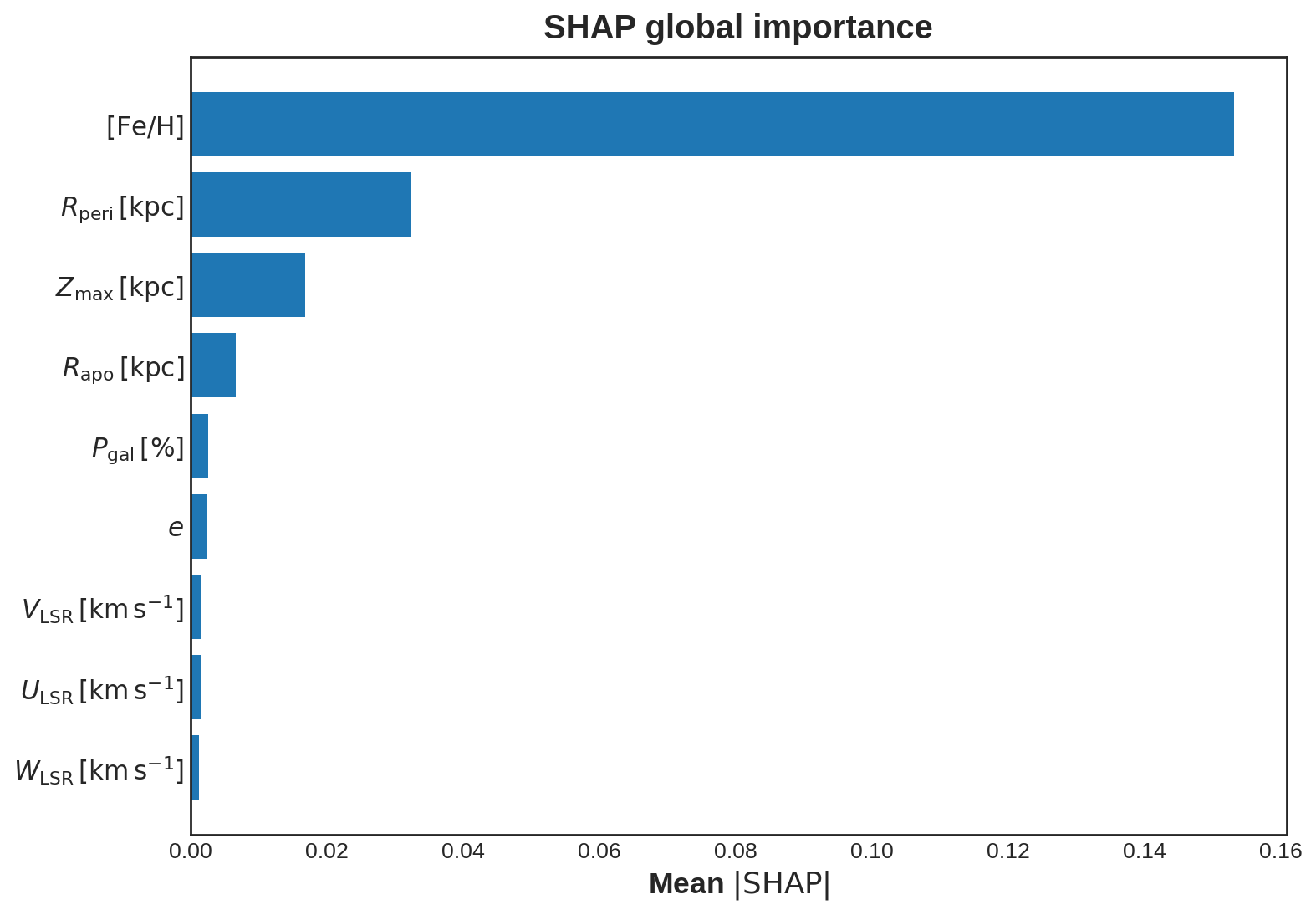}
\includegraphics[width=0.24\textwidth]{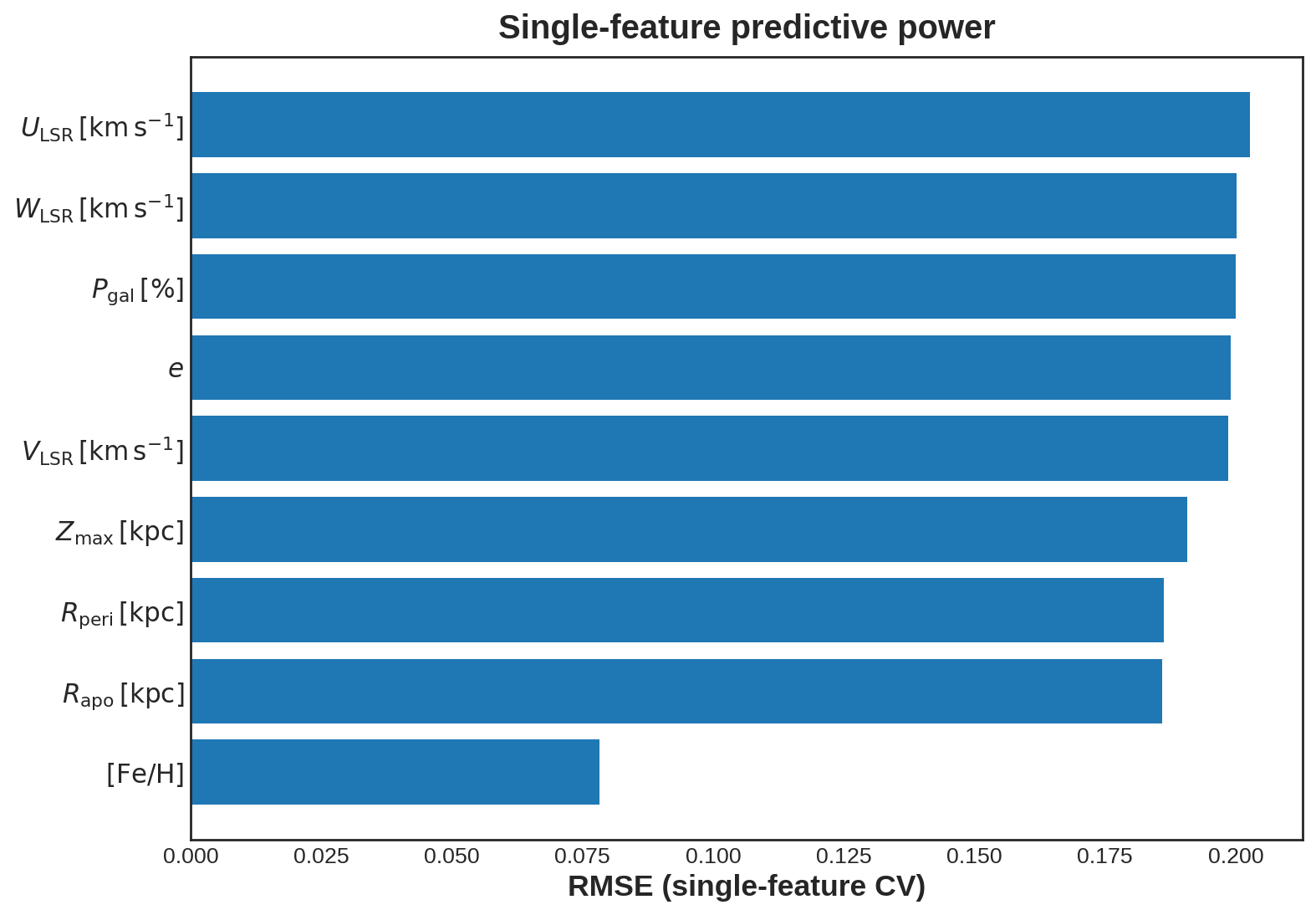}
\caption{Same as Fig.\ref{fig:feature_importance_ch} but for $\mathrm{[Mg/H]}$}
\label{fig:feature_importance_mgh}
\end{figure*}

\begin{figure*}[ht]
\centering
\includegraphics[width=0.24\textwidth]{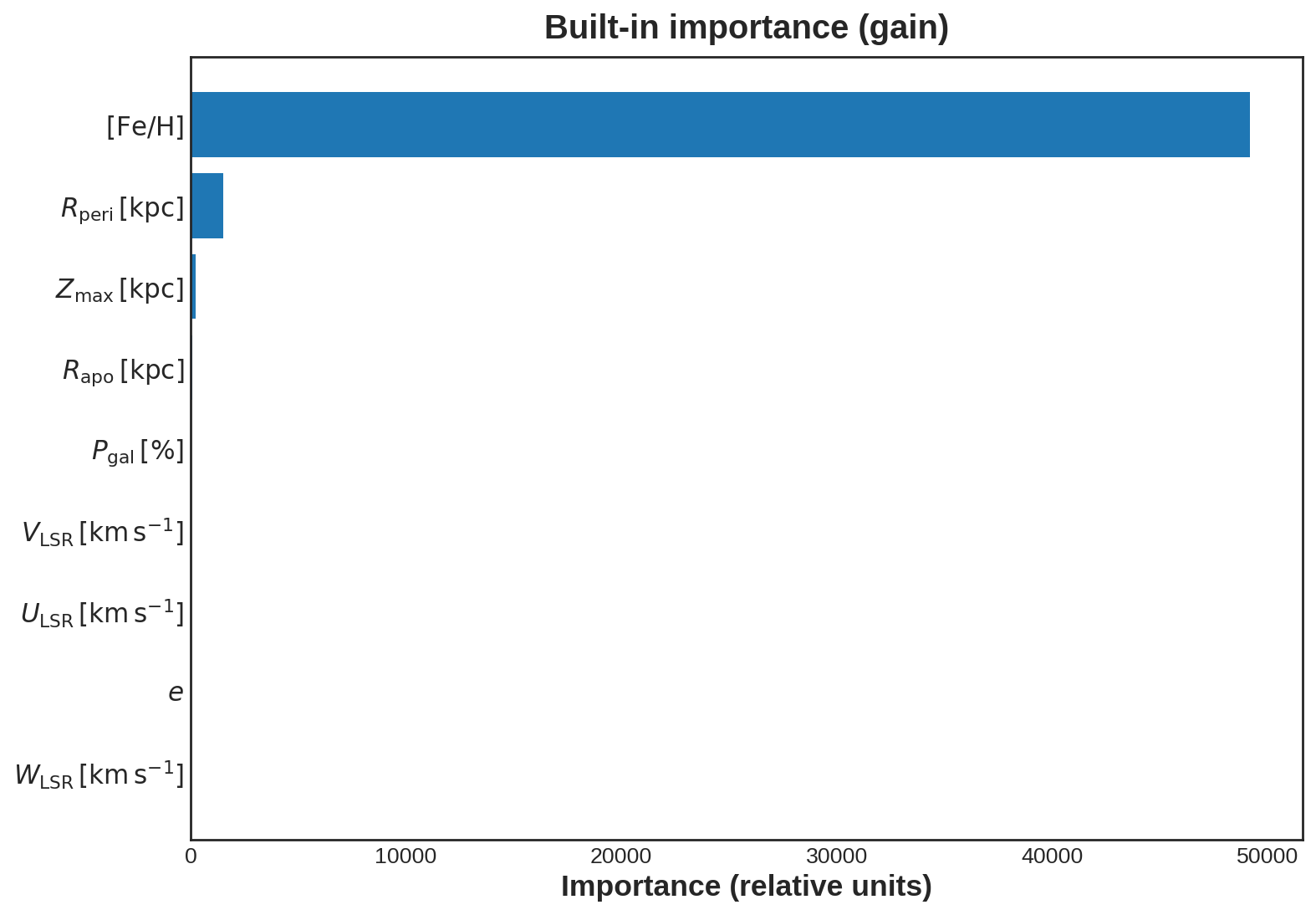}
\includegraphics[width=0.24\textwidth]{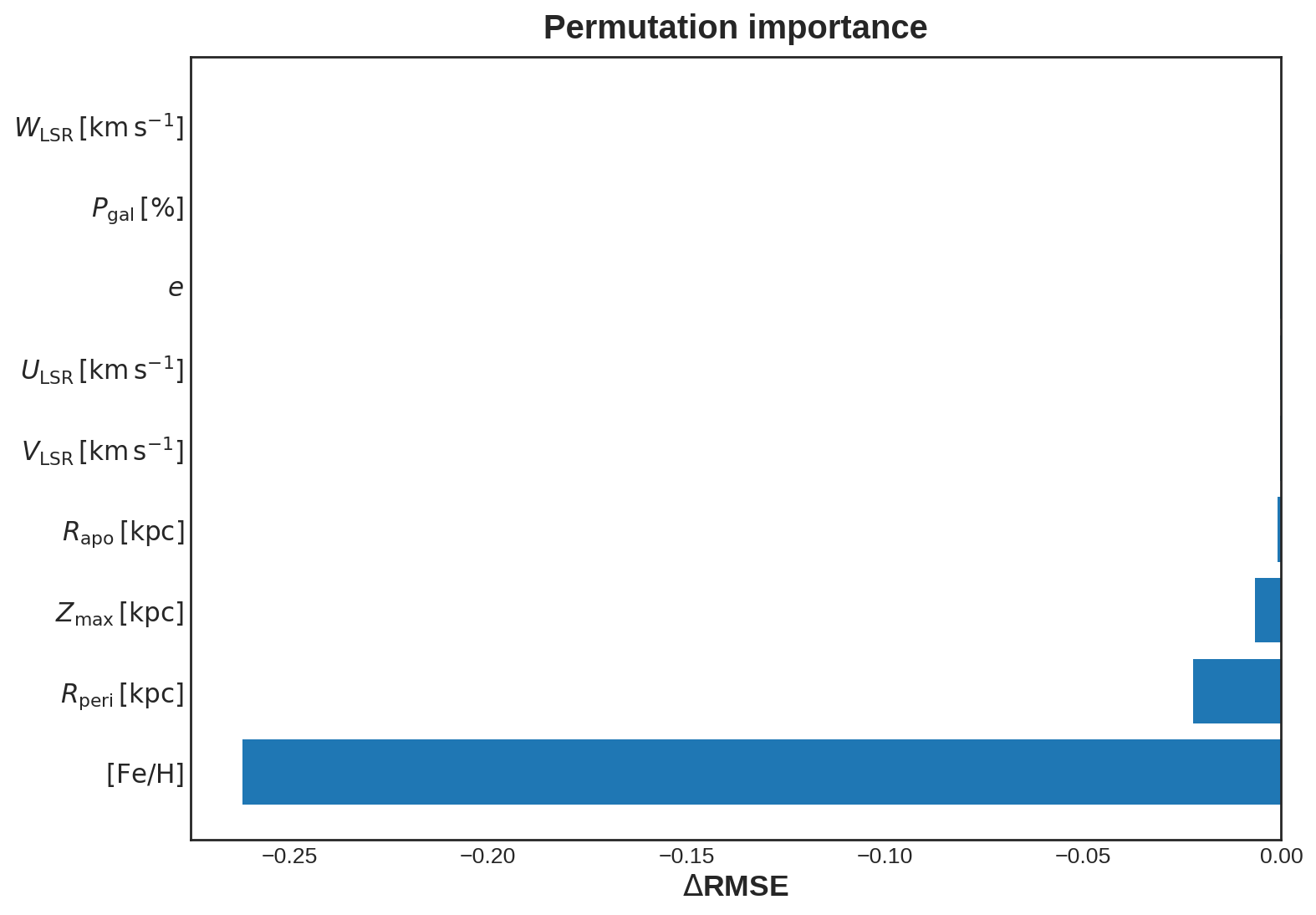}
\includegraphics[width=0.24\textwidth]{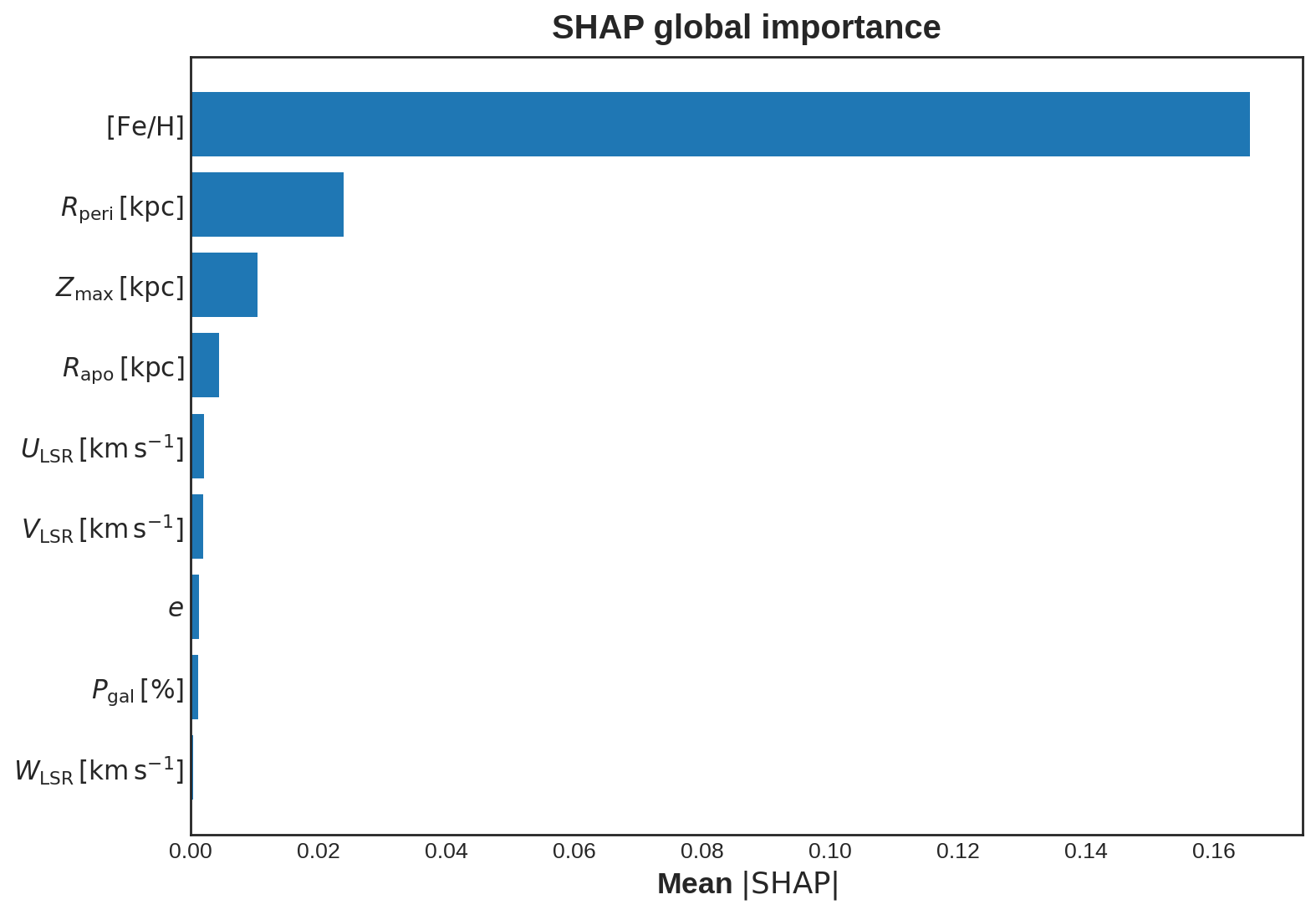}
\includegraphics[width=0.24\textwidth]{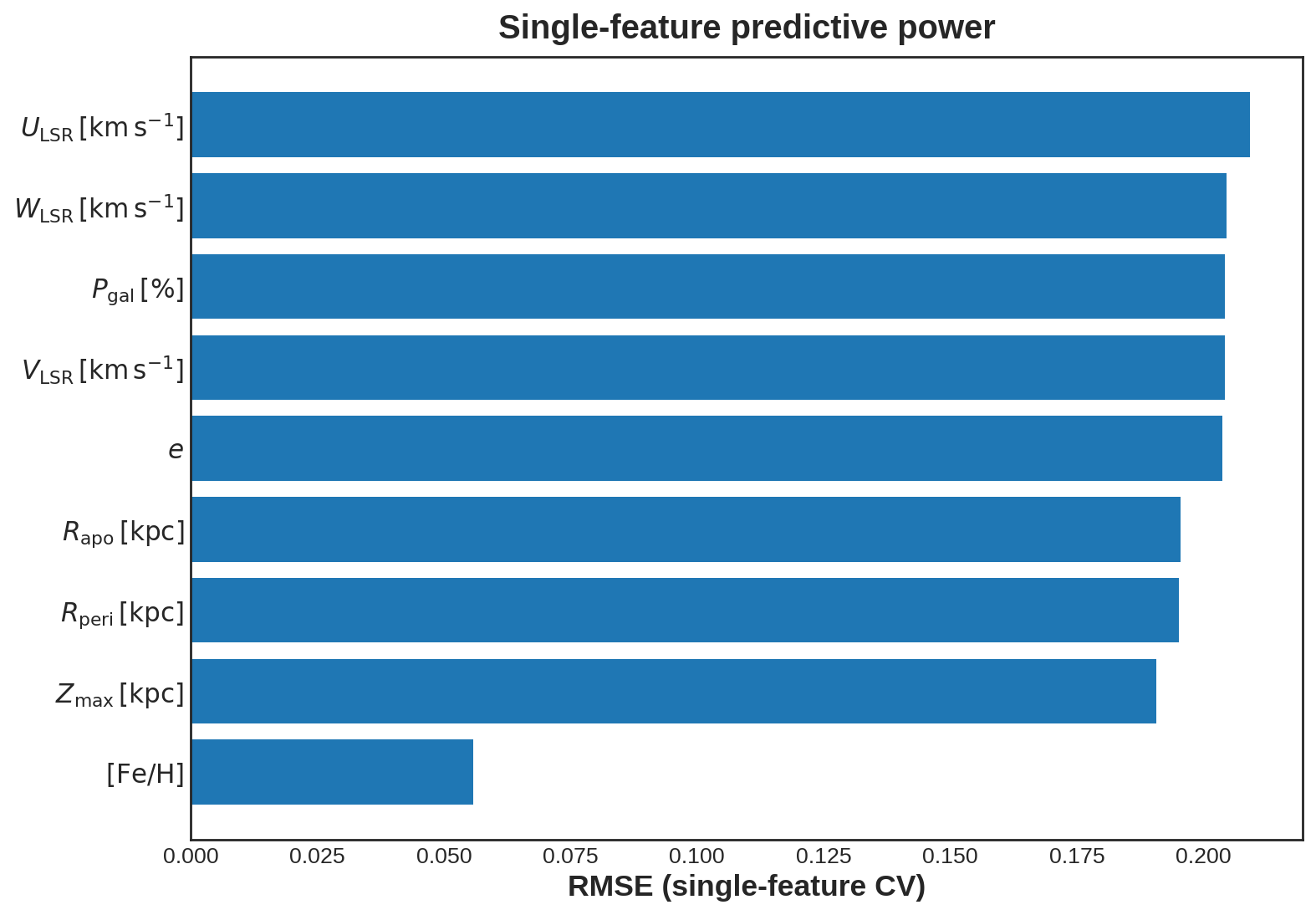}
\caption{Same as Fig.\ref{fig:feature_importance_ch} but for $\mathrm{[Si/H]}$.}
\label{fig:feature_importance_sih}
\end{figure*}

\section{Predicting C and O abundances from Mg, Si and Fe}
\label{appendix:CO_from_MgSiFe}

This appendix presents the full set of feature-importance figures discussed in Sect.~\ref{main_CO_from_MgSiFe}. For completeness, we show the results of the built-in gain importance, permutation importance, SHAP global importance, and single-feature predictive performance for the LightGBM model.

\begin{figure*}[ht]
\centering
\includegraphics[width=0.24\textwidth]{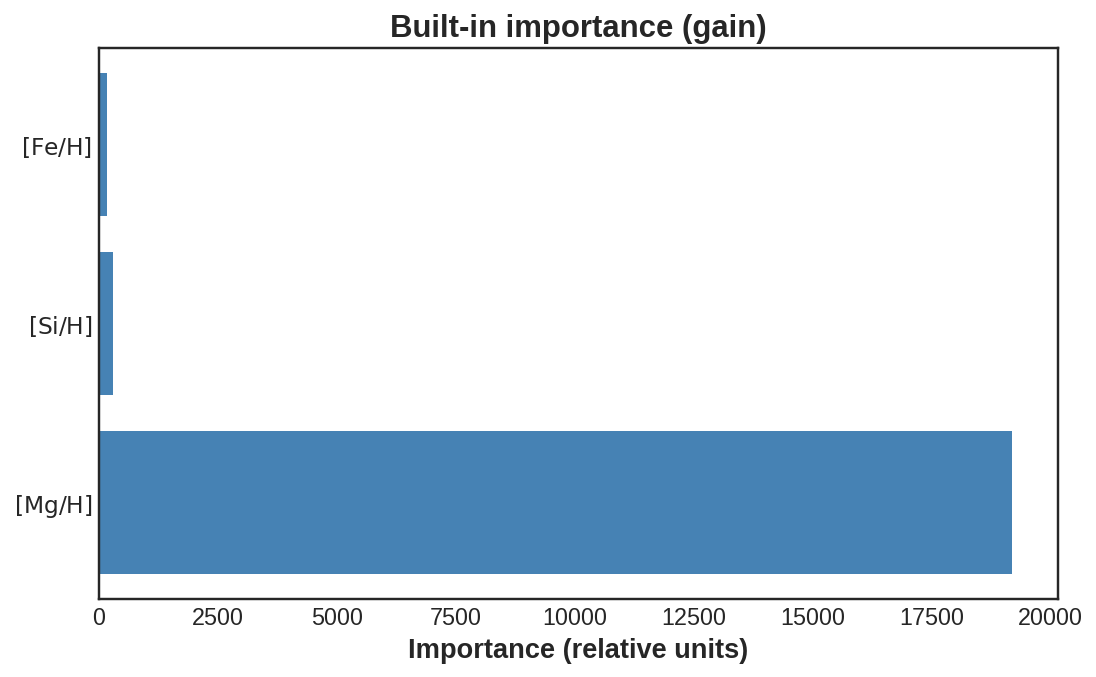}
\includegraphics[width=0.24\textwidth]{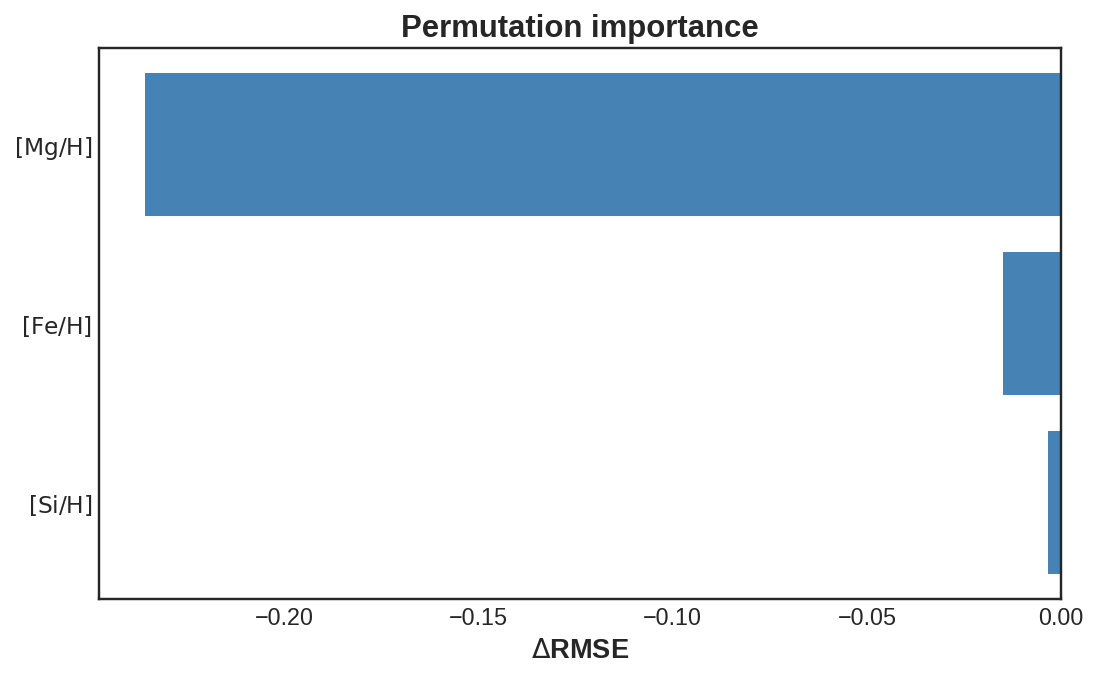}
\includegraphics[width=0.24\textwidth]{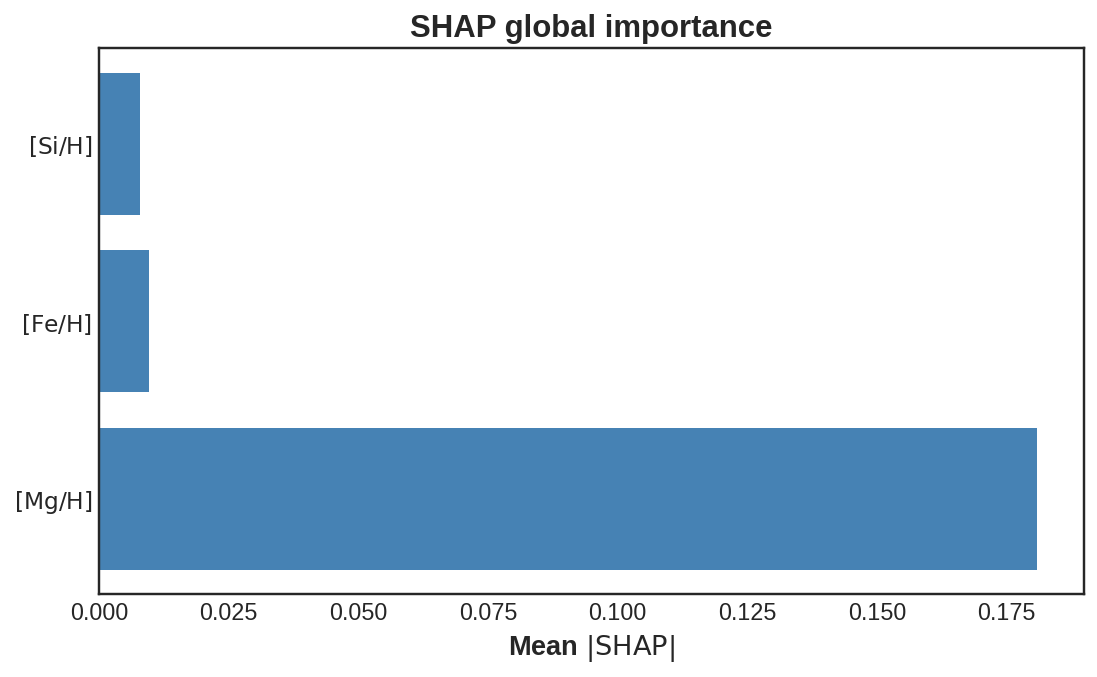}
\includegraphics[width=0.24\textwidth]{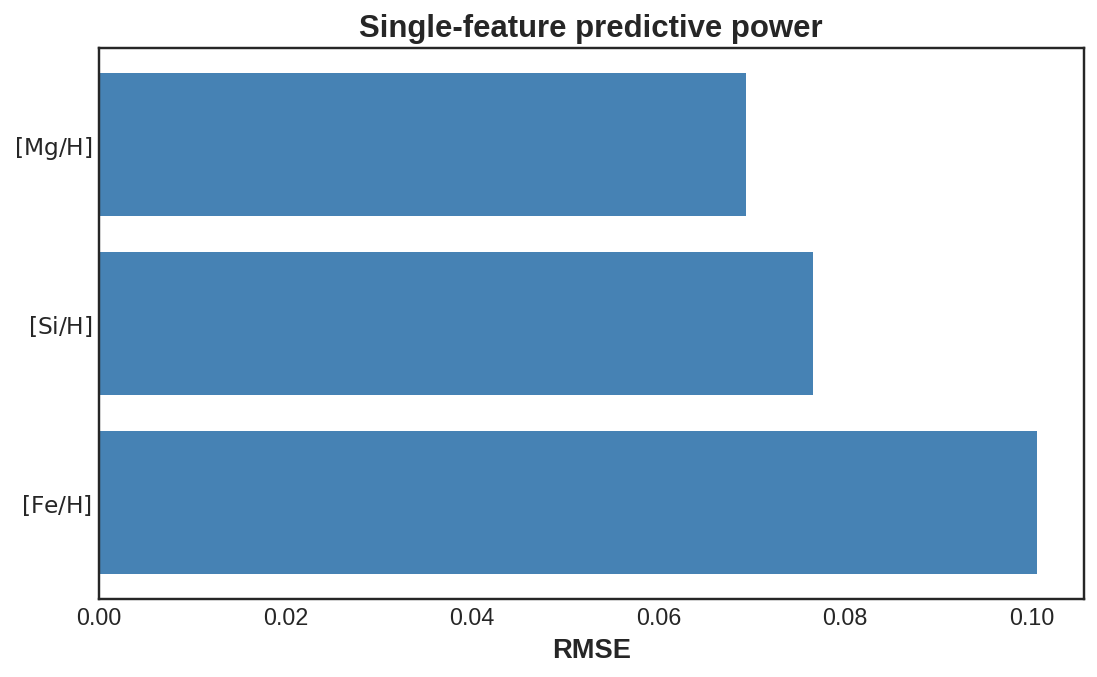}
\caption{Feature-importance diagnostics for the tuned LightGBM model predicting $\mathrm{[C/H]}$ using Mg, Si, and Fe abundances. Built-in gain feature importance derived from the gradient boosting trees (top left). Permutation feature importance, measured as the increase in prediction error ($\Delta$RMSE) after randomly shuffling each feature (top right). Global SHAP importance, showing the mean absolute SHAP value for each feature (bottom left). Predictive performance (RMSE from cross-validation) obtained when training the model using only a single feature at a time (bottom right).}
\label{fig:feature_importance_ch_frommgsife}
\end{figure*}

\begin{figure*}[ht]
\centering
\includegraphics[width=0.24\textwidth]{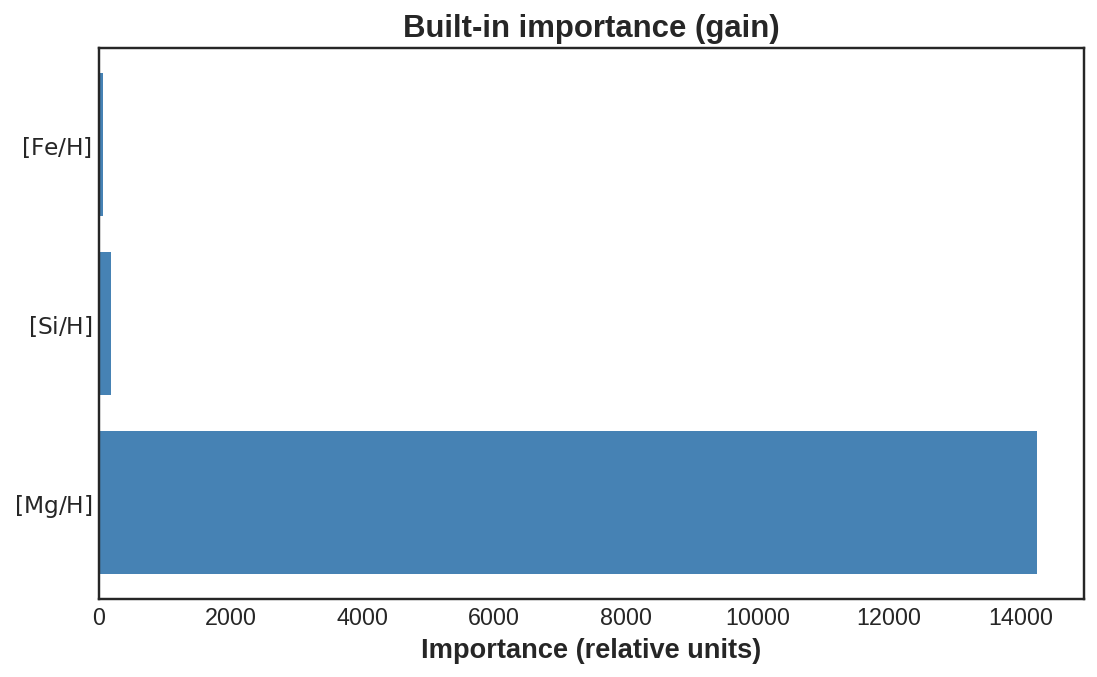}
\includegraphics[width=0.24\textwidth]{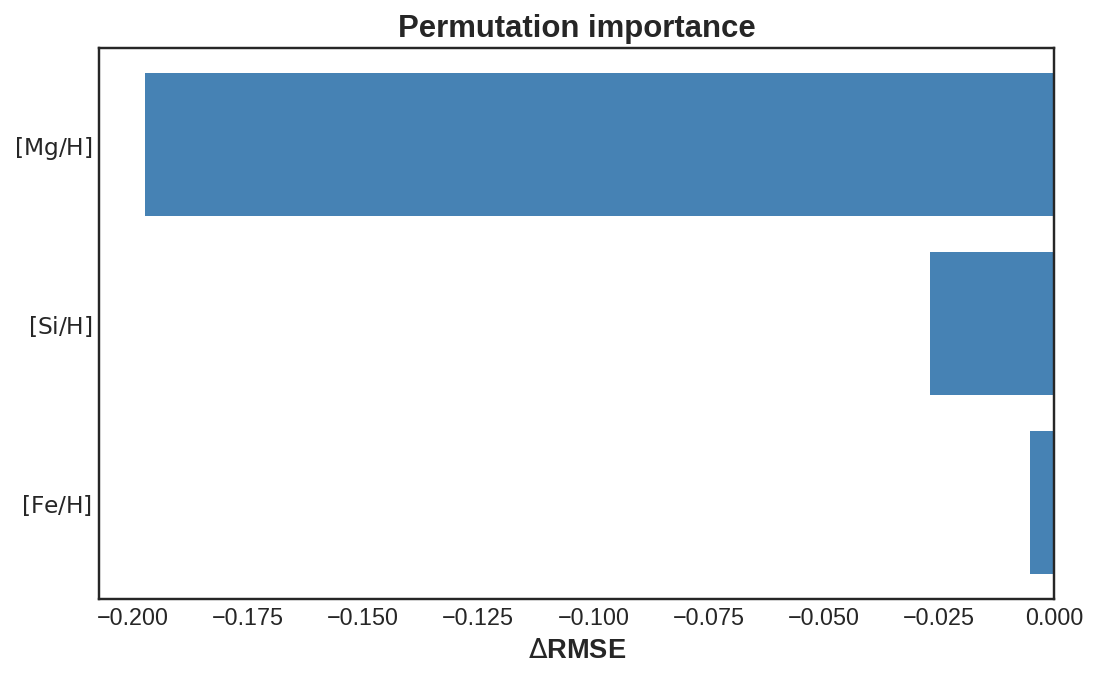}
\includegraphics[width=0.24\textwidth]{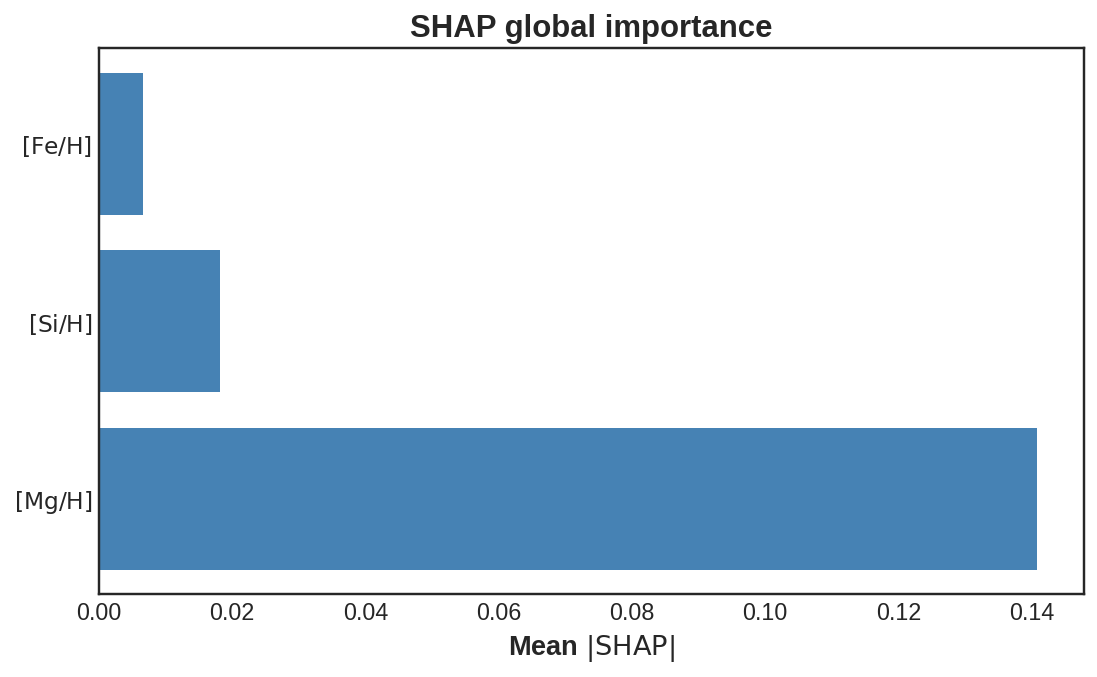}
\includegraphics[width=0.24\textwidth]{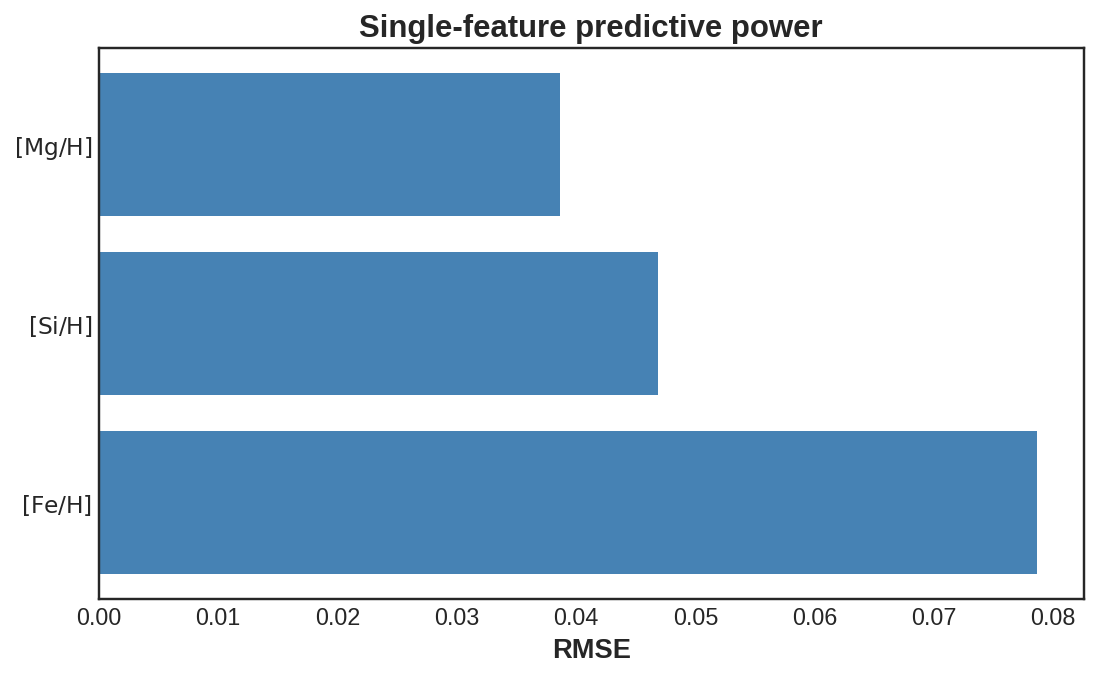}
\caption{Same as Fig.~\ref{fig:feature_importance_ch_frommgsife}, but for $\mathrm{[O/H]}$.}
\label{fig:feature_importance_oh_frommgsife}
\end{figure*}

\end{appendix}
\end{document}